\documentclass[10pt]{article}
 \usepackage{graphicx,amsfonts,amsmath,amssymb,latexsym}
 \renewcommand{\title}[1] {%
 \begingroup\begin{center}\vspace{0.0cm}\bf\Large
 \addtolength{\baselineskip}{1mm} #1 \end{center}\endgroup}

 \renewcommand{\author}[1] {%
 \begingroup\begin{center}\vspace{0.2cm}\bf #1 \vspace{0.2cm}
 \end{center}\endgroup}

 \newcommand{\address}[1] {%
 \begingroup\begin{center} #1 \end{center}\endgroup}

 \newtheorem{thm}{Theorem}[section]
 \newtheorem{defin}[thm]{Definition}
 
 \newtheorem{prop}[thm]{Proposition}

 \newcommand\sigmamn{\left(\begin{array}{cc}
 0 & 0 \\ 1 & 0
 \end{array}\right)}
 \newcommand\sigmapl{\left(\begin{array}{cc}
 0 & 1 \\ 0 & 0
 \end{array}\right)}

 \newcommand\bz{\bar{z}}
 \newcommand\ds{\displaystyle}
 \newcommand\hanu{\hat{H}^{(a,\nu)}}
 
 \newcommand\Zb{\mathbb{Z}}
 \newcommand\Rb{\mathbb{R}}
 \newcommand\Cb{\mathbb{C}}

 \newcommand\ben{\begin{equation*}}
 \newcommand\ebn{\end{equation*}}
 \newcommand\be{\begin{equation}}
 \newcommand\eb{\end{equation}}

 \numberwithin{equation}{section}

\begin{document}
 \title{On Painlev\'e VI transcendents \\
 related to the Dirac operator on the hyperbolic disk}
 \author{O. Lisovyy\footnote{On leave from Bogolyubov Institute for Theoretical Physics, 03680, Kyiv, Ukraine}}
 \address{Laboratoire de Math\'ematiques et Physique Th\'eorique CNRS/UMR 6083,\\
 Universit\'e de Tours, Parc de Grandmont, 37200 Tours, France}
 \date{}

 \begin{abstract}
 Dirac hamiltonian on the Poincar\'e disk in the presence of
 an Aha\-ro\-nov-Bohm flux and a uniform magnetic field admits a one-parameter
 family of self-adjoint extensions. We determine the spectrum and calculate the
 resolvent for each element of this family.
 Explicit expressions for  Green functions are then used to find
 Fredholm determinant representations for
 the tau function of the Dirac operator with two branch points on the Poincar\'e disk.
 Isomonodromic deformation theory for the Dirac equation relates this tau function
 to a one-parameter class of solutions of the Painlev\'e~VI equation with $\gamma=0$. We analyze long
 distance behaviour of the tau function, as well as the asymptotics of the corresponding
 Painlev\'e~VI transcendents as $s\rightarrow 1$.
 Considering the
 limit of flat space, we also obtain a class of
 solutions of the Painlev\'e~V equation with $\beta=0$.
 \end{abstract}

 \tableofcontents

 \section{Introduction}
 It has  been known since \cite{mtw,wmtb} that the two-point correlation
 function of the 2D Ising model in the scaling limit is expressible in
 terms of a solution of a Painlev\'e~III equation. This remarkable
 result turned out to be a special case of a more general phenomena,
 described by Sato-Miwa-Jimbo (SMJ) theory of holonomic
 quantum fields and monodromy preserving deformations of the Dirac
 equation \cite{smj}. One of the central
 objects in SMJ theory is the $\tau$-function of the Dirac
 operator acting on a suitable class of multivalued functions on
 the Euclidean plane.

 The SMJ $\tau$-function admits a geometric
 interpretation \cite{pacific,ising_book}. Loosely speaking, it can be
 obtained by trivializing the $\mathrm{det}^*$-bundle over an
 infinite-dimensional grassmannian, composed of the spaces of boundary values of
 certain local solutions of the Dirac equation, where different
 points of the grassmannian correspond to different positions of
 the
 branch points on the plane. The same idea was earlier used in
 \cite{cauchy} to show that the $\tau$-function of the
 Schlesinger system can be interpreted as a determinant of a singular
 Cauchy-Riemann operator. A simple finite-dimensional
 example of this construction arises in the study of
 a one-dimensional Laplacian with $\delta$-interactions \cite{lisovyy_LMP}.
 The most important thing about the geometric picture is that it
 allows to find an explicit representation of the SMJ
 $\tau$-function in terms of a Fredholm determinant, thereby
 giving
 a solution of the deformation equations.

 The simplest way to generalize the above setup is to replace the
 plane by an infinite cylinder. In this case, the deformation
 equations and a Fredholm determinant representation for the
 $\tau$-function of the corresponding Dirac operator were obtained
 in \cite{lisovyy_CMP}. These results provide a shortcut
 derivation of the PDEs satisfied by the scaled Ising correlation functions on the
 cylinder \cite{lisovyy_ATMP} and of the exact expressions for the  one- and two-particle
 finite-volume
 form factors of the Ising spin and disorder field \cite{PLA,TMF,fz} (and, more
 generally, of twist fields \cite{doyon_ff}).

 In \cite{beatty}, Palmer, Beatty and Tracy (PBT) have extended
 the SMJ analysis of isomonodromic deformations to the case of a
 Dirac operator on the Poincar\'e disk (see also earlier works \cite{narayanan,pt}
 on this subject). The associated
 $\tau$-function in the simplest nontrivial case of two branch points
 was shown to be related \cite{beatty} to a solution of
 the Painlev\'e~VI equation
 \begin{eqnarray}
 \label{pvi}\frac{d^2w}{ds^2}&=&\frac{1}{2}\left(\frac{1}{w}+\frac{1}{w-1}+
  \frac{1}{w-s}\right)\left(\frac{dw}{ds}\right)^2 -
  \left(\frac{1}{s}+\frac{1}{s-1}+\frac{1}{w-s}\right)\frac{dw}{ds}\,+\\
 \nonumber &\;&+\,
  \frac{w(w-1)(w-s)}{s^2(s-1)^2}\left(\alpha+\frac{\beta s}{w^2}+\gamma\frac{s-1}{(w-1)^2}
  +\delta\frac{s(s-1)}{(w-s)^2}\right)\,
  \end{eqnarray}
 with only one fixed parameter.
 Before stating the PBT result in more detail, it is useful to reformulate it in
 a slightly different way. The Dirac operator, considered in
 \cite{beatty},  is simply related to the hamiltonian of a massive
 Dirac particle moving on the Poincar\'e disk in the superposition
 of a uniform magnetic field $B$ and the field of two
 non-integer Aharonov-Bohm (AB) fluxes $\Phi_{1,2}=2\pi\nu_{1,2}$
 located at the points $a_1$ and $a_2$. Without any loss of generality, one may choose
 $-1<\nu_{1,2}<0$. It is preferable
 to work with the hamiltonian, as it is a symmetric
 operator that can be made self-adjoint after a proper
 specification of the domain, and many assertions of \cite{beatty}
 (e.~g. symmetry of the Green function, non-existence of certain
 global solutions of the Dirac equation, etc.) immediately follow from the
 self-adjointness.

  Write the disk curvature as $-4/R^2$, denote by
 $m$ and $E$ the particle mass and energy, and introduce two
 dimensionless parameters $ b=\frac{BR^2}{4}$, $\mu=\frac{\sqrt{(m^2-E^2)R^2+4b^2}}{2}$.
 It turns out that the $\tau$-function associated to the above hamiltonian depends
 only on the geodesic distance $d(a_1,a_2)$  between the
 points $a_1$ and $a_2$.
  If we further introduce $s=\tanh^2\frac{d(a_1,a_2)}{R}$, then
  it can be expressed  \cite{beatty} in terms of a solution of the Painlev\'e~VI equation (\ref{pvi})
 \be\label{taurelpvi}
 \frac{d\;}{ds}\ln\tau(s)=\frac{s(1-s)}{4w(1-w)(w-s)}
 \left(\frac{dw}{ds}-\frac{1-w}{1-s}\right)^2-
 \frac{1-w}{1-\,s\,}\left(\frac{\lambda^2}{4s}-\frac{\tilde{\lambda}^2}{4w}+\frac{\mu^2}{w-s}\right)
 \eb
  where
 $\lambda=\nu_2-\nu_1$, $\tilde{\lambda}=2+\nu_1+\nu_2-2b$, and the values of the PVI parameters
 are given by
 \be\label{pvipars}
 \alpha=\frac{\lambda^2}{2},\qquad \beta=-\frac{(\tilde{\lambda}-1)^2}{2},\qquad
 \gamma=0, \qquad \delta= \frac{1-4\mu^2}{2}.
 \eb
 Actually, the paper \cite{beatty} is concerned with the case
 $E=0$ (the mass term in the Dirac operator is not of the most
 general form). We include this parameter from the very beginning
 because it will be shown below that the final answer for the $\tau$-function
 depends on $E$ only via the variable $\mu$.

  The aim of the present study is to solve the remaining part of the problem
  --- that is, to compute the PBT $\tau$-function and to
 investigate its asymptotic behaviour, which can be used  to specify the
 appropriate initial conditions for the equation (\ref{pvi}).
 We summarize our results in the following theorem:
 \begin{thm}\label{mainthm}
 The  PBT tau function admits Fredholm determinant representation
 \be\label{frere}
 \tau(s)=
 \mathrm{det}\Bigl(\mathbf{1}-L_{\nu_2,s}L'_{\nu_1,s}\Bigr),
 \eb
 where  the kernels of integral operators $L_{\nu,s}$, $L'_{\nu,s}$
 are
 \begin{eqnarray}
 \label{lnus01}
 \nonumber L_{\nu,s}(p,q)&=&e^{i(p-q)l_s/2}\;\sqrt{\rho(p)\rho(q)}\;\mathcal{F}_{\nu}(p,q)\,,\\
 \label{lnus02}
  \nonumber L'_{\nu,s}(p,q)&=&L_{\nu,s}(-p,-q)\,,
 \end{eqnarray}
 $l_s=\mathrm{arctanh}\sqrt{s}=\frac{d(a_1,a_2)}{R}$ and $ p,q\in\Rb$.
 The functions $\rho(p)$, $\mathcal{F}_{\nu}(p,q)$ are given by
  \begin{eqnarray}
 \label{ff001a}\nonumber\rho(p)&=&\frac{2^{2\mu}\;\Gamma(1+2\mu)}{\Gamma\left(\mu+\frac{1}{2}+\frac{ip}{2}\right)
 \Gamma\left(\mu+\frac{1}{2}-\frac{ip}{2}\right)}\,, \\
 \label{ff001b}\nonumber \mathcal{F}_{\nu}(p,q)&=&\frac{\sin\pi\nu}{2\pi^2}
 \int\limits_{-\infty}^{\infty}d\theta
 \int\limits_{0}^{\pi/2}dx
 \int\limits_{0}^{\pi/2}dy\;\;\frac{
 e^{\left(1+\nu+\frac{1-2b}{2}\right)\left(\theta-2i(x-y)\right)}}{
 \left(e^{\theta-2i(x-y)}+1\right)\left(2+2\cosh\theta\right)^{\frac{1+2\mu}{2}}}\;
 \times
 \\
 \nonumber&\;&\qquad\qquad\times\;  (\sin x)^{\mu+\frac{ip}{2}-\frac12}
 (\cos x)^{\mu-\frac{ip}{2}-\frac12}
 (\sin y)^{\mu-\frac{iq}{2}-\frac12}
 (\cos y)^{\mu+\frac{iq}{2}-\frac12}.
 \end{eqnarray}
 To leading order the long-distance ($s\rightarrow1$) asymptotics of $\tau(s)$ is
 \be
 \nonumber 1-\tau(s)\simeq  A_{\tau}(1-s)^{1+2\mu}+O\left((1-s)^{2+2\mu}\right)\;\text{ as
 }s\rightarrow1,
 \eb
 where the coefficient $A_{\tau}$ is given by
 \begin{eqnarray}
 \label{tauas}A_{\tau}&=&\frac{\sin\pi\nu_1\sin\pi\nu_2}{\pi^2}
 \frac{
 \Gamma(\mu+2+\nu_1-b)
 \Gamma(\mu-\nu_1+b)
 \Gamma(\mu+2+\nu_2-b)
 \Gamma(\mu-\nu_2+b)}{\bigl[\Gamma(2+2\mu)\bigr]^2}.
 \end{eqnarray}
 \end{thm}
 A few remarks are in order. The formula (\ref{tauas}) implies
 that the asymptotic behaviour of the corresponding PVI transcendent
 for $\mu>1/2$ is
 \begin{eqnarray}
 \nonumber &\;&1-w(s)\simeq A (1-s)^{1+2\mu}+
 O\left((1-s)^{2+2\mu}\right)\;\text{ as
 }s\rightarrow1,\\
 \nonumber &\;& A=\frac{(1+2\mu)^2}{(\mu+1+\nu_1-b)(\mu+1+\nu_2-b)}\,A_{\tau}.
 \end{eqnarray}
 The transformation $\nu_1\mapsto\nu_1+c$,
 $\nu_2\mapsto\nu_2+c$, $b\mapsto b+c$, $\mu\mapsto\mu$ does not change the
 values (\ref{pvipars}) of the PVI parameters $\alpha$, $\beta$, $\gamma$, $\delta$.
 However, our solution nontrivially depends on all four variables $\nu_1$,
 $\nu_2$, $b$, $\mu$ as can be seen from the above asymptotics.
 Thus we have constructed a
 one-parameter family of solutions of the Painlev\'e~VI equation with
 $\gamma=0$.


 We also note that  in the case $b=0$ the PBT $\tau$-function
 was conjectured to coincide with the correlation function
 $\langle \mathcal{O}_{1+\nu_1}(a_1)\mathcal{O}_{1+\nu_2}(a_2)\rangle$
 of $U(1)$ twist fields in the theory of free massive Dirac fermions on the Poincar\'e
 disk \cite{doyon} (particular case of Ising monodromy was later studied
 in more detail in \cite{doyon2,doyon_fonseca}). The asymptotics of $\tau(s)$ as $s\rightarrow0$ then
 follows from the known flat-space OPEs for twist fields. Long-distance
 behaviour  is determined by a form factor
 expansion of $\langle
 \mathcal{O}_{1+\nu_1}(a_1)\mathcal{O}_{1+\nu_2}(a_2)\rangle$.
 The method of angular quantization, employed in
 \cite{doyon} for the calculation of form factors,
 does not seem to work quite well, as it leads to formally divergent expressions.
 A sensible answer for the infrared asymptotics
 was nevertheless extracted from them after a number of
 regularization procedures. The formula (\ref{tauas}), specialized
 to the case $b=0$, proves the latter result.

  The main technical problem arising in the direct computation of $\tau(s)$
 is the unknown formula for the Green function of the Dirac hamiltonian
 on the disk in the presence of a uniform magnetic field and one AB
 vortex. Such hamiltonian can always be made commuting with the
 angular momentum operator by a suitable choice of the gauge. Partial Green
 functions are then
 calculated relatively easily in each channel with fixed angular momentum;
 the difficult part is the summation of these
 partial contributions to a closed-form expression.
 We have solved this problem
 by writing radial solutions of the Dirac equation
 as Sommerfeld-type superpositions of horocyclic waves, similarly
 to a simpler scalar case \cite{lisovyy_JMP}.
 This allows to perform the summation and to obtain a simple integral representation for
 the one-vortex Green function.

 This paper is organized as follows. In Section~2, after introducing
 basic notation, we describe the solutions of the radial Dirac
 equation and compute radial Green functions. Spectrum,
 self-adjointness and admissible boundary conditions for the full
 Dirac hamiltonian are also briefly analyzed. Finally, we write contour
 integral representations for the radial solutions and obtain
 a compact formula for the one-vortex resolvent. The
 PBT $\tau$-function is studied in Section~3. We start by giving a general
 definition of the $\tau$-function in terms of the projections on
 some boundary spaces. Next we introduce coordinates in these
 spaces, using the solutions of the Dirac equation on the
 Poincar\'e strip. Explicit formulas for the projections in these
 coordinates are obtained in Subsection~\ref{fdras} by analyzing the
 asymptotics of the Green function, computed in Section~2. These formulas
 give the  kernels of integral operators in the Fredholm determinant
 representation of $\tau(s)$. In Subsection~\ref{pbtpvi}, we recall the
 relation of the PBT $\tau$-function
 to Painlev\'e VI equation.
 Subsection~\ref{ldass} deals with the derivation of the long-distance
 asymptotics of $\tau(s)$. The analogs of the above results in the
 limit of flat space, where Painlev\'e~VI equation transforms into
 Painlev\'e~V, are established in Section~4.
 The Appendix contains a proof of the fact that the
 $\tau$-function depends only on the geodesic distance.

 \section{\label{onevortex} One-vortex Dirac hamiltonian on the Poincar\'e disk}
 \subsection{Preliminaries}
 Let us first establish our notations. We denote by $D$
 the unit disk $|z|^2<1$ in the complex $z$-plane, endowed with
 the Poincar\'e metric
 \be\label{pmetric}
 ds^2=g_{z\bz}\,dz\,
 d\bz=R^2\frac{dz\,d\bz}{\left(1-|z|^2\right)^2}\,.
 \eb
 This metric has a constant negative Gaussian curvature $-4/R^2$ and
 is invariant with respect to the natural $SU(1,1)$-action on $D$:
  \be\label{su11}
 z\mapsto z_g(z)=\frac{\alpha z+\beta}{\bar{\beta}z+\bar{\alpha}}\,,\qquad
 g=\left(\begin{array}{cc}
 \alpha & \beta \\ \bar{\beta} & \bar{\alpha}
 \end{array}\right)\in SU(1,1).
 \eb

 The hamiltonian of a Dirac particle of unit charge moving on the
 Poincar\'e disk in an external magnetic field has the form
 \be\label{dirac_ham}
 \hat{H}=
 \left(\begin{array}{cc}
 m & K \\ K^* & \!\!-m
 \end{array}\right),
 \eb
 where the operator $K$ and its formal adjoint $K^*$
 are given by
 \begin{eqnarray}
 \label{dirac_k1}
 K&=&\frac{1}{\sqrt{g_{z\bz}}}\,\left\{2D_z+
 \frac12\,\partial_z\ln g_{z\bz}\right\}, \\
 \label{dirac_k2}
 K^*&=&-\frac{1}{\sqrt{g_{z\bz}}}\,\left\{2D_{\bz}+
 \frac12\,\partial_{\bz}\ln g_{z\bz}\right\},
 \end{eqnarray}
 and $D_z=\partial_z+iA_z$, $D_{\bz}=\partial_{\bz}+iA_{\bz}$
 denote the covariant derivatives.

 Connection 1-form
 $\mathcal{A}=A_z\,dz+A_{\bz}\,d\bz$, which is considered in the present
 section, consists of two
 parts. Namely, we put
 $\mathcal{A}= \mathcal{A}^{(B)}+\mathcal{A}^{(\nu)}$, where
 \begin{eqnarray}
 \label{vect_b}
 \mathcal{A}^{(B)}&=&-\frac{i{BR}^{\,2}}{4}\,\frac{\bz\, dz -
 z\,d\bz}{1-|z|^2}\,,\\
 \label{vect_nu}
 \mathcal{A}^{(\nu)}&=&-\frac{i\nu}{2}\left(\frac{dz}{z}-\frac{d\bz}{\bz}\right).
 \end{eqnarray}
 The first contribution describes a uniform magnetic field of
 intensity $B$, since $d\mathcal{A}^{(B)}$
 is proportional to the volume form
 $\displaystyle d\mu=\frac{i}{2}\,g_{z\bz}\,dz\wedge d\bz$. The
 second part corresponds to the vector potential of an
 AB flux $\Phi=2\pi\nu$, situated at the disk center.

 Introducing polar coordinates  $z=re^{i\varphi}$,
 $\bz=re^{-i\varphi}$, one can explicitly rewrite the operators
 $K$ and $K^*$ as follows:
 \begin{eqnarray}
 \label{k1_polar}
 K&=&\;\frac{\;e^{-i\varphi}}{R}\left[(1-r^2)\left(
 \partial_r-\frac{i}{r}\partial_{\varphi}+\frac{\nu}{r}\right)+(1+2b)r\right],\\
 \label{k2_polar}
 K^*\!&=&-\frac{\;e^{i\varphi}}{R}\left[(1-r^2)\left(
 \partial_r+\frac{i}{r}\partial_{\varphi}-\frac{\nu}{r}\right)+(1-2b)r\right].
 \end{eqnarray}
 Here, we have introduced a dimensionless parameter $\displaystyle
 b=\frac{BR^2}{4}$ characterizing the ratio of magnetic field and
 the disk curvature.

 \subsection{Radial hamiltonians and self-adjointness}
 Since the formal hamiltonian (\ref{dirac_ham}), corresponding to the vector
 potential $ \mathcal{A}^{(B)}+\mathcal{A}^{(\nu)}$, commutes with
 the angular momentum operator
 $\displaystyle \hat{L}=-i\partial_{\varphi}+\frac12\,\sigma_{z}$,
 we will attempt to diagonalize them simultaneously. The
 eigenvalues of $\hat{L}$ are half-integer numbers
 $\displaystyle l_0+\frac12$ ($l_0\in\Zb$) and the appropriate
 eigenspaces are spanned by the spinors of the form
 \ben
 w_{l_0}(r,\varphi)=\left(\begin{array}{c}
 w_{l_0,1}(r)e^{il_0\varphi} \\
 w_{l_0,2}(r)e^{i(l_0+1)\varphi}\end{array}\right).
 \ebn
 The action of
 $\hat{H}$ leaves these eigenspaces invariant. One has
 \be\label{formal_Hl}
 \left(\begin{array}{c} w_{l_0,1}(r) \\ w_{l_0,2}(r)\end{array}\right)\mapsto
 \hat{H}_{l_0+\nu}  \left(\begin{array}{c} w_{l_0,1}(r) \\
 w_{l_0,2}(r)\end{array}\right),\qquad
 \hat{H}_l= R^{-1}\left(\begin{array}{cc}
 mR & K_l \\ K^*_l & \!\!-mR
 \end{array}\right),
 \eb
 where the operators $K_l$
 and $K^*_l$  are explicitly given by
 \begin{eqnarray}
 \label{kl} K_l&=&(1-r^2)\Bigl(
 \partial_r+\frac{l+1}{r}\Bigr)+(1+2b)r,\\
 \label{kel} K^*_l\!&=&-(1-r^2)\Bigl(
 \partial_r-\frac{l}{r}\Bigr)-(1-2b)r.
 \end{eqnarray}

 Let us make several remarks concerning the solutions of the
 radial Dirac equation
 \be\label{rad_de}
 (\hat{H}_l-E)w_l(r)=0\,.
 \eb
 It will be assumed that $E\in \Cb\backslash\Bigl((-\infty,-m]\cup [m,\infty)\Bigr)$ and $l$ is an
 arbitrary real parameter. We also introduce for further convenience the following
 quantities:
 \ben
 \mu=\frac{\sqrt{(m^2-E^2)R^2+4b^2}}{2}\,,
 \ebn
 \ben
 C_{\pm}=\left(\frac{m-E}{m+E}\right)^{1/4}
 \left(\frac{\mu+b}{\mu-b}\right)^{\pm 1/4}.
 \ebn
 All fractional powers in these formulas are defined so that $\mu$
 and $C_{\pm}$ are real and positive for real values of $E$
 satisfying $|E|<m$.

 Let us first look at the space of solutions of (\ref{rad_de})
 on the open unit interval $I=(0,1)$,
 which are square integrable  in the
 vicinity of the point $r=1$ with respect to the measure
 $\displaystyle d\mu_r=R^2\frac{rdr}{(1-r^2)^2}$, induced by the Poincar\'e
 metric. It is a simple matter to check that for any $l\in\Rb$
 this space is one-dimensional (i.~e. the singular point $r=1$ is
 of the limit point type) and is generated by the function
 \begin{eqnarray}
 \nonumber
 w_l^{(I)}(r)
 &=&\frac{\sqrt{\mu^2-b^2}}{2\mu}\,
 \frac{\Gamma(\mu-b)\Gamma(\mu+b)}{\Gamma(2\mu)}\;\;\times
 \\
 \label{wsqis0}&\times&(1-r^2)^{\frac{1+2\mu}{2}}\left(
 \begin{array}{c}
 C_+^{-1}\, r^{-l}\,{}_2F_1\left(\mu-b+1,\mu+b-l,1+2\mu,1-r^2\right)\vspace{0.1cm}\\
 C_+ r^{-l-1}\,{}_2F_1\left(\mu-b,\mu+b-l,1+2\mu,1-r^2\right)
 \end{array}\right)=\qquad\\
 \nonumber&=& \frac{\sqrt{\mu^2-b^2}}{2\mu}\,
 \frac{\Gamma(\mu-b)\Gamma(\mu+b)}{\Gamma(2\mu)}\;\;\times
 \\
 \label{wsqis}&\times& (1-r^2)^{\frac{1+2\mu}{2}} \left(
 \begin{array}{c}
 C_+^{-1}\, r^l\,{}_2F_1\left(\mu+b,\mu-b+1+l,1+2\mu,1-r^2\right)\vspace{0.1cm}\\
 C_+ r^{l+1}\,{}_2F_1\left(\mu+b+1,\mu-b+1+l,1+2\mu,1-r^2\right)
 \end{array}\right).
 \end{eqnarray}
 If $l\in(-\infty,-1]\cup[0,\infty)$, the limit point case is also
 realized at $r=0$. The solution that satisfies the condition of square integrability
 in the vicinity of $r=0$ can be written as
  \begin{eqnarray}
  \nonumber
 w_l^{(II,+)}(r)
 &=& \left(\begin{array}{cc}
 \displaystyle \frac{\Gamma(\mu-b+1+l)}{\Gamma(1+l)\Gamma(\mu-b+1)} & 0 \\
 0 &  -\displaystyle \frac{\Gamma(\mu-b+1+l)}{\Gamma(2+l)\Gamma(\mu-b)}
 \end{array}\right)\times
 \\
 \label{wplus}&\times& (1-r^2)^{\frac{1+2\mu}{2}}\left(
 \begin{array}{c}
 C_+^{-1}\, r^{l}\,{}_2F_1\left(\mu+b,\mu-b+1+l,1+l,r^2\right)\vspace{0.1cm}\\
 C_+ r^{l+1}\,{}_2F_1\left(\mu+b+1,\mu-b+1+l,2+l,r^2\right)
 \end{array}\right),\\
 \nonumber
 w_l^{(II,-)}(r)
 &=& \left(\begin{array}{cc}
 \displaystyle \frac{\Gamma(\mu+b-l)}{\Gamma(1-l)\Gamma(\mu+b)} & 0 \\
 0 &  -\displaystyle \frac{\Gamma(\mu+b-l)}{\Gamma(-l)\Gamma(\mu+b+1)}
 \end{array}\right)\times
 \\
 \label{wmoins}&\times&(1-r^2)^{\frac{1+2\mu}{2}}\left(
 \begin{array}{c}
 C_+^{-1}\, r^{-l}\,{}_2F_1\left(\mu-b+1,\mu+b-l,1-l,r^2\right)\vspace{0.1cm}\\
 C_+ r^{-l-1}\,{}_2F_1\left(\mu-b,\mu+b-l,-l,r^2\right)
 \end{array}\right),
 \end{eqnarray}
 where the first formula corresponds to the case $l\geq0$  and
 the second one to $l\leq -1$. For $l\in(-1,0)$ both solutions (\ref{wplus}) and
  (\ref{wmoins}) are square integrable at $r=0$ (the limit circle case).

 The functions $w_l^{(I)}(r)$ and $w_l^{(II,+)}(r)$ are linearly
 independent for $l>-1$, and the functions $w_l^{(I)}(r)$ and $w_l^{(II,-)}(r)$
 are linearly independent for $l<0$. One can show this, for
 example, by computing the determinant of the fundamental matrix
 constructed from these solutions:
 \be\label{wronsk1}
 \mathrm{det}\left(w_{l}^{(I)}(r),w_{l}^{(II,\pm)}(r)\right)=
 -\frac{1}{\sqrt{\mu^2-b^2}}\,
 \frac{1-r^2}{r}\,.
 \eb
 This implies that for $E\in \Cb\backslash\Bigl((-\infty,-m]\cup
 [m,\infty)\Bigr)$ and $l\in (-\infty,-1]\cup[0,\infty)$ the
 equation (\ref{rad_de}) has no square integrable solutions on the
 whole interval $I$. For $l\in(-1,0)$, however, there is a
 one-dimensional space of such solutions, generated
 by the function (\ref{wsqis0})--(\ref{wsqis}).

 We now examine the issue of
 self-adjointness of the operators $\hat{H}_l$. The above remarks
 can be summarized as follows:
 \begin{prop}
 Let us restrict the domain of the formal radial hamiltonian
 $\hat{H}_l$, defined by (\ref{formal_Hl})--(\ref{kel}), to smooth functions
 with compact support in $I$. Then
 \begin{itemize}
 \item  $\hat{H}_l$ is essentially self-adjoint for $l\in
 (-\infty,-1]\cup[0,\infty)$;
 \item  for $l\in (-1,0)$ the operator $\hat{H}_l$ has deficiency indices $(1,1)$
 and admits a one-parameter family of self-adjoint extensions (SAEs).
 \end{itemize}
 \end{prop}

 Assume that $l\in (-1,0)$. Deficiency subspaces
 $\mathcal{K}_{\pm}=\mathrm{ker}\left(\hat{H}_l^{\dag}\mp im\right)$
 are generated by the elements
 \be\label{wpm}
 w_{\pm}(r)=w^{(I)}_{l}(r)\Bigl|_{E=\pm im}\Bigr..
 \eb
 Different SAEs $\hat{H}_{l}^{(\gamma)}$ are in one-to-one
 correspondence with the isometries between $\mathcal{K}_+$
 and~$\mathcal{K}_-$. They  may be labeled by a parameter
 $\gamma\in[0,2\pi)$
 and characterized by the domains
 \be\label{sae}
 \mathrm{dom}\,\hat{H}_{l}^{(\gamma)}=\left\{w_0+c\left(w_+ +e^{i\gamma}w_-\right)|
 w_0\in\mathrm{dom}\,\hat{H}_{l},\,c\in\Cb\right\}.
 \eb
 It is also conventional to characterize the functions from the domain
 of the closure of $\hat{H}_{l}^{(\gamma)}$ by their asymptotic
 behaviour near the point $r=0$. Namely, it follows from (\ref{wsqis}) and (\ref{wpm})--(\ref{sae}) that
  for $w\in\mathrm{dom}\left(\overline{\hat{H}_{l}^{(\gamma)}}\right)$
 one should have
 \be\label{asym0}
 \lim_{r\rightarrow0}\,
 \cos\left(\frac{\Theta}{2}+\frac{\pi}{4}\right)
 \left(mRr\right)^{-l}w_1(r)=
 -\lim_{r\rightarrow0}\,
 \sin\left(\frac{\Theta}{2}+\frac{\pi}{4}\right)
 \left(mRr\right)^{1+l}w_2(r).
 \eb
 Here, we have introduced instead of $\gamma$ a new self-adjoint
 extension parameter $\Theta\in[0,2\pi)$, defined by
 \begin{eqnarray}
 \nonumber\tan\left(\frac{\Theta}{2}+\frac{\pi}{4}\right)&=&
 \frac{2^{-l}}{\tan\left(\frac{\gamma}{2}-\frac{\pi}{8}\right)-1}\,
 \frac{\Gamma(-l)}{\Gamma(1+l)}\times
 \\
 \label{deftheta}&\;&\!\!\!\!\times\;
 \frac{\Gamma(\tilde{\mu}+b+1)\Gamma(\tilde{\mu}-b+l+1)}{\Gamma(\tilde{\mu}-b+1)
 \Gamma(\tilde{\mu}+b-l)}\,
 \sqrt{\frac{\tilde{\mu}-b}{\tilde{\mu}+b}}\,
 \left(\frac{mR}{\sqrt{2}}\right)^{-1-2l},
 \end{eqnarray}
 where $\displaystyle \tilde{\mu}=\mu\bigl|_{E=\pm
 im}\bigr.=\frac{\sqrt{2m^2R^2+4b^2}}{2}\,$. Note that the choice
 $\displaystyle \Theta=\frac{\pi}{2}$ ($\displaystyle -\frac{\pi}{2}$) is equivalent to
 requiring the regularity of the lower (resp. upper) component of the
 Dirac spinor at $r=0$.

 Let us now consider full Dirac hamiltonian. Since the shift of the
 AB flux by any integer number is equivalent to a
 unitary transformation of $\hat{H}$, hereafter we will assume that $-1<\nu\leq 0$.
 \begin{prop}
 Suppose that $\mathrm{dom}\,\hat{H}=C^{\infty}_0(D\backslash\{0\})$. Then
 \begin{itemize}
 \item for $\nu=0$ the operator $\hat{H}$ is essentially
 self-adjoint;
 \item for $\nu\in(-1,0)$ it has deficiency indices $(1,1)$ and
 admits a one-parameter family of SAEs, henceforth denoted by $\hat{H}^{(\gamma)}$,
 which correspond to those of the
 radial mode with $l_0=0$.
 \end{itemize}
 \end{prop}

 \subsection{\label{sect23}Radial Green functions}
 Let us begin with the case $l\in(-\infty,-1]\cup[0,\infty)$.
 Green function $G_{E,l}(r,r')$ of the radial hamiltonian
 $\hat{H}_l$ can be viewed as the solution of the equation
 \be\label{gfeq}
 \Bigl(\hat{H}_l(r)-E\Bigr)G_{E,l}(r,r')=
 \frac{(1-r^2)^2}{R^2 r\;}\,\delta(r-r')\,\mathbf{1}_2,
 \eb
 which is square integrable in the vicinity of the boundary points $r=0$ and
 $r=1$. Standard ansatz
 \be\label{gfans}
 G_{E,l}(r,r')=
  \begin{cases}
    A^{\,\pm}_{E,l}\;w^{(II,\pm)}_l(r)\otimes
    \Bigl( w^{(I)}_l(r')\Bigr)^T & \text{for}\;\; 0<r<r'<1, \\
    A^{\,\pm}_{E,l}\;w^{(I)}_l(r)\;\otimes
    \Bigl( w^{(II,\pm)}_l(r')\Bigr)^T & \text{for}\;\;
    0<r'<r<1
  \end{cases}
 \eb
 solves (\ref{gfeq}) for $r\neq r'$ and meets the requirements of
 square integrability. The sign ``$+$'' (``$-$'') in the above formula
 should be chosen for $l\geq0$ (resp. $l\leq-1$).
 Prescribed singular behaviour of the Green function at
 the point $r=r'$ is equivalent to
 the condition
 \ben
 G_{E,l}(r+0,r)-G_{E,l}(r-0,r)=-\frac{i}{R}\frac{1-r^2}{r}\;\sigma_y\,.
 \ebn
 Substituting (\ref{gfans}) into the last relation, we may rewrite
 it as follows
 \ben
 A^{\,\pm}_{E,l}\cdot\det
 W\left(w_{l}^{(I)}(r),w_{l}^{(II,\pm)}(r)\right)=-\frac{1-r^2}{Rr}\;.
 \ebn
 Finally, using (\ref{wronsk1}) one finds that
 \ben
 \displaystyle
 A_{E,l}^{\,+}=A_{E,l}^{\,-}=\frac{\sqrt{m^2-E^2}}{2}\,.
 \ebn

 Now suppose that $l\in(-1,0)$. In order to find the resolvent of the radial hamiltonian
 $\hat{H}_{l}^{(\gamma)}$, we need a solution of the radial
 Dirac equation $\displaystyle (\hat{H}_{l}^{(\gamma)}-E)w^{(\gamma)}_{l}=0$, which
 satisfies the boundary condition (\ref{asym0}) at the point $r=0$
 (square integrability near the point $r=1$ is not
 required). Such a solution can always be represented as a linear combination of the
 functions $w^{(II,\pm)}_{l}(r)$, defined by
 (\ref{wplus})--(\ref{wmoins}). These functions have the following
 asymptotic behaviour as $r\rightarrow0$:
 \begin{eqnarray}
 \label{asympt1} w^{(II,+)}_{l}(r)&=&
 \frac{\Gamma(\mu-b+1+l)}{\Gamma(1+l)\Gamma(\mu-b+1)}
 \left(
 \begin{array}{c}
 C_+^{-1} r^l \\ 0
 \end{array}\right)+O\left(r^{1+l}\right), \\
 \label{asympt2} w^{(II,-)}_{l}(r)&=&
 -\frac{\Gamma(\mu+b-l)}{\Gamma(-l)\Gamma(\mu+b+1)}
 \left(
 \begin{array}{c}
 0 \\ C_+ r^{-l-1}
 \end{array}\right)+O\Bigl(r^{-l}\Bigr).
 \end{eqnarray}
 Therefore, the solution $w^{(\gamma)}_{l}(r)$ can be written as
 \be\label{wgamma}
 w^{(\gamma)}_{l}(r)=\cos\eta\, w^{(II,+)}_{l}(r)+\sin\eta\,
 w^{(II,-)}_{l}(r),
 \eb
 Note that for special values of SAE parameter,
 $\displaystyle \Theta=\frac{\pi}{2}$ ($\displaystyle -\frac{\pi}{2}$)
 we have $\eta=0$ (resp. $\displaystyle \eta=\frac{\pi}{2}$).
 Explicit dependence of $\eta$ on $\Theta$,
 $l$, $\mu$, $b$ in the general case
 can be easily found from (\ref{asym0}) and (\ref{asympt1})--(\ref{asympt2}).
 Now, analogously to the above, consider the
 following ansatz for the Green function:
  \be\label{gfgans}
 G_{E,l}^{(\gamma)}(r,r')=
  \begin{cases}
    {A}^{(\gamma)}_{E,l}\;w^{(\gamma)}_{l}(r)\otimes \Bigl( w^{(I)}_{l}(r')\Bigr)^T & \text{for}\;\; 0<r<r'<1, \\
    {A}^{(\gamma)}_{E,l}\;w^{(I)}_{l}(r)\otimes \Bigl( w^{(\gamma)}_{l}(r')\Bigr)^T & \text{for}\;\;
    0<r'<r<1.
  \end{cases}
 \eb
 This ansatz automatically solves the equation (\ref{gfeq}) for $r\neq
 r'$ and satisfies the appropriate boundary conditions at the
 points $r=0$ and $r=1$. The jump condition at
 $r=r'$ will be satisfied provided we have
 \ben
 {A}^{(\gamma)}_{E,l}\cdot\det
 W\left(w_{l}^{(I)}(r),w_{l}^{(\gamma)}(r)\right)=-\frac{1-r^2}{Rr}\;.
 \ebn
 The last condition trivially holds if one chooses
 \be\label{cgamma}
 {A}^{(\gamma)}_{E,l}=\frac{\sqrt{m^2-E^2}}{2}\,\frac{1}{\sqrt{2}
 \sin\left(\eta+\frac{\pi}{4}\right)}.
 \eb
 Hence the formulas (\ref{wgamma})--(\ref{cgamma}) give the radial
 Green function for $l\in(-1,0)$.

 \subsection{Spectrum}



  The spectrum of the full Dirac hamiltonian $\hat{H}^{(\gamma)}$ consists of several
 parts:
 \begin{itemize}
 \item a continuous spectrum: $\displaystyle |E|^2\geq
 m^2+4b^2/R^2$;
 \item a finite number of infinitely degenerate Landau levels,
 given by
 \ben
 \left|E^{(0)}_{n}\right|^2=m^2+\frac{4}{R^2}\left[b^2-(|b|-n)^2\right],
 \ebn
 where $n=1,2,\ldots n_{max}<|b|$. The allowed eigenvalues of
 angular momentum correspond to $l_0=-1,-2,\ldots$ for $b>0$ and
 $l_0=1,2,\ldots $ for $b<0$.
 \item a finite number of bound states with finite degeneracy,
 whose form depends on the sign of the magnetic field. Namely, for
 $b>0$ one has
 \ben
 \left|E^{(\nu,+)}_{n}\right|^2=m^2+\frac{4}{R^2}\left[b^2-(b-n-(1+\nu))^2\right],
 \ebn
 where $n=1,2,\ldots,n'_{max}<b-(1+\nu)$, and for $b<0$ we obtain
 \ben
 \left|E^{(\nu,-)}_{n}\right|^2=m^2+\frac{4}{R^2}\left[b^2-(|b|-n+\nu)^2\right],
 \ebn
 with $n=1,2,\ldots,n''_{max}< |b|+\nu$. The allowed angular
 momenta are given by $l_0=1,2,\ldots, n'_{max}$ (for $b>0$) and
 $l_0=-1,-2,\ldots,-n''_{max}$ (for $b<0$).
 \item a finite number of non-degenerate bound states, corresponding to
 the mode with $l_0=0$. These energy levels are determined as real roots of the
 equation
 \ben
 \frac{(m+E)R}{2}\;\left(\frac{2}{mR}\right)^{1+2\nu}
 \frac{\Gamma(\mu+b)\Gamma(\mu-b+\nu+1)}{\Gamma(\mu-b+1)\Gamma(\mu+b-\nu)}=
 \ebn
 \ben
 =-\frac{\Gamma(\nu+1)}{\Gamma(-\nu)}\;2^{1+2\nu}
 \tan\left(\frac{\Theta}{2}+\frac{\pi}{4}\right)\substack{def \\ = \\ \; }
 -A(\Theta,\nu).
 \ebn
 Note that for $A(\Theta,\nu)<0$ we can have a solution of this
 equation satisfying $|E|<m$. This is in constrast with the
 previous cases, where all energy levels lie in the interval
 $m^2<E^2<m^2+4b^2/R^2$.
 \end{itemize}

 \subsection{\label{full1v}Full one-vortex Green function}
 Once the Green function of a particular SAE is found, one
 can also obtain it for any other extension using Krein's formula.
 This fact very much simplifies the analysis of the
 $\delta$-interaction hamiltonians (see, e.~g.~\cite{albeverio}), since in that
 case the family of SAEs usually includes free Laplacian/Dirac
 operator, whose Green function can be computed relatively
 easily (for example, in the planar case one has just to apply
 Fourier transform). The situation with AB hamiltonians is
 different. Here, the calculation of the resolvent constitutes a
 non-trivial problem even for distinguished values of the extension
 parameters.

 In the present subsection, we obtain integral representations for
 the Green function
 \be\label{full_GF}
 G(z,z')=\frac{1}{2\pi}\sum\limits_{l_0\in\Zb}
 \left(\begin{array}{cc}
 e^{il_0\varphi} & 0 \\ 0 & e^{i(l_0+1)\varphi}
 \end{array}\right)
 G_{E,l_0+\nu}(r,r')
  \left(\begin{array}{cc}
 e^{-il_0\varphi'} & 0 \\ 0 & e^{-i(l_0+1)\varphi'}
 \end{array}\right)
 \eb
 of the full hamiltonian $H^{(\gamma)}$ for two values of  SAE
 parameter, namely, for $\displaystyle\Theta=\pm\frac{\pi}{2}\,$.
 The outline of the calculation is similar to \cite{lisovyy_JMP}
 and the reader is referred to this paper for more details.

 We begin by introducing two classes of solutions of the Dirac
 equation on the disk without AB field:
 \ben
 \Psi_{\pm}(z,\theta)=\left(\begin{array}{c}
 \displaystyle C_{\pm}^{-1} e^{-\theta/2}\,
 \frac{(1-|z|^2)^{\frac{1\pm 2\mu}{2}}}{
 (1+ze^{-\theta})^{1\pm\mu-b}(1+\bz e^{\,\theta})^{\pm\mu+b}}\vspace{0.2cm} \\
 \displaystyle \pm\, C_{\pm} e^{\,\theta/2}\,
 \frac{(1-|z|^2)^{\frac{1\pm 2\mu}{2}}}{
 (1+ze^{-\theta})^{\pm\mu-b}(1+\bz e^{\,\theta})^{1\pm\mu+b}}
 \end{array}\right).
 \ebn
 These functions are delimited by two families of branch cuts in
 the $\theta$-plane:
 $
 \displaystyle
 \Bigl(-\infty+i(\varphi+\pi+2\pi\Zb),\ln r+i(\varphi+\pi+2\pi\Zb)\Bigr]
 $
 and
 $
 \displaystyle
 \Bigl[-\ln
 r+i(\varphi+\pi+2\pi\Zb),\infty+i(\varphi+\pi+2\pi\Zb)\Bigr)
 $,
 with the arguments of $1+ze^{-\theta}$ and $1+\bz e^{\theta}$
 equal to zero on the line $\mathrm{Im}\,\theta=\varphi$. It is
 also convenient to introduce the ``conjugates'' of these
 solutions, defined by
 \be\label{psihat}
 \hat{\Psi}_{\pm}(z,\theta)=\Psi_{\pm}(z\longleftrightarrow\bz,\theta\longleftrightarrow-\theta).
 \eb
 The relation $\displaystyle \hat{L}(z){\Psi}_{\pm}(z,\theta)=
 -\partial_{\theta}{\Psi}_{\pm}(z,\theta)$ allows to construct
 multivalued radial solutions of the Dirac equation
 with specified monodromy as
 superpositions of ${\Psi}_{\pm}(z,\theta)$. One can check that
 \be\label{radsol1}
 \mathrm{w}_l^{(I)}(z)\;\substack{def \\ =\\ \;}\;\int\nolimits_{C_0(z)}
 e^{\left(l+1/2\right)\theta}\,\Psi_-(z,\theta)\,d\theta=
 e^{i\pi l}
 \left(\begin{array}{cc}
 e^{il\varphi} & 0 \\ 0 & e^{i(l+1)\varphi}
 \end{array}\right) w_l^{(I)}(r),
 \eb
 \be\label{radsol2}
 \mathrm{w}_l^{(II,\pm)}(z)\;\substack{def \\ =\\ \;}\;\pm\int\nolimits_{C_\pm(z)}
 e^{\left(l+1/2\right)\theta}\,\Psi_+(z,\theta)\,d\theta=
 2\pi i\,e^{i\pi l}
 \left(\begin{array}{cc}
 e^{il\varphi} & 0 \\ 0 & e^{i(l+1)\varphi}
 \end{array}\right) w_l^{(II,\pm)}(r),
 \eb
 where the contour $C_+(z)$ ($C_-(z)$) goes counterclockwise
 around the branch cut $\Bigl(-\infty+i(\varphi+\pi),\ln
 r+i(\varphi+\pi)\Bigr]$ (resp. $\Bigr[-\ln
 r+i(\varphi+\pi),\infty+i(\varphi+\pi)\Bigr)$), and the contour
 $C_0(z)$ is the line segment joining the branch points $\pm\ln
 r+i(\varphi+\pi)$. Conjugate solutions are obtained analogously:
 \be\label{radsol3}
 \hat{\mathrm{w}}_l^{(I)}(z)\;\substack{def \\ =\\ \;}\;\int\nolimits_{C_0(z)}
 e^{-\left(l+1/2\right)\theta}\,\hat{\Psi}_-(z,\theta)\,d\theta=
 e^{-i\pi l}
 \left(\begin{array}{cc}
 e^{-il\varphi} & 0 \\ 0 & e^{-i(l+1)\varphi}
 \end{array}\right) w_l^{(I)}(r),
 \eb
  \be\label{radsol4}
 \hat{\mathrm{w}}_l^{(II,\pm)}(z)\;\substack{def \\ =\\ \;}\;\mp\int\nolimits_{C_\mp(z)}
 e^{-\left(l+1/2\right)\theta}\,\hat{\Psi}_+(z,\theta)\,d\theta=
 2\pi i \,e^{-i\pi l}
 \left(\begin{array}{cc}
 e^{-il\varphi} & 0 \\ 0 & e^{-i(l+1)\varphi}
 \end{array}\right) w_l^{(II,\pm)}(r).
 \eb

 Let us assume the regularity of the upper component
 of the Dirac wave function, i.~e. $\displaystyle \Theta=-\frac{\pi}{2}\,$.
 Then the Green
 function (\ref{full_GF}) can be conveniently expressed in terms
 of the radial solutions (\ref{radsol1})--(\ref{radsol4}) as
 follows:
 \be\label{sumgf}
 G(z,z')=\,\frac{\sqrt{m^2-E^2}}{8i\pi^2}\,e^{-i\nu(\varphi-\varphi')}\Bigl[\mathcal{G}^{(+)}(z,z')+
 \mathcal{G}^{(-)}(z,z')\Bigr],
 \eb
 where
  \begin{eqnarray}
 \label{gpm1}
 \mathcal{G}^{(\pm)}(z,z')&=&
 \sum\limits_{{l\in\Zb+\nu,\; l\gtrless0}}
 \mathrm{w}_l^{(I)}(z)\otimes\left(\hat{\mathrm{w}}_l^{(II,\pm)}(z')\right)^T\qquad \text{for }|z|>|z'|,\\
 \label{gpm2}
 \mathcal{G}^{(\pm)}(z,z')&=&
 \sum\limits_{{l\in\Zb+\nu,\; l\gtrless0}}
 \mathrm{w}_l^{(II,\pm)}(z)\otimes\left(\hat{\mathrm{w}}_l^{(I)}(z')\right)^T\qquad \text{for }|z|<|z'|.
 \end{eqnarray}
 \textbf{Remark}. For $\displaystyle \Theta=\frac{\pi}{2}$ one
 obtains similar representations, but in this case the summation
 in $\mathcal{G}^{(\pm)}(z,z')$ is over $l\gtrless-1$. Further
 calculation is also
 completely analogous, so we will continue with $\displaystyle \Theta=-\frac{\pi}{2}$,
 and present only the final result for  $\displaystyle \Theta=\frac{\pi}{2}$
 at the end of this section.\vspace{0.2cm}

 Following \cite{lisovyy_JMP}, we substitute into
 (\ref{gpm1})--(\ref{gpm2}) instead of $\mathrm{w}_l^{(I)}(z)$,
 $\hat{\mathrm{w}}_l^{(I)}(z')$, $\mathrm{w}_l^{(II,\pm)}(z)$,
 $\hat{\mathrm{w}}_l^{(II,\pm)}(z')$ their contour integral
 representations (\ref{radsol1})--(\ref{radsol4}). After interchanging
 the order of summation and integration the sums
 over $l$ are reduced to geometric series. For example, in the
 case $|z|>|z'|$ this gives
  \be\label{intergf1}
 \mathcal{G}^{(+)}_k(z,z')+\mathcal{G}^{(-)}_k(z,z')=
 \int\limits_{C_0(z)}\!\!d\theta_1\!\!\!\!
 \int\limits_{C_+(z')\cup C_-(z')}\!\!\!\!\!\!\!\! d\theta_2
 \;\;\;\Psi_-(z,\theta_1)\otimes\hat{\Psi}^T_+(z',\theta_2)\;
 \frac{e^{\left(1+\nu+\frac12\right)(\theta_1-\theta_2)}}{e^{\theta_1-\theta_2}-1}\,,
 \eb
 where the contours $C_{\pm}(z')$ satisfy additional constraints:
 $\displaystyle \mathrm{Re}(\theta_1-\theta_2)<0$ for all
 $\theta_1\in C_0(z)$, $\theta_2\in C_-(z')$ and
 $\displaystyle \mathrm{Re}(\theta_1-\theta_2)>0$ for all
 $\theta_1\in C_0(z)$, $\theta_2\in C_+(z')$. After a suitable
 deformation of
 integration contours one can obtain a representation, which is valid not
 only for $|z|>|z'|$, but for all $z$, $z'$ such that $\varphi-\varphi'\neq \pm
 \pi$. It has the following form (cf. with (2.24)--(2.26) in \cite{pacific}):
  \be\label{ans}
 G(z,z')=\left\{
 \begin{array}{rl}
 e^{- i \nu(\varphi-\varphi'+2\pi)}\,G^{(0)}(z,z')+\Delta(z,z') & \text{for }\varphi-\varphi'\in(-2\pi,-\pi),\\
 e^{- i \nu(\varphi-\varphi')}\; G^{(0)}(z,z')+\Delta(z,z') & \text{for }\varphi-\varphi'\in(-\pi,\pi),\\
 e^{- i \nu(\varphi-\varphi'-2\pi)}\,G^{(0)}(z,z')+\Delta(z,z') & \text{for
 }\varphi-\varphi'\in(\pi,2\pi)\,,
 \end{array}\right.
 \eb
 with
 \be\label{g0v1}
 G^{(0)}(z,z')=\frac{\sqrt{m^2-E^2}}{4\pi}\int\limits_{C_0(z)}\!\!d\theta\;
 \Psi_-(z,\theta)\otimes\hat{\Psi}^T_+(z',\theta),
 \eb
 \be\label{deltav1}
 \Delta(z,z')=\sqrt{m^2-E^2}\,e^{-i\nu(\varphi-\varphi')}
 \frac{1-e^{-2\pi i
 \nu}}{8i\pi^2}\;
  \int\limits_{C_0(z)}\!\!d\theta_1\!\!\!\!
 \int\limits_{\mathrm{Im}\,\theta_2=\varphi'}\!\!\!\!\!\!\!\! d\theta_2
 \;\;\;\Psi_-(z,\theta_1)\otimes\hat{\Psi}^T_+(z',\theta_2)\;
 \frac{e^{\left(1+\nu+\frac12\right)(\theta_1-\theta_2)}}{e^{\theta_1-\theta_2}-1}\,.
 \eb
 \textbf{Remark}. Starting from (\ref{gpm2}),
 similar results can be obtained. More precisely, one
 finds again the formula (\ref{ans}), but with
 \be\label{g0v2}
 G^{(0)}(z,z')=\frac{\sqrt{m^2-E^2}}{4\pi}\int\limits_{C_0(z')}\!\!d\theta\;
 \Psi_+(z,\theta)\otimes\hat{\Psi}^T_-(z',\theta),
 \eb
 \be\label{deltav2}
 \Delta(z,z')=\sqrt{m^2-E^2}\,e^{-i\nu(\varphi-\varphi')}
 \frac{1-e^{2\pi i
 \nu}}{8i\pi^2}\;
  \int\limits_{C_0(z')}\!\!d\theta_1\!\!\!\!
 \int\limits_{\mathrm{Im}\,\theta_2=\varphi}\!\!\!\!\!\!\!\! d\theta_2
 \;\;\;\Psi_+(z,\theta_2)\otimes\hat{\Psi}^T_-(z',\theta_1)\;
 \frac{e^{\left(1+\nu+\frac12\right)(\theta_2-\theta_1)}}{1-e^{\theta_2-\theta_1}}\,.
 \eb
 The proof of equivalence of the representations (\ref{g0v1}),
 (\ref{deltav1}) and (\ref{g0v2}), (\ref{deltav2}) for $ G^{(0)}(z,z')$
 and $\Delta(z,z')$  is left to the
 reader as an exercise.\vspace{0.2cm}

 Since for $\nu=0$ the second term in (\ref{ans}) vanishes,
 $ G^{(0)}(z,z')$ coincides with the Green function of the Dirac hamiltonian
 without AB field, which we will denote by $\hat{H}^{(0)}$.
 Using the technique described in the Appendix~A of
 \cite{lisovyy_JMP}, one may compute the integral for
 $ G^{(0)}(z,z')$ in terms of hypergeometric functions. The result reads
 \be\label{g0fan}
 G^{(0)}(z,z')=\left(\frac{1-\bz z'}{1-z\bz'}\right)^{-b}
 \left(\begin{array}{cc}
 \displaystyle
 \left(\frac{1-\bz z'}{1-z\bz'}\right)^{1/2}
 \zeta_{11}\Bigl(u(z,z')\Bigr) &
 \displaystyle
 \!\!\frac{|1-\bz z'|}{z'-z}\; \zeta_{12}\Bigl(u(z,z')\Bigr) \\
 \displaystyle
 -\frac{|1-\bz z'|}{\bz'-\bz}\; \zeta_{21}\Bigl(u(z,z')\Bigr) &
 \displaystyle
 \!\!\!-\left(\frac{1-\bz z'}{1-z\bz'}\right)^{-1/2}
 \zeta_{22}\Bigl(u(z,z')\Bigr)
 \end{array}\right),
 \eb
 where $\displaystyle u(z,z')=\left|\frac{z'-z}{1-\bz z'}\right|^2 $ and
 \be\label{zetadef}
 \zeta(u)=\frac{1}{2\pi R}
 \frac{\Gamma(\mu-b+1)\Gamma(\mu+b+1)}{\Gamma(1+2\mu)}\;(1-u)^{\frac{1+2\mu}{2}}\;\times
 \eb
 \ben
 \times
 \left(
 \begin{array}{cc}
 C_+^{-2}{}_2F_1\left(\mu-b+1,\mu+b,1+2\mu,1-u\right) &
 {}_2F_1\left(\mu-b,\mu+b,1+2\mu,1-u\right) \vspace{0.1cm} \\
 {}_2F_1\left(\mu-b,\mu+b,1+2\mu,1-u\right) &
 C_+^{2}\;{}_2F_1\left(\mu-b,\mu+b+1,1+2\mu,1-u\right)
 \end{array}\right).
 \ebn
 One may also obtain a more explicit expression for
 $\Delta(z,z')$:
 \be\label{deltafanm}
 \Delta(z,z')=\frac{\sin\pi\nu}{\pi}\int\limits_{-\infty}^{\infty}d\theta\;
 \frac{e^{(1+\nu)\theta+i(\varphi-\varphi')}}{e^{\theta+i(\varphi-\varphi')}+1}
 \left(\frac{1+rr'e^{\theta}}{1+rr'e^{-\theta}}\right)^{-b}\;\times
 \eb
 \ben
 \times\;\left(\begin{array}{cc}
 \displaystyle
 \left(\frac{1+rr'e^{\theta}}{1+rr'e^{-\theta}}\right)^{1/2}
 \zeta_{11}\Bigl(v(r,r',\theta)\Bigr) &
 \displaystyle
 \frac{\left(1+r^2r'^2+2r r'\cosh\theta\right)^{1/2}}{re^{-\theta}+r'}\,
 e^{-i\varphi'}\zeta_{12}\Bigl(v(r,r',\theta)\Bigr) \\
 \displaystyle
 \frac{\left(1+r^2r'^2+2r r'\cosh\theta\right)^{1/2}}{re^{\theta}+r'}\,
 e^{\theta+i\varphi}\zeta_{21}\Bigl(v(r,r',\theta)\Bigr) &
 \displaystyle
 e^{\theta+i(\varphi-\varphi')}
 \left(\frac{1+rr'e^{\theta}}{1+rr'e^{-\theta}}\right)^{-1/2}
 \zeta_{22}\Bigl(v(r,r',\theta)\Bigr)
 \end{array}\right),
 \ebn
 where we have introduced the notation
 \be\label{vrrpr}
 v(r,r',\theta)=\frac{r^2+r'^2+2rr'\cosh\theta}{1+r^2r'^2+2rr'\cosh\theta}\,.
 \eb
 \textbf{Remark}. In the case $\displaystyle \Theta=\frac{\pi}{2}$ one
 finds
 essentially the
 same answer (\ref{ans}). The only difference is that we should replace $\nu$ by $\nu-1$ in the
 double integrals entering the definitions (\ref{deltav1}) and (\ref{deltav2})
 of $\Delta(z,z')$. This finally gives
  \be\label{deltafanp}
 \Delta(z,z')=\frac{\sin\pi\nu}{\pi}\int\limits_{-\infty}^{\infty}d\theta\;
 \frac{{-}e^{{\nu}{\theta}}}{e^{\theta+i(\varphi-\varphi')}+1}
 \left(\frac{1+rr'e^{\theta}}{1+rr'e^{-\theta}}\right)^{-b}\;\times
 \eb
 \ben
 \times\;\left(\begin{array}{cc}
 \displaystyle
 \left(\frac{1+rr'e^{\theta}}{1+rr'e^{-\theta}}\right)^{1/2}
 \zeta_{11}\Bigl(v(r,r',\theta)\Bigr) &
 \displaystyle
 \frac{\left(1+r^2r'^2+2r r'\cosh\theta\right)^{1/2}}{re^{-\theta}+r'}\,
 e^{-i\varphi'}\zeta_{12}\Bigl(v(r,r',\theta)\Bigr) \\
 \displaystyle
 \frac{\left(1+r^2r'^2+2r r'\cosh\theta\right)^{1/2}}{re^{\theta}+r'}\,
 e^{\theta+i\varphi}\zeta_{21}\Bigl(v(r,r',\theta)\Bigr) &
 \displaystyle
 e^{\theta+i(\varphi-\varphi')}
 \left(\frac{1+rr'e^{\theta}}{1+rr'e^{-\theta}}\right)^{-1/2}
 \zeta_{22}\Bigl(v(r,r',\theta)\Bigr)
 \end{array}\right).\vspace{0.2cm}
 \ebn
 Representations (\ref{deltafanm}) and (\ref{deltafanp}) for the
 vortex-dependent part of the Green function constitute the main
 technical result of this section which will be used later in
 the construction of Painlev\'e~VI transcendents.

 \section{Two-point tau function}
 \subsection{\label{taudef}General setting}
 In this section, we consider Dirac hamiltonian
 (\ref{dirac_ham}) with two AB vortices located at the points
 $a_1,a_2\in D$. Corresponding vector potential has the form
 \be\label{vp2ab}
 \mathcal{A}=-\frac{i{BR}^{\,2}}{4}\,\frac{\bz\, dz -
 z\,d\bz}{1-|z|^2}
 -\frac{i\nu_1}{2}\left(\frac{dz}{z-a_1}-\frac{d\bz}{\bz-\bar{a}_1}\right)
 -\frac{i\nu_2}{2}\left(\frac{dz}{z-a_2}-\frac{d\bz}{\bz-\bar{a}_2}\right),
 \eb
 and it is assumed that $-1<\nu_{1,2}<0$. Let us make a
 singular gauge transformation $\displaystyle\hat{H}\mapsto
 \hat{H}^{(a,\nu)}=U\hat{H}U^{\dag}$ with
 \be\label{sgtr}
 (U\psi)(z)=
 \left(\frac{z-a_1}{\bz-\bar{a}_1}\right)^{\nu_1/2}
 \left(\frac{z-a_2}{\bz-\bar{a}_2}\right)^{\nu_2/2}
 \psi(z).
 \eb
 It is easy to check that the local action of $\hanu$ coincides
 with that of the free hamiltonian $\hat{H}^{(0)}$ (i.~e., in the absence of AB
 fluxes). However, the functions from the domain of $\hanu$ are
 multivalued: they pick up a phase given by $\displaystyle e^{2\pi i \nu_j}$
 when continued around $a_j$ ($j=1,2$). One should then introduce
 two branch cuts $\ell_1$, $\ell_2$  on $D$ as shown in Fig.~1. We do not fix the
 branches of fractional powers in (\ref{sgtr}), since this will be
 implicitly done later.
 \begin{figure}[h]
 \begin{center}
 \resizebox{5cm}{!}{
 \includegraphics{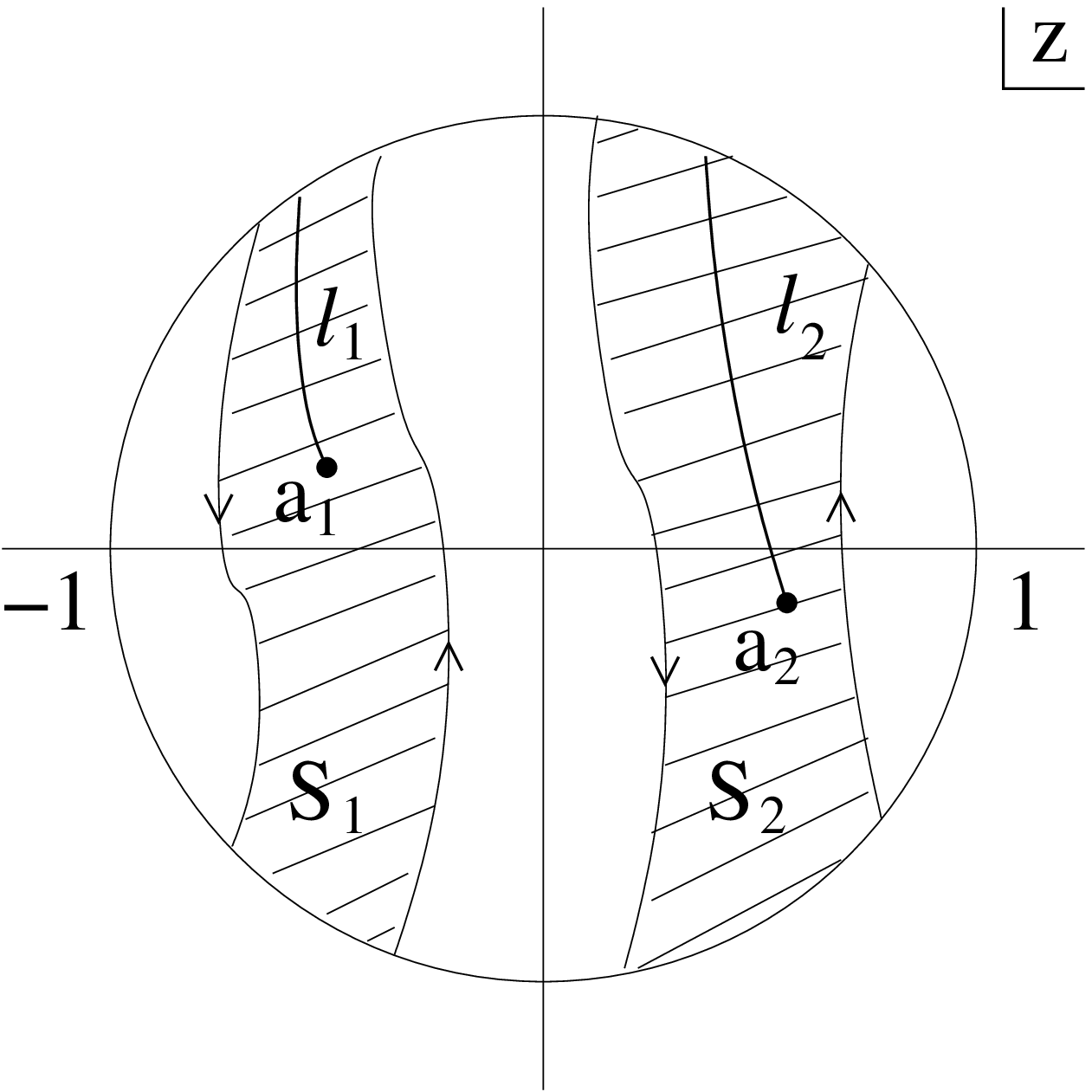}} \\ Fig. 1\\
 \end{center}
 \end{figure}

 Let us isolate the branch cuts in the union $S$ of two open strips $S_1$ and $S_2$
 (see Fig.~1) and denote by $\mathcal{H}(S)$ the set of all hamiltonians
 $\hanu$ satisfying $a_1\in S_1$, $a_2\in S_2$. In other words,
 the elements of $\mathcal{H}(S)$ are parametrized by the
 positions of AB vortices, their fluxes being fixed. Consider the localizations of
 different elements of $\mathcal{H}(S)$ to $D\backslash S$. Since all of them are given by the same
 differential operator, all dependence on $a$ is encoded into the
 spaces of boundary values on $\partial S$ of the functions from
 their domains. Several such boundary spaces will be considered:
 \begin{itemize}
 \item a suitably chosen space $W=H^{1/2}(\partial S)$ of $\Cb^2$-valued functions on
 $\partial S$.
 \item a subspace $\ds W^{int}(a)\subset W$, which is composed of
 boundary values of functions $\psi\in H^1(S\backslash(\ell_1\cup\ell_2))$ solving the equation $\ds
 (\hanu-E)\psi=0$ on $S$. Note that
 $\ds W^{int}(a)=W^{int}_1(a_1)\oplus W^{int}_2(a_2)$,
 where $W^{int}_j(a_j)$ ($j=1,2$)
 is composed of the boundary values on $\partial S_j$ of local solutions to the Dirac
 equation on the strip $S_j$.
 \item similarly, $\ds W^{ext}\subset W$ is defined to be the
 space of boundary values of $H^1$-solutions of  $\ds (\hanu-E)\psi=0$
 on $D\backslash \overline{S}$. This subspace clearly does not depend on
 $a$.
 \item it is also convenient to fix two points $a_1^0\in S_1$, $a_2^0\in S_2$
 and to introduce a reference subspace $\ds W^{int}(a^0)$.
 \end{itemize}
 It will be assumed that near each branch point one of the components of the Dirac spinor
 is regular, and the other is square integrable
 (i.~e. we consider four possible types of boundary conditions
 for $\hanu$). With this choice of the domain,
 the spaces $W^{int}(a)$ and $W^{ext}$ can be shown to be transverse in $W$ for
 $E\in\Cb\backslash\Bigl((-\infty,-m]\cup[m,\infty)\Bigr)$. Then,
 given any subspace $V\subset W$, one can define the projection
 $\ds P(a):\,V \stackrel{W^{ext}}{\longrightarrow}W^{int}(a)$ of $V$ on $W^{int}(a)$
 along~$W^{ext}$.
 \begin{prop} Let $f\in W$ and consider the function $f_{int}\in
 H^1(S\backslash(\ell_1\cup\ell_2))$ defined by
 \be\label{projp}
 f_{int}(z)=-\int\nolimits_{\partial S}\ddot{G}^{(a,\nu)}(z,z')
 \frac{iR}{1-|z'|^2}
 \left\{
 \sigma_-\,f(z')\,dz'+
 \sigma_+\,f(z')\,d\bz'
 \right\},
 \eb
 \ben
 \sigma_+=\sigmapl,\qquad\sigma_-=\sigmamn,
 \ebn
 where $ z\in S\backslash(\ell_1\cup\ell_2)$,
 $\partial S$ is oriented counterclockwise and $\ds \ddot{G}^{(a,\nu)}(z,z')$
 denotes the Green function of $\hanu$.
 Then the boundary value of $f_{int}$ on $\partial S$
 coincides with the projection $P(a)f\in W^{int}(a)$.
 \end{prop}
 $\blacksquare$ First remark that $f_{int}$ satisfies Dirac equation on $S$.
 Therefore, we may write
  \ben
  f_{int}(z)=\int\nolimits_{D\backslash
  \overline{S}}\ddot{G}^{(a,\nu)}(z,z')\left(\hanu-E\right)  f_{int}(z')
  \;d\mu_{z'}.
  \ebn
  Integrating once by parts and using Stokes theorem, one obtains
  \be\label{auxf1}
   f_{int}(z)=-\int\nolimits_{\partial S}\ddot{G}^{(a,\nu)}(z,z')
 \frac{iR}{1-|z'|^2}
 \left\{
 \sigma_-\,f_{int}(z')\,dz'+
 \sigma_+\,f_{int}(z')\,d\bz'\right\},
  \eb
  and thus the map (\ref{projp}) is indeed a projection on
  $W^{int}(a)$. Using similar arguments, one can show that its
  kernel coincides with $W^{ext}$.
  $\square$
  \begin{defin}
  Let $f_j\in H^{1/2}(\partial S_j)$ ($j=1,2$) and consider the function $f_{int,j}\in
  H^1(S_j\backslash\ell_j)$ defined by
  \be\label{proj1p}
 f_{int,j}(z)=-\int\nolimits_{\partial S_j}\dot{G}^{(a_j,\nu_j)}(z,z')
 \frac{iR}{1-|z'|^2}
 \left\{
 \sigma_-\,f_j(z')\,dz'+
 \sigma_+\,f_j(z')\,d\bz'
 \right\},
 \eb
 where $ z\in S_j\backslash \ell_j$ and $\ds
 \dot{G}^{(a_j,\nu_j)}(z,z')$ is the Green function of the
 Dirac hamiltonian on the disk with only one branch point~$a_j$. Passing to
 boundary values of $f_{int,j}$ on $\partial S_j$, one obtains a
 projection $\ds P_j(a_j):\,H^{1/2}(\partial S_j)\longrightarrow
 W^{int}_j(a_j)$. We will denote by $F(a)$ the
 direct sum of such one-point projections:
 \ben
 \ds F(a)\stackrel{def}{\;=\;}P_1(a_1)\oplus P_2(a_2):\,W\longrightarrow
 W^{int}(a).
 \ebn
  \end{defin}
  \begin{defin}
  $\tau$-function of $\hanu$ is defined as follows:
  \be\label{taudefin}
  \tau(a,a^0)=\mathrm{det}\left(\Bigl[P_1(a_1)\oplus
  P_2(a_2)\Bigr]_{W^{int}(a^0)}
  \stackrel{W^{ext}}{\longrightarrow}W^{int}(a^0)\right)=
  \mathrm{det}_{\,W^{int}(a^0)}\left(P^{-1}(a)F(a)\right).
  \eb
  \end{defin}
  \textbf{Remark}. For this definition to make sense, the restriction
  of the map $P^{-1}(a)F(a)$ to $W^{int}(a^0)$ should be
  a trace class perturbation of the identity on $W^{int}(a^0)$.
  This can be shown analogously to Proposition~4.2 of
  \cite{pacific}. Factorization formulas for the derivatives of one-point
  Green functions $\dot{G}^{(a_j,\nu_j)}(z,z')$, needed for the proof, are presented in
  Subsection~\ref{ssstau}.\vspace{0.2cm}

  We are  interested in explicit calculation of the
  $\tau$-function (\ref{taudefin}). In order to do this, it is
  convenient to introduce yet other boundary maps.
  \begin{defin} Let $\gamma$ be a smooth curve dividing Poincar\'e
  disk into two disconnected parts $D_1$ and $D_2$ as shown in Fig.~2. For any $f\in
  H^{1/2}(\gamma)$ define the functions
  \be
  \label{fplusminus}
  f_{\pm}(z)=\pm\int\nolimits_{\gamma}{G}^{(0)}(z,z')
 \frac{iR}{1-|z'|^2}
 \left\{
 \sigma_-\,f(z')\,dz'+
 \sigma_+\,f(z')\,d\bz'
 \right\},
 \eb
 where $z\in D_2$ for $f_+(z)$, $z\in D_1$ for $f_-(z)$ and ${G}^{(0)}(z,z')$
 is the Green function of the Dirac hamiltonian without
 AB field, explicitly given by (\ref{g0fan})--(\ref{zetadef}).
 With some abuse of notation, the boundary values of $f_{\pm}(z)$
 on $\gamma$ will also be denoted by $f_{\pm}(z)$.
  \end{defin}
   \begin{figure}[h]
 \begin{center}
 \resizebox{4cm}{!}{
 \includegraphics{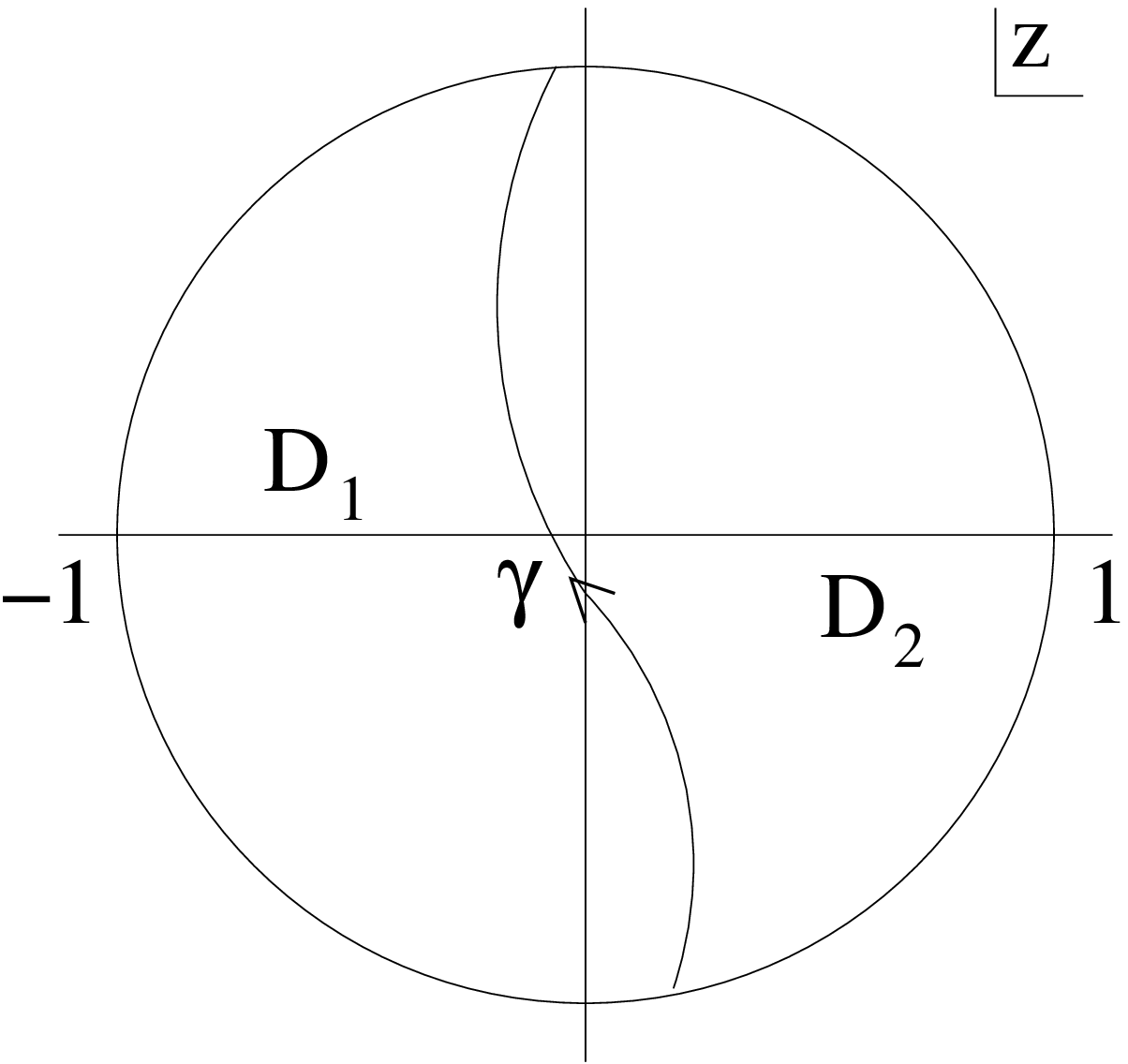}}\vspace{0.2cm} \\ Fig. 2\\
 \end{center}
 \end{figure}
  Note that $f_{\pm}$ have the properties of projections:
  \ben
  \left(f_{\pm}\right)_{\pm}=f_{\pm},\qquad
  \left(f_{\pm}\right)_{\mp}=0,
  \ebn
  and, in addition, $ f_++f_-=f$. One can thus uniquely write any $f\in
  H^{1/2}(\gamma)$ as a sum of two functions,
  where the first term ($f_+$) may be continued from $\gamma$
  to $D_2$ as a solution of the Dirac equation without AB field, and
  the second one ($f_-$) is the boundary value of a solution on
  $D_1$.

  Let us represent any function $\psi^{(j)}\in
  H^{1/2}(\partial S_j)$ ($j=1,2$) in the following form:
  \be\label{diskproj}
  \psi^{(j)}=\left(
  \begin{array}{c}
 \psi^{(j)}_{L,+} \vspace{0.1cm} \\ \psi^{(j)}_{R,-}
  \end{array}\right) \oplus
  \left(
  \begin{array}{c}
 \psi^{(j)}_{L,-} \vspace{0.1cm} \\ \psi^{(j)}_{R,+}
  \end{array}\right),
  \eb
  where the indices $L$ and $R$ correspond to the left and right
  boundary of the strip $S_j$.
  \begin{prop}\label{propwithoutab} If $\psi^{(j)}$ is the boundary
  value of a solution of the Dirac equation on $S_j$ without AB
  vortex, then its components satisfy the relation
  \be\label{strip0}
   \left(
  \begin{array}{c}
 \psi^{(j)}_{L,-} \vspace{0.1cm} \\ \psi^{(j)}_{R,+}
  \end{array}\right)=
  \left(
  \begin{array}{cc}
  0 & \hat{\omega}\left(\partial S^L_j|\partial S^R_j\right) \\
  \hat{\omega}\left(\partial S^R_j|\partial S^L_j\right) & 0
  \end{array}
  \right)
  \left(
  \begin{array}{c}
 \psi^{(j)}_{L,+} \vspace{0.1cm} \\ \psi^{(j)}_{R,-}
  \end{array}\right),
  \eb
  where the integral operators $\hat{\omega}\left(\partial S^L_j|\partial S^R_j\right)$,
  $ \hat{\omega}\left(\partial S^R_j|\partial S^L_j\right)$
  are defined as follows:
  \begin{eqnarray}
  \label{omega12}
  \left(\hat{\omega}\left(\partial S^L_j|\partial S^R_j\right)
  \psi_{R,-}^{(j)}\right)(z)&=&
  -\int\nolimits_{\partial S_j^R}{G}^{(0)}(z,z')
 \frac{iR}{1-|z'|^2}
 \left\{
 \sigma_-\,\psi_{R,-}^{(j)}(z')\,dz'+
 \sigma_+\,\psi_{R,-}^{(j)}(z')\,d\bz'
 \right\}, \\
  \left(\hat{\omega}\left(\partial S^R_j|\partial S^L_j\right)
  \psi_{L,+}^{(j)}\right)(z)&=&
  - \int\nolimits_{\partial S^L_j}{G}^{(0)}(z,z')
 \frac{iR}{1-|z'|^2}
 \left\{
 \sigma_-\,\psi_{L,+}^{(j)}(z')\,dz'+
 \sigma_+\,\psi_{L,+}^{(j)}(z')\,d\bz'
 \right\}.
  \end{eqnarray}
   \end{prop}
 $\blacksquare$ If $\psi^{(j)}$ can be continued to
 a solution on $S_j$,
 then, similarly to (\ref{auxf1}), we have
 \begin{eqnarray}
 \nonumber  \psi^{(j)}(z)&=&-\int\nolimits_{\partial S_j}{G}^{(0)}(z,z')
 \frac{iR}{1-|z'|^2}
 \left\{
 {\sigma}_-\,\psi^{(j)}(z')\,dz'+
 {\sigma}_+\,\psi^{(j)}(z')\,d\bz'\right\}=\\
 &=&-\int\nolimits_{\partial S^R_j}{G}^{(0)}(z,z')
 \frac{iR}{1-|z'|^2}
 \left\{
 {\sigma}_-\,\psi^{(j)}_{R,-}(z')\,dz'+
 {\sigma}_+\,\psi^{(j)}_{R,-}(z')\,d\bz'\right\}\\
 \label{auxf2}&\;&-\int\nolimits_{\partial S^L_j}{G}^{(0)}(z,z')
 \frac{iR}{1-|z'|^2}
 \left\{
 {\sigma}_-\,\psi^{(j)}_{L,+}(z')\,dz'+
 {\sigma}_+\,\psi^{(j)}_{L,+}(z')\,d\bz'\right\},
  \end{eqnarray}
 where the second equality follows from the fact that it is not possible
 to construct a solution on the whole disk $D$, belonging to the domain of
 the free Dirac hamiltonian $\hat{H}^{(0)}$.
 Passing in (\ref{auxf2}) to boundary values and taking the projections,
 we obtain the result (\ref{strip0}).
 $\square$

  \begin{prop}\label{propwithab}
  If $\psi^{(j)}$ is the boundary
  value of a solution of the Dirac equation on $S_j\backslash\ell_j$
  with only one branching point $a_j$,
  then its components satisfy the relation
  \be\label{strip1}
   \left(
  \begin{array}{c}
 \psi^{(j)}_{L,-} \vspace{0.1cm} \\ \psi^{(j)}_{R,+}
  \end{array}\right)=
  \left(
  \begin{array}{cc}
  \hat{\alpha}_{S_j}(a_j) & \hat{\beta}_{S_j}(a_j) \\
  \hat{\gamma}_{S_j}(a_j) & \hat{\delta}_{S_j}(a_j)
  \end{array}
  \right)
  \left(
  \begin{array}{c}
 \psi^{(j)}_{L,+} \vspace{0.1cm} \\ \psi^{(j)}_{R,-}
  \end{array}\right),
  \eb
  where the integral operators $\hat{\alpha}_{S_j}(a_j)$, $\hat{\beta}_{S_j}(a_j)$
  $\hat{\gamma}_{S_j}(a_j)$, $\hat{\delta}_{S_j}(a_j)$  are defined as follows:
   \begin{eqnarray}
  \label{alpha}
   \left(\hat{\alpha}_{S_j}(a_j)\,\psi_{L,+}^{(j)}\right)(z)&=&
  - \int\nolimits_{\partial S^L_j}\dot{\Delta}^{(a_j,\nu_j)}(z,z')
 \frac{iR}{1-|z'|^2}
 \left\{
 \sigma_-\,\psi_{L,+}^{(j)}(z')\,dz'+
 \sigma_+\,\psi_{L,+}^{(j)}(z')\,d\bz'
 \right\}, \\
     \label{beta}
  \left(\hat{\beta}_{S_j}(a_j)\,\psi_{R,-}^{(j)}\right)(z)&=&
  -\int\nolimits_{\partial S_j^R}\dot{G}^{(a_j,\nu_j)}(z,z')
 \frac{iR}{1-|z'|^2}
 \left\{
 \sigma_-\,\psi_{R,-}^{(j)}(z')\,dz'+
 \sigma_+\,\psi_{R,-}^{(j)}(z')\,d\bz'
 \right\}, \\
  \label{gamma}
  \left(\hat{\gamma}_{S_j}(a_j)\,\psi_{L,+}^{(j)}\right)(z)&=&
  -\int\nolimits_{\partial S_j^L}\dot{G}^{(a_j,\nu_j)}(z,z')
 \frac{iR}{1-|z'|^2}
 \left\{
 \sigma_-\,\psi_{L,+}^{(j)}(z')\,dz'+
 \sigma_+\,\psi_{L,+}^{(j)}(z')\,d\bz'
 \right\}, \\
  \label{delta}
   \left(\hat{\delta}_{S_j}(a_j)\,\psi_{R,-}^{(j)}\right)(z)&=&
  - \int\nolimits_{\partial S^R_j}\dot{\Delta}^{(a_j,\nu_j)}(z,z')
 \frac{iR}{1-|z'|^2}
 \left\{
 \sigma_-\,\psi_{R,-}^{(j)}(z')\,dz'+
 \sigma_+\,\psi_{R,-}^{(j)}(z')\,d\bz'
 \right\}.
  \end{eqnarray}
  Here, we denote $\dot{\Delta}^{(a_j,\nu_j)}(z,z')=
  \dot{G}^{(a_j,\nu_j)}(z,z')-{G}^{(0)}(z,z')$.
  \end{prop}
  $\blacksquare$ The proof is completely analogous to the previous one.
  $\square$\vspace{0.2cm}

  We now choose $\psi^{(j)}_{L,+}$ and $\psi^{(j)}_{R,-}$ to be
  the ``coordinates'' in $W^{int}_j(a_j)$ ($j=1,2$). It is not difficult to
  check that in these coordinates the map $F(a):\,W^{int}(a^0)\rightarrow W^{int}(a)$
  is given by the identity operator. In order to find the
  representation of $P(a)^{-1}$, one should be able to decompose any $g\in
  W^{int}(a)$ as $g=f-h$, with $f\in W^{int}(a^0)$ and $h\in
  W^{ext}$. More precisely, we are interested in the relation
  between $f$ and $g$. This calculation is very
  similar to the proof of the Theorem~3.1 in \cite{pacific}:
  \begin{itemize}
  \item First note that $h^{(1)}_L$ can be continued as a solution
  to the left of $\partial S_1^L$ and, analogously, $h^{(2)}_R$
  can be continued to the right of $\partial S_2^R$. Then one has
  $h^{(1)}_{L,+}=h^{(2)}_{R,-}=0$ and, therefore,
  \be\label{rels01}
  g^{(1)}_{L,+}=f^{(1)}_{L,+},\qquad g^{(2)}_{R,-}=f^{(2)}_{R,-}.
  \eb
  \item Next recall that the boundary values $h^{(1)}_R$ and
  $h^{(2)}_L$ are not independent. They are related by the
  formulas
  \ben
  \left(
  \begin{array}{c}
  h^{(1)}_{R,-}\vspace{0.1cm} \\ h^{(2)}_{L,+}
  \end{array}
  \right)=
  \left(
  \begin{array}{cc}
  0 & \hat{\omega}_{12} \\
  \hat{\omega}_{21} & 0
  \end{array}
  \right)
  \left(
  \begin{array}{c}
  h^{(1)}_{R,+}\vspace{0.1cm} \\ h^{(2)}_{L,-}
  \end{array}
  \right),
  \ebn
  \ben
   \hat{\omega}_{12}=\hat{\omega}\left(\partial S^R_1|\partial
   S^L_2\right),\qquad
    \hat{\omega}_{21}=\hat{\omega}\left(\partial S^L_2|\partial
    S^R_1\right),
  \ebn
  which follow from the Proposition~\ref{propwithoutab}. Then one
  finds
  \begin{eqnarray}
  \label{rels02a}
  g^{(1)}_{R,-}-\hat{\omega}_{12}g^{(2)}_{L,-}&=&
  f^{(1)}_{R,-}-\hat{\omega}_{12}f^{(2)}_{L,-},
  \\
  \label{rels02b}
   g^{(2)}_{L,+}-\hat{\omega}_{21}g^{(1)}_{R,+}&=&
    f^{(2)}_{L,+}-\hat{\omega}_{21}f^{(1)}_{R,+}.
  \end{eqnarray}
  \item Finally, according to the Proposition~\ref{propwithab} one
  has
  \begin{eqnarray}
  \label{rels03a}
    \left(
  \begin{array}{c}
 f^{(j)}_{L,-} \vspace{0.1cm} \\ f^{(j)}_{R,+}
  \end{array}\right)=
  \left(
  \begin{array}{cc}
  \hat{\alpha}_{S_j}(a_j^0) & \hat{\beta}_{S_j}(a_j^0) \vspace{0.1cm}\\
  \hat{\gamma}_{S_j}(a_j^0) & \hat{\delta}_{S_j}(a_j^0)
  \end{array}
  \right)
  \left(
  \begin{array}{c}
 f^{(j)}_{L,+} \vspace{0.1cm} \\ f^{(j)}_{R,-}
  \end{array}\right),\\
 \label{rels03b}
     \left(
  \begin{array}{c}
 g^{(j)}_{L,-} \vspace{0.1cm} \\ g^{(j)}_{R,+}
  \end{array}\right)=
  \left(
  \begin{array}{cc}
  \hat{\alpha}_{S_j}(a_j) & \hat{\beta}_{S_j}(a_j) \vspace{0.1cm}\\
  \hat{\gamma}_{S_j}(a_j) & \hat{\delta}_{S_j}(a_j)
  \end{array}
  \right)
  \left(
  \begin{array}{c}
 g^{(j)}_{L,+} \vspace{0.1cm} \\ g^{(j)}_{R,-}
  \end{array}\right).
  \end{eqnarray}
  \end{itemize}
  If we now find $g^{(1)}_{R,+}$, $f^{(1)}_{R,+}$,
  $g^{(2)}_{L,-}$, $f^{(2)}_{L,-}$ from
  (\ref{rels03a})--(\ref{rels03b}) and substitute the corresponding expressions
  into (\ref{rels02a})--(\ref{rels02b}), two more relations between the coordinates
  of $f$ and $g$ can be obtained. Together with (\ref{rels01}),
  they may be written as follows:
  \begin{eqnarray}
  \label{rels04}
  &\;&\left(
  \begin{array}{cccc}
  \mathbf{1} & 0 & 0 & 0 \\
  0 & \mathbf{1} & -\hat{\omega}_{12}\hat{\alpha}_{S_2}(a_2) & -\hat{\omega}_{12}\hat{\beta}_{S_2}(a_2) \\
  -\hat{\omega}_{21}\hat{\gamma}_{S_1}(a_1) &
    -\hat{\omega}_{21}\hat{\delta}_{S_1}(a_1) & \mathbf{1} & 0 \\
  0 & 0 & 0 & \mathbf{1}
  \end{array}
  \right)
  \left(
  \begin{array}{c}
   g^{(1)}_{L,+}  \\ g^{(1)}_{R,-} \\
      g^{(2)}_{L,+} \\ g^{(2)}_{R,-}
  \end{array}
  \right)\;=\\
    \nonumber
    &\;&\left(
  \begin{array}{cccc}
  \mathbf{1} & 0 & 0 & 0 \\
  0 & \mathbf{1} & -\hat{\omega}_{12}\hat{\alpha}_{S_2}(a_2^0) & -\hat{\omega}_{12}\hat{\beta}_{S_2}(a_2^0) \\
  -\hat{\omega}_{21}\hat{\gamma}_{S_1}(a_1^0) &
    -\hat{\omega}_{21}\hat{\delta}_{S_1}(a_1^0) & \mathbf{1} & 0 \\
  0 & 0 & 0 & \mathbf{1}
  \end{array}
  \right)
  \left(
  \begin{array}{c}
   f^{(1)}_{L,+}  \\ f^{(1)}_{R,-} \\
      f^{(2)}_{L,+} \\ f^{(2)}_{R,-}
  \end{array}
  \right),
  \end{eqnarray}
  or, in a more compact form,
  \be\label{rels05}
  \left(\mathbf{1}+M(a)\right)g=\left(\mathbf{1}+M(a^0)\right)f.
  \eb
  Therefore, the $\tau$-function is given by
  \be\label{tau01}
  \tau(a,a^0)=\mathrm{det}_{W^{int}(a^0)}
  \left(\mathbf{1}+M(a^0)\right)^{-1}\left(\mathbf{1}+M(a)\right).
  \eb
  \textbf{Remark}. One-point Green function $\ds
  \dot{G}^{(a_j,\nu_j)}(z,z')$ is related to the one-vortex Green
  function $G(z,z')$, calculated in the previous section, by a
  simple unitary ($+$ singular gauge) transformation. For example,
  one has
  \be\label{sgtr2}
  \dot{\Delta}^{(a_j=0,\nu_j=\nu)}(z,z')=e^{i\nu(\varphi-\varphi')}\Delta(z,z').
  \eb
  Thus the action of all operators in (\ref{rels04}) is known.
  However, in order to obtain from (\ref{tau01}) an explicit
  formula for the $\tau$-function, one still needs to introduce some coordinates in the
  infinite-dimensional spaces $\ds \left(H^{1/2}(\partial S^{L,R}_{1,2})\right)_{\pm}$.
  It turns out, though, (see Theorem~6.3 in \cite{beatty} and Subsection~\ref{ssstau} of the present paper)
  that the logarithmic derivative of the $\tau$-function
  (\ref{taudefin}) does not depend on localization (i.~e. on the choice
  of $S$) and on $a^0$.
  Moreover, since $\tau(a^0,a^0)=1$, one
  should have
  \ben
  \tau(a,a^0)=\frac{\tau(a)\;}{\tau(a^0)}\,.
  \ebn
  Therefore, one may fix the coordinates in
  $\ds \left(H^{1/2}(\partial S^{L,R}_{1,2})\right)_{\pm}$ and calculate the $\tau$-function
  for a convenient choice of $S$. These problems are addressed in the
  following subsections.



 \subsection{Dirac equation on the Poincar\'e strip}
 Coordinate change $z=\tanh\xi$ maps the Poincar\'e disk onto the
 strip $\ds\mathcal{U}=\left\{\xi: |\xi_y|<\frac{\pi}{4}\right\}$ in the complex $\xi$-plane. In
 these coordinates, the Poincar\'e metric is given by
 \ben
 ds^2=g_{\xi\bar{\xi}}\,d\xi d\bar{\xi}=R^2\frac{d\xi
 d\bar{\xi}}{\cosh^2\left(\xi-\bar{\xi}\right)},
 \ebn
 and the vector potential of the uniform magnetic
 field $B$ can be chosen in the form
 \be\label{gauge02}
 \mathcal{A}^{(B)}_{\text{strip}}=-2b\tan2\xi_y\,d\xi_x.
 \eb
 Corresponding Dirac hamiltonian $\ds \hat{H}^{(0)}_{\text{strip}}$
 is explicitly given by the formula (\ref{dirac_ham}) with
  \begin{eqnarray}
 \label{dirac_k1_strip}
 K&=&\frac{1}{R}\,\Bigl\{2\cos2\xi_y\,\partial_{\xi}-i(1+2b)\sin2\xi_y\Bigr\}, \\
 \label{dirac_k2_strip}
 K^*&=&-\frac{1}{R}\,\Bigl\{2\cos2\xi_y\,\partial_{\bar{\xi}}+i(1-2b)\sin2\xi_y\Bigr\}.
 \end{eqnarray}
 It is related to the free Dirac hamiltonian on the disk by a unitary
 transformation
 \be\label{unitrans}
 \hat{H}^{(0)}_{\text{strip}}(\xi)=U_{DS}(\xi)\;\hat{H}^{(0)}_{\text{disk}}(z\mapsto\tanh\xi)\;
 U_{DS}^{\dag}(\xi),
 \eb
 \be\label{unitrans2}
 U_{DS}(\xi)=\left(
 \begin{array}{cc}
 \ds \left(\frac{\cosh\xi}{\cosh\bar{\xi}}\right)^{-\frac{1-2b}{2}} &
 0 \\ \ds 0 & \ds
 \left(\frac{\cosh\xi}{\cosh\bar{\xi}}\right)^{\frac{1+2b}{2}}
 \end{array}
 \right).
 \eb
 The main advantage of the gauge (\ref{gauge02}) is that $\hat{H}^{(0)}_{\text{strip}}$
 commutes  with the  $\xi_x$-momentum operator $\ds
 \hat{P}_x=-i\partial_{\xi_x}$. The eigenspace of $\hat{P}_x$,
 characterized by the momentum $p\in\Rb$, is composed of the
 spinors of the form ${g}(p,\xi_y)e^{ip\xi_x}$. Being restricted to
 this eigenspace, the hamiltonian
 $\hat{H}^{(0)}_{\text{strip}}$ acts as follows:
 \be\label{strippartham}
 {g}(p,\xi_y)\mapsto \hat{H}_p\,  {g}(p,\xi_y),\qquad
 \hat{H}_p=R^{-1}\left(
 \begin{array}{cc}
 mR & K_p \\ K^*_p & -mR
 \end{array}\right),
 \eb
 where the operators $ K_p$ and $ K_p^*$ are given by
 \begin{eqnarray}
 \label{k1strip}
 K_p&=&-i\Bigl[\cos2\xi_y\,(\partial_{\xi_y}-p)+(1+2b)\sin
 2\xi_y\Bigr],\\
 \label{k2strip}
  K_p^*&=&-i\Bigl[\cos2\xi_y\,(\partial_{\xi_y}+p)+(1-2b)\sin
 2\xi_y\Bigr].
 \end{eqnarray}

 Let us consider the partial Dirac equation
 \be\label{strip_de}
 (\hat{H}_p-E){g}(p,\xi_y)=0,
 \eb
 where it is assumed that $E$ is real and $|E|<m$. Two linearly
 independent solutions of (\ref{strip_de}) can be chosen in the
 following way:
 \begin{eqnarray}
 \label{gpm}
 {\Phi}^{(\pm)}(p,\xi_y) &=&
 \left(2\cos2\xi_y\right)^{\frac{1+2\mu}{2}}\sqrt{\chi(p)}\;\times
 \\
 \nonumber &\times&\left(\begin{array}{c}
 \ds
 C_+^{-1}e^{\pm i(2\mu+2b\pm ip)(\xi_y\mp\frac{\pi}{4})}
 {}_2\tilde{F}_1\left(\mu+b,\mu+\frac12\pm\frac{ip}{2},1+2\mu,1+e^{\pm 4i\xi_y}\right)\vspace{0.1cm}
 \\
 \ds
 \pm i\,C_+\,e^{\pm i(2\mu-2b\mp ip)(\xi_y\mp\frac{\pi}{4})}
 {}_2\tilde{F}_1\left(\mu-b,\mu+\frac12\mp\frac{ip}{2},1+2\mu,1+e^{\pm 4i\xi_y}\right)
 \end{array}\right),
 \end{eqnarray}
 with
 \ben
 \chi(p)= \frac{
 \Gamma(\mu-b+1)\Gamma(\mu+b+1)
 \Gamma\left(\mu+\frac12-\frac{ip}{2}\right)
 \Gamma\left(\mu+\frac12+\frac{ip}{2}\right)}{4\pi\left(\Gamma(1+2\mu)\right)^2}\,.
 \ebn
 The ``$\sim$'' in ${}_2\tilde{F}_1$ indicates that the
 hypergeometric function is defined not on its principal branch,
 but on the cut plane $\Cb\backslash(-\infty,1]$ with
 $\ds\lim_{z\rightarrow0,\mathrm{Im}\, z>0}{}_2\tilde{F}_1(a,b,c,z)=1$,
 as in the formula (3.5) of
 \cite{doyon}. Notice that the first solution, $
 {\Phi}^{(+)}(p,\xi_y)$, satisfies the condition of square
 integrability on the upper edge of the strip $\mathcal{U}$ ($\ds \xi_y=\frac{\pi}{4}$)
 and the second one, $ {\Phi}^{(-)}(p,\xi_y)$, does so on the lower edge ($\ds
 \xi_y=-\frac{\pi}{4}$):
 \be\label{asymptfpm}
 \Phi^{(\pm)}\left(p,\xi_y\rightarrow\pm\frac{\pi}{4}\right)=
 \left(2\cos2\xi_y\right)^{\frac{1+2\mu}{2}}\sqrt{\chi(p)}
 \left(\begin{array}{c}
 C_+^{-1} \vspace{0.1cm}\\ \pm iC_+
 \end{array}\right)+
 O\left(\left(\cos2\xi_y\right)^{\frac{3+2\mu}{2}}\right).
 \eb
 These two solutions verify simple
 symmetry relations:
 \be\label{relsgstrip}
  \Phi^{(+)}(p,\xi_y)=\sigma_z  \,{\Phi}^{(-)}(-p,-\xi_y),\qquad\qquad
   \Phi^{(\pm)}(p,\xi_y)=\sigma_z  \,\overline{\Phi^{(\pm)}(p,\xi_y)}\,.
 \eb
 It is also worthwhile to give a formula for the determinant of
 the fundamental matrix built from $ \Phi^{(+)}(p,\xi_y)$ and $
 \Phi^{(-)}(p,\xi_y)$. Using transformation formulas for hypergeometric
 functions, one obtains
 \be\label{detgstrip}
 \mathrm{det}\left(\Phi^{(+)}(p,\xi_y),\Phi^{(-)}(p,\xi_y)\right)=
 -i\cos2\xi_y \,.
 \eb

 Green function $G_{E,p}(\xi_y,\xi_y')$ of the partial hamiltonian
 $\hat{H}_p$ satisfies the equation
 \be\label{stripge}
 \left(\hat{H}_p(\xi)-E\right)G_{E,p}(\xi_y,\xi_y')=\frac{\cos^2
 2\xi_y}{R^2}\,\delta(\xi_y-\xi_y')\,\mathbf{1}_2.
 \eb
 Analogously to  Subsection~\ref{sect23}, consider the ansatz
 \be\label{partgfstrip}
 G_{E,p}(\xi_y,\xi_y')=  \begin{cases}
    {C}_{E,p}\;\Phi^{(-)}(p,\xi_y)\otimes \Bigl( \Phi^{(+)}(p,\xi_y')\Bigr)^{\dag} &
    \text{for}\;\; -\frac{\pi}{4}<\xi_y<\xi_y'<\frac{\pi}{4}\,, \\
    {C}_{E,p}\;\Phi^{(+)}(p,\xi_y)\otimes \Bigl( \Phi^{(-)}(p,\xi_y')\Bigr)^{\dag} & \text{for}\;\;
    -\frac{\pi}{4}<\xi_y'<\xi_y<\frac{\pi}{4}\,.
  \end{cases}
 \eb
 It solves (\ref{stripge}) for $\xi_y\neq \xi_y'$ and satisfies
 the appropriate boundary conditions as $\ds\xi_y,\xi_y'\rightarrow
 \pm \frac{\pi}{4}\,$. Required singular behaviour at
 $\xi_y=\xi_y'$ is equivalent to the condition
 \be\label{jumpstrip}
 G_{E,p}(\xi_y+0,\xi_y)-
 G_{E,p}(\xi_y-0,\xi_y)=i\,\frac{\cos2\xi_y}{R}\,\sigma_x.
 \eb
 Substituting (\ref{partgfstrip}) into the last relation and using
 symmetry properties (\ref{relsgstrip}), one may show that
 (\ref{jumpstrip}) holds true if we choose
 ${C}_{E,p}=R^{-1}$.

 \subsection{\label{fdras}Fredholm determinant representations}
  \subsubsection{\label{bpr}Boundary projections revisited}
 Let us fix a line $\ds
 \mathcal{L}_{\xi_y^{(0)}}=\left\{\xi\in\mathcal{U}\,|\,\xi_y=\xi_y^{(0)}\right\}$
 and consider a $\Cb^2$-valued function $g_{\xi_y^{(0)}}(\xi_x)\in
 H^{1/2}(\mathcal{L}_{\xi_y^{(0)}})$, represented by its Fourier decomposition
 \be\label{fourier_dec}
 g_{\xi_y^{(0)}}(\xi_x)=\int\nolimits_{-\infty}^{\infty}dp\;\;
 g(p,\xi_y^{(0)})\,e^{ip\xi_x}.
 \eb
 Next we introduce two operators, $Q_{\pm}(\xi_y^{(0)})$,
 whose action on Fourier transform is given by a matrix
 multiplication
 \ben
 Q_{\pm}(\xi_y^{(0)})g(p,\xi_y^{(0)})=Q_{\pm}(p,\xi_y^{(0)})g(p,\xi_y^{(0)}),
 \ebn
 with
 \be\label{qplmn}
 Q_{\pm}(p,\xi_y^{(0)})=\mp
 \frac{i}{\cos2\xi_y^{(0)}}\;\Phi^{(\pm)}(p,\xi_y^{(0)})\otimes
 \left(\Phi^{(\mp)}(p,\xi_y^{(0)})\right)^{\dag}\sigma_x.
 \eb
 Symmetry properties (\ref{relsgstrip}) and the formula
 (\ref{detgstrip}) imply the following result:
 \begin{prop}
 The operators $Q_{\pm}(p,\xi_y^{(0)})$ satisfy the relations
 \be\label{stripproj}
 \left(Q_{\pm}(p,\xi_y^{(0)})\right)^2=Q_{\pm}(p,\xi_y^{(0)}),\qquad
 Q_{+}(p,\xi_y^{(0)})+Q_{-}(p,\xi_y^{(0)})=\mathbf{1}_2.
 \eb
 \end{prop}
 Thus one has a decomposition $\ds H^{1/2}(\mathcal{L}_{\xi_y^{(0)}})=
 H_+(\xi_y^{(0)})\oplus
 H_-(\xi_y^{(0)})$, where $\ds
 H_{\pm}(\xi_y^{(0)})=Q_{\pm}(\xi_y^{(0)})H^{1/2}(\mathcal{L}_{\xi_y^{(0)}})$.
 We remark at once that $\ds H_{\pm}(\xi_y^{(0)})$ will play the role of
 the subspaces $\ds \left(H^{1/2}(\partial
 S^{L,R}_{1,2})\right)_{\pm}$ from  Subsection~\ref{taudef},
 since they are composed of the boundary values of functions
 $\psi$
 satisfying the Dirac equation $\left(\hat{H}^{(0)}_{\text{strip}}-E\right)\psi=0$
 and the condition of square integrability in the region
 $\xi_y>\xi_y^{(0)}$ (in the case of $H_+$) or  $\xi_y<\xi_y^{(0)}$ (for
 $H_-$).

 Using the formulas (\ref{qplmn})--(\ref{stripproj}),
 one may write Fourier transform of $\ds
 g_{\xi_y^{(0)}}(\xi_x)$ as follows:
 \ben
 g(p,\xi_y^{(0)})=\tilde{g}_+(p,\xi_y^{(0)})\Phi^{(+)}(p,\xi_y^{(0)})+
 \tilde{g}_-(p,\xi_y^{(0)})\Phi^{(-)}(p,\xi_y^{(0)}),
 \ebn
 where
 \be\label{hpmcoords}
 \tilde{g}_{\pm}(p,\xi_y^{(0)})=
 \mp\frac{i}{\cos2\xi_y^{(0)}}\;
 \left(\Phi^{(\mp)}(p,\xi_y^{(0)})\right)^{\dag}\sigma_x\;
 g(p,\xi_y^{(0)})\,.
 \eb
 It is convenient to think of $\tilde{g}_{\pm}(p,\xi_y^{(0)})$ as
 coordinates of $g_{\xi_y^{(0)}}$ in $H_{\pm}(\xi_y^{(0)})$. It should be noted
 that $\tilde{g}_{\pm}(p,\xi_y^{(0)})$ are ordinary functions, in contrast to (\ref{diskproj}), where
 a similar notation was used for 2-columns. In
 such
 coordinates, the Propositions~\ref{propwithoutab} and \ref{propwithab} have
 the following form:
   \begin{figure}[h]
 \begin{center}
 \resizebox{5.5cm}{!}{
 \includegraphics{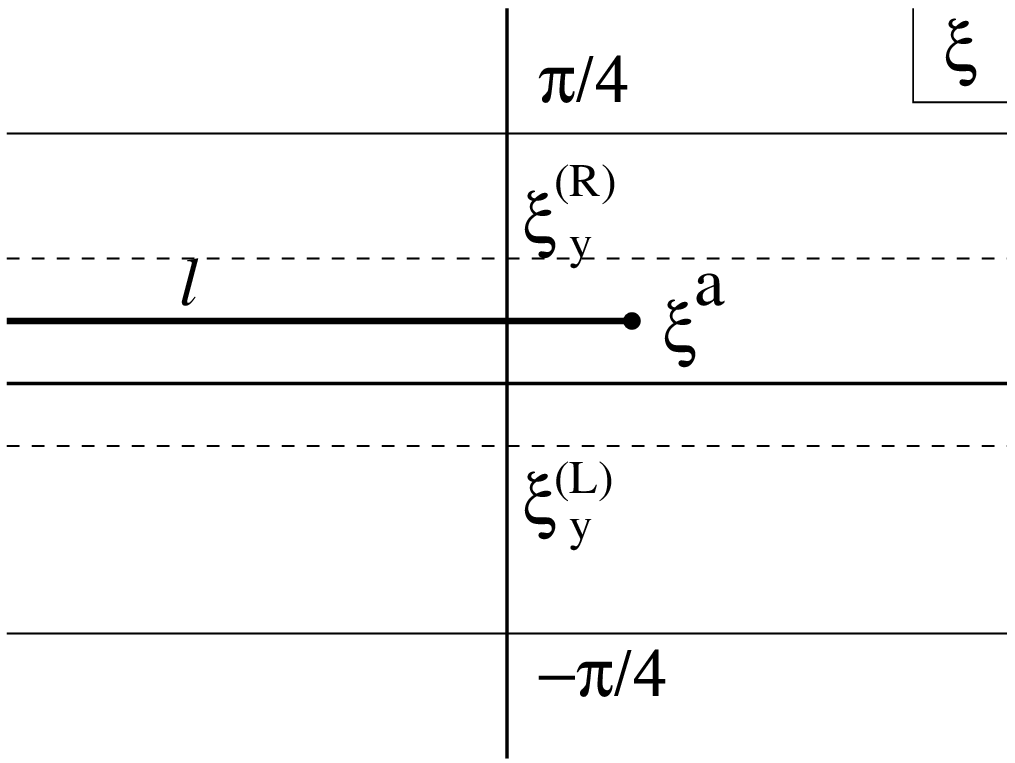}} \\ Fig. 3\\
 \end{center}
 \end{figure}
 \begin{prop}\label{ppp1}
 Let us consider a strip $\ds \mathcal{S}=\left\{\xi\in\mathcal{U}\,|\,
 \xi_y^{(L)}<\xi_y<\xi_y^{(R)}\right\}$. Suppose that a $\Cb^2$-valued function
 $\psi\in H^{1/2}(\partial\mathcal{S})$ can  be continued to
 $\mathcal{S}$ as a solution of the Dirac equation
 $\left(\hat{H}^{(0)}_{\text{strip}}-E\right)\psi=0$. Then one has
 \be\label{stripprojwithout}
    \left(
  \begin{array}{c}
 \tilde{\psi}_{L,-}(p)  \\ \tilde{\psi}_{R,+}(p)
  \end{array}\right)=
  \left(
  \begin{array}{cc}
  0 & 1 \\
  1 & 0
  \end{array}
  \right)
  \left(
  \begin{array}{c}
 \tilde{\psi}_{L,+}(p)  \\ \tilde{\psi}_{R,-}(p)
  \end{array}\right).
 \eb
 \end{prop}
 \begin{prop}\label{ppp2}
 Let us now assume that the strip $\mathcal{S}$ contains one branching
 point $\xi^{a}$ and introduce a horizontal branch cut
 $\ell\in\mathcal{S}$ as shown in Fig.~3. Suppose
 that $\psi\in  H^{1/2}(\partial\mathcal{S})$ is the boundary value
 of a multivalued solution of the Dirac equation on
 $\mathcal{S}\backslash\ell$, characterized by the monodromy $e^{2\pi i
 \nu}$ at the point $\xi^{a}$. Then  \ben
    \left(
  \begin{array}{c}
 \tilde{\psi}_{L,-}(p)  \\ \tilde{\psi}_{R,+}(p)
  \end{array}\right)=
  \left(
  \begin{array}{cc}
  \hat{\alpha}_{\mathcal{S}}(\xi^{a}) & \hat{\beta}_{\mathcal{S}}(\xi^{a}) \\
  \hat{\gamma}_{\mathcal{S}}(\xi^{a}) & \hat{\delta}_{\mathcal{S}}(\xi^{a})
  \end{array}
  \right)
  \left(
  \begin{array}{c}
 \tilde{\psi}_{L,+}(p)  \\ \tilde{\psi}_{R,-}(p)
  \end{array}\right),
 \ebn
 where
 \begin{eqnarray}
 \label{alphast}
 \left(\hat{\alpha}_{\mathcal{S}}(\xi^{a})\tilde{\psi}_{L,+}\right)(p)&=&
 R\int\nolimits_{-\infty}^{\infty}
 \dot{\Delta}_-^{(\xi^{a},\nu)}(p,q)\;
 \tilde{\psi}_{L,+}(q)\;dq,\\
 \label{betast}
 \left(\hat{\beta}_{\mathcal{S}}(\xi^{a})\tilde{\psi}_{R,-}\right)(p)&=&
 R\int\nolimits_{-\infty}^{\infty}
 \dot{G}^{(\xi^{a},\nu)}_{+}(p,q)\;
 \tilde{\psi}_{R,-}(q)\;dq,\\
 \label{gammast}
 \left(\hat{\gamma}_{\mathcal{S}}(\xi^{a})\tilde{\psi}_{L,+}\right)(p)&=&
 R\int\nolimits_{-\infty}^{\infty}
 \dot{G}^{(\xi^{a},\nu)}_{-}(p,q)\;
 \tilde{\psi}_{L,+}(q)\;dq,\\
 \label{deltast}
 \left(\hat{\delta}_{\mathcal{S}}(\xi^{a})\tilde{\psi}_{R,-}\right)(p)&=&
 R\int\nolimits_{-\infty}^{\infty}
 \dot{\Delta}_+^{(\xi^{a},\nu)}(p,q)\;
 \tilde{\psi}_{R,-}(q)\;dq.
 \end{eqnarray}
 and
 \begin{eqnarray}
 \label{deltapmstrip1}
 &\;&\dot{\Delta}_{\pm}^{(\xi^{a},\nu)}(p,q)\;=\;
 \frac{1}{2\pi\cos2\xi_y\cos2\xi_y'}\times\\
 \nonumber &\; &\;\times \;
 \int\limits_{-\infty}^{\infty}\int\limits_{-\infty}^{\infty}d\xi_x\,d\xi_x'\;
 e^{-ip\xi_x+iq\xi_x'}\left(\Phi^{(\mp)}(p,\xi_y)\right)^{\dag}\sigma_x\;
 \dot{\Delta}^{(\xi^a,\nu)}_{\text{strip}}(\xi,\xi')\Bigl|_{\xi_y,\xi_y'\gtrless\xi_y^a}\sigma_x
 \Phi^{(\mp)}(q,\xi_y'),\\
  \label{gpmstrip2}
 &\;&\dot{G}_{\pm}^{(\xi^{a},\nu)}(p,q)\;=\;
 -\frac{1}{2\pi\cos2\xi_y\cos2\xi_y'}\times\\
 \nonumber &\;&\;\times \;
 \int\limits_{-\infty}^{\infty}\int\limits_{-\infty}^{\infty}d\xi_x\,d\xi_x'\;
 e^{-ip\xi_x+iq\xi_x'}\left(\Phi^{(\pm)}(p,\xi_y)\right)^{\dag}\sigma_x\;
 \dot{G}^{(\xi^a,\nu)}_{\text{strip}}(\xi,\xi')\Bigl|_{\xi_y\lessgtr\xi_y^a,\xi_y'\gtrless\xi_y^a}\sigma_x
 \Phi^{(\mp)}(q,\xi_y').
 \end{eqnarray}
 Here, $ \dot{G}^{(\xi^a,\nu)}_{\text{strip}}(\xi,\xi')$ denotes the
 Green function of the Dirac hamiltonian on the strip with one branching point
 $\xi^a$, and $\dot{\Delta}^{(\xi^a,\nu)}_{\text{strip}}(\xi,\xi')=
  \dot{G}^{(\xi^a,\nu)}_{\text{strip}}(\xi,\xi')-
  {G}^{(0)}_{\text{strip}}(\xi,\xi')$.
 \end{prop}
 $\blacksquare$ We illustrate the idea by deriving (\ref{alphast}). On the strip, the formula
 (\ref{alpha}) is replaced by
 \ben
   \Bigl(\hat{\alpha}_{\mathcal{S}}(\xi^a)\,\psi_{L,+}\Bigr)(\xi)=
  - \frac{iR}{\cos2\xi_y^{(L)}}\int\nolimits_{-\infty}^{\infty}
  \dot{\Delta}^{(\xi^a,\nu)}_{\text{strip}}(\xi,\xi')\,
  \sigma_x\,\psi_{L,+}(\xi')\;d\xi_x'\,,
 \ebn
 with $\xi_y=\xi_y'=\xi_y^{(L)}$. Then after Fourier transform one obtains
 \begin{eqnarray}
  \label{auxf111}&\;& \Bigl(\hat{\alpha}_{\mathcal{S}}(\xi^a)\,\tilde{\psi}_{L,+}\Bigr)(p)
   \cdot\Phi^{(-)}(p,\xi_y^{(L)})=\\
   \nonumber&=&
  -\frac{iR}{\cos2\xi_y^{(L)}}\int\nolimits_{-\infty}^{\infty}dq
  \left\{\frac{1}{2\pi}\int\limits_{-\infty}^{\infty}
  \int\limits_{-\infty}^{\infty}d\xi_x d\xi_x'\;
  \dot{\Delta}^{(\xi^a,\nu)}_{\text{strip}}(\xi,\xi')\,e^{-ip\xi_x+iq\xi_x'}
  \right\}\sigma_x\,\Phi^{(+)}(q,\xi_y^{(L)})\,\tilde{\psi}_{L,+}(q)\;.
 \end{eqnarray}
 The columns of the expression in curly brackets satisfy partial
 Dirac equation $(\hat{H}_p(\xi)-E)\psi=0$ and, in addition, for $\xi_y<\xi_y^a$ they
 are square
 integrable as $\ds\xi_y\rightarrow-\frac{\pi}{4}$. Hence
 they are both proportional to $\Phi^{(-)}(p,\xi_y)$. Similarly,
 the rows of this expression are proportional to
 $\left(\Phi^{(-)}(q,\xi_y')\right)^{\dag}$. Therefore, for
 $\xi_y,\xi_y'<\xi_y^a$ one has
 \be\label{deltapmstrip2}
 \frac{1}{2\pi}\int\limits_{-\infty}^{\infty}
  \int\limits_{-\infty}^{\infty}d\xi_x d\xi_x'\;
  \dot{\Delta}^{(\xi^a,\nu)}_{\text{strip}}(\xi,\xi')\,e^{-ip\xi_x+iq\xi_x'}=
  \dot{\Delta}_{-}^{(\xi^{a},\nu)}(p,q)\;
  \Phi^{(-)}(p,\xi_y)\otimes
  \left(\Phi^{(-)}(q,\xi_y')\right)^{\dag},
 \eb
 and it is a simple matter to verify that the coefficient
 $\dot{\Delta}_{-}^{(\xi^{a},\nu)}(p,q)$ is indeed given by
 (\ref{deltapmstrip1}). Substituting (\ref{deltapmstrip2}) into
 (\ref{auxf111}), we obtain (\ref{alphast}).
 $\square$\vspace{0.2cm}

 Finally, suppose that the Poincar\'e strip $\mathcal{U}$ contains
 two branching points, $\xi^{a_1}$ and $\xi^{a_2}$, such that
 $\tanh\xi^{a_j}=a_j$ ($j=1,2$) and $\xi^{a_1}_y<\xi^{a_2}_y$. Using (\ref{unitrans}) and
 going through the definitions of Subsection~\ref{taudef}, one finds that the
 two-point tau function $\tau(a)$ can be written as a Fredholm
 determinant:
 \be\label{tauf2}
 \tau(a)=\mathrm{det}\left(\mathbf{1}-\hat{\alpha}(\xi^{a_2})\hat{\delta}(\xi^{a_1})\right),
 \eb
 The kernels (\ref{alphast}) and (\ref{deltast}) of the integral operators
 $\hat{\alpha}(\xi^{a_2})$
 and $\hat{\delta}(\xi^{a_1})$ do not depend on the choice of the strips
 $\mathcal{S}_1$ and $\mathcal{S}_2$, hence
 the corresponding indices will be omitted from now on.\vspace{0.2cm}\\
 \textbf{Remark}. It turns out
 that the tau function depends only on the geodesic distance
 $d(a_1,a_2)$
 between the points $a_1$ and $a_2$ (see Appendix). Therefore, to
 simplify (\ref{tauf2}), we may choose $\xi^{a_1}=0$,
 $\xi^{a_2}=l_s+i0$, $l_s=\frac{d(a_1,a_2)}{R}\in\Rb^+$. In addition, the invariance
 of $\hat{H}^{(0)}_{\text{strip}}$ with respect
 to $\xi_x$-translations implies that
 \ben
 \dot{\Delta}_{\pm}^{(\xi_x^a+l_s+i\xi_y^a,\nu)}(p,q)=
 e^{i(p-q)l_s}\dot{\Delta}_{\pm}^{(\xi_x^a+i\xi_y^a,\nu)}(p,q).
 \ebn
 Thus it remains to compute only the quantities
 $\dot{\Delta}_{\pm}^{(0,\nu)}(p,q)$, henceforth referred to as
 ``form factors''. This problem is solved in the next subsection.

 \subsubsection{\label{aas}Asymptotics of $\dot{\Delta}(\xi,\xi')$}
 Let us first compute  $\dot{\Delta}_{-}^{(0,\nu)}(p,q)$.
%
 The idea is to use instead of
 (\ref{deltapmstrip1}) the equivalent formula
 (\ref{deltapmstrip2}). Since it is valid for all
 $\xi_y,\xi_y'<0$, one may obtain
 $\dot{\Delta}_{-}^{(0,\nu)}(p,q)$ by exploring the asymptotics of
 both sides of
 this relation as $\ds\xi_y,\xi_y'\rightarrow-\frac{\pi}{4}$. The
 asymptotics of the RHS  of (\ref{deltapmstrip2}) is determined by the formula
 (\ref{asymptfpm}), which gives
 \be\label{asymptRHS}
 \begin{array}{c}
 \text{leading behaviour} \\
 \text{of the RHS of (\ref{deltapmstrip2}) as
 }\xi_y,\xi_y'\rightarrow-\frac{\pi}{4}
 \end{array}\simeq\dot{\Delta}_{-}^{(0,\nu)}(p,q)
 \,\sqrt{\chi(p)\chi(q)}\,\left(4\cos2\xi_y\cos2\xi_y'\right)^{\frac{1+2\mu}{2}}
 \left(\begin{array}{cc}
 C_+^{-2} & i \\ -i & C_+^2
 \end{array}\right).
 \eb

 To find the asymptotics of the LHS, we will use the results
 of Section~\ref{onevortex}. Set for definiteness
 $\ds\Theta=-\frac{\pi}{2}$.  The formula (\ref{deltafanm}) then
 implies that to leading order
 \begin{eqnarray}
 \label{asymptLHS1}
 \Delta(z,z')\Bigl|_{r,r'\rightarrow
 1,|\varphi-\varphi'|<\pi}&\simeq&
 \frac{\sin\pi\nu}{2\pi^2 R }\;
 \frac{\Gamma(\mu-b+1)\Gamma(\mu+b+1)}{\Gamma(1+2\mu)}
 \left( \begin{array}{cc}
 C_+^{-2} & e^{-i\varphi'} \\ e^{i\varphi} & C_+^2
 e^{i(\varphi-\varphi')}
 \end{array}\right)\times\\
 \nonumber
 &\times&\left[(1-r^2)(1-r'^2)\right]^{\frac{1+2\mu}{2}}
 \int\limits_{-\infty}^{\infty}d\theta\;
 \frac{e^{(1+\nu)\theta+i(\varphi-\varphi')}}{e^{\theta+i(\varphi-\varphi')}+1}\;
 \frac{e^{\frac{1-2
 b}{2}\,\theta}}{\left(2+2\cosh\theta\right)^{\frac{1+2\mu}{2}}}\,.
 \end{eqnarray}
 In order to obtain from (\ref{asymptLHS1}) the asymptotics of
 $\dot{\Delta}^{(0,\nu)}_{\text{strip}}(\xi,\xi')$ as
 $\ds\xi_y,\xi'_y\rightarrow -\frac{\pi}{4}$, one should take into
 account the following comments:
    \begin{figure}[h]
 \begin{center}
 \resizebox{10cm}{!}{
 \includegraphics{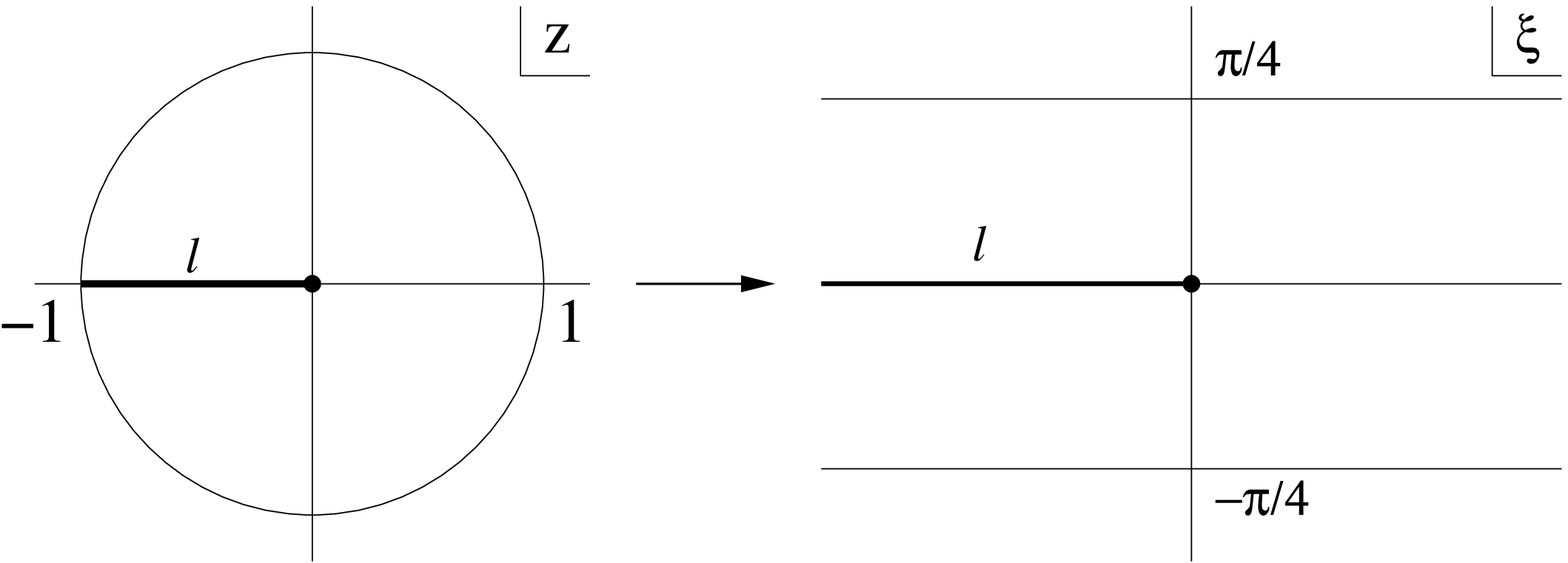}} \\ Fig. 4\\
 \end{center}
 \end{figure}
 \begin{itemize}
 \item Dirac hamiltonian on the disk with one AB vortex is related
 to the free hamiltonian on the disk with one branching point by a
 singular gauge transformation, see (\ref{sgtr2}). The choice of
 the branch cut represented on the Fig.~4 is equivalent to saying
 that $\varphi$ and $\varphi'$ vary between $-\pi$ and $\pi$.
 \item Free hamiltonians on the disk and on the strip are related
 by the unitary transformation (\ref{unitrans2}). Also notice that
 \ben
 U_{DS}(\xi\mapsto \mathrm{arctanh}\,z)\Bigl|_{\xi_y\rightarrow -\frac{\pi}{4}}\simeq
 \left(\begin{array}{cc}
 e^{i\frac{1-2b}{2}\left(\varphi+\frac{\pi}{2}\right)} & 0 \\
 0 & e^{-i\frac{1+2b}{2}\left(\varphi+\frac{\pi}{2}\right)}
 \end{array}\right),\qquad\varphi\in(-\pi,0).
 \ebn
 \item
 The integrals over $\xi_x$ and $\xi_x'$ in (\ref{asymptLHS1})
 can be written as integrals over $\varphi$ and $\varphi'$ using
 the formulas
 \ben
 e^{2\xi_x}\Bigl|_{\xi_y\rightarrow-\frac{\pi}{4}}\simeq
 \mathrm{ctg}\left(-\frac{\varphi}{2}\right),\qquad
 2\cos{2\xi_y}\Bigl|_{\xi_y\rightarrow-\frac{\pi}{4}}\simeq
 \frac{1-r^2}{\sin(-\varphi)}\,,\qquad
 d\xi_x\Bigl|_{\xi_y\rightarrow-\frac{\pi}{4}}\simeq
 \frac{d\varphi}{2\sin(-\varphi)}\,,
 \ebn
 where, as above, $\varphi\in(-\pi,0)$.
 \end{itemize}
 Once the leading behaviour of
 $\dot{\Delta}^{(0,\nu)}_{\text{strip}}(\xi,\xi')$ is found, one
 should substitute it into (\ref{deltapmstrip2}). Then, comparing the
 LHS of (\ref{deltapmstrip2})
 with (\ref{asymptRHS}), we obtain a triple integral
 representation for the form factors:
 \be\label{ff01}
 R\,\dot{\Delta}_{-}^{(0,\nu)}(p,q)=\sqrt{\rho(p)\rho(q)}\;\mathcal{F}_{\nu}(p,q)\,,
 \eb
 where
 \begin{eqnarray}
 \label{ff01a}\rho(p)&=&\frac{2^{2\mu}\;\Gamma(1+2\mu)}{\Gamma\left(\mu+\frac{1}{2}+\frac{ip}{2}\right)
 \Gamma\left(\mu+\frac{1}{2}-\frac{ip}{2}\right)}\,, \\
 \label{ff01b}\mathcal{F}_{\nu}(p,q)&=&\frac{\sin\pi\nu}{2\pi^2}
 \int\limits_{-\infty}^{\infty}d\theta
 \int\limits_{0}^{\pi/2}dx
 \int\limits_{0}^{\pi/2}dy\;\;\frac{
 e^{\left(1+\nu+\frac{1-2b}{2}\right)\left(\theta-2i(x-y)\right)}}{
 \left(e^{\theta-2i(x-y)}+1\right)\left(2+2\cosh\theta\right)^{\frac{1+2\mu}{2}}}\;
 \times
 \\
 \nonumber&\;&\qquad\qquad\qquad\qquad\times\;  (\sin x)^{\mu+\frac{ip}{2}-\frac12}
 (\cos x)^{\mu-\frac{ip}{2}-\frac12}
 (\sin y)^{\mu-\frac{iq}{2}-\frac12}
 (\cos y)^{\mu+\frac{iq}{2}-\frac12}.
 \end{eqnarray}
 One may deduce from (\ref{ff01})--(\ref{ff01b}) the following symmetry properties:
 \be\label{ffsymmetry}
 \dot{\Delta}_{-}^{(0,\nu)}(p,q)=
 \overline{\dot{\Delta}_{-}^{(0,\nu)}(p,q)}=
 \dot{\Delta}_{-}^{(0,\nu)}(q,p).
 \eb

 Another useful representation for $\dot{\Delta}_{-}^{(0,\nu)}(p,q)$ can be found by decomposing the
 integral over $\theta$ in (\ref{ff01b}) into two pieces, $\int\nolimits_{-\infty}^0$
 and $\int\nolimits_{0}^{\infty}$, and  replacing $\left(e^{\theta-2i(x-y)}+1\right)^{-1}$
 by the appropriate geometric series in each piece. Then the
 integrals over $x$, $y$ and $\theta$ can be computed in terms of hypergeometric functions
 and we obtain
 \be\label{ff02}
 R\,\dot{\Delta}_{-}^{(0,\nu)}(p,q)=2^{2\mu-1}\;\frac{\sin\pi\nu}{\pi^2}\;
 \frac{\Bigl[
 \Gamma\left(\mu+\frac{1}{2}+\frac{ip}{2}\right)
 \Gamma\left(\mu+\frac{1}{2}-\frac{ip}{2}\right)
 \Gamma\left(\mu+\frac{1}{2}+\frac{iq}{2}\right)
 \Gamma\left(\mu+\frac{1}{2}-\frac{iq}{2}\right)
 \Bigr]^{1/2}}{\Gamma(1+2\mu)} \;
 \times
 \eb
 \begin{eqnarray*}
 \times\left\{\sum\limits_{n=1}^{\infty}\;\frac{(-1)^{n-1} e^{-\pi(p+q)/4}}{\mu+1+\nu-b+n}
 \!\!\!\!\!\!\!\!\!\!\!\!\right.&\;&{}_2F_1\left(\mu+1+\nu-b+n,\mu+\frac12+\frac{ip}{2},1+2\mu,2-i0\right)\\
 &\;&{}_2F_1\left(\mu+1+\nu-b+n,\mu+\frac12-\frac{iq}{2},1+2\mu,2+i0\right)\\
 &\;&{}_2F_1\,\biggl(\mu+1+\nu-b+n,1+2\mu,\mu+2+\nu-b+n,-1\biggr)\;+\\
 +\sum\limits_{n=0}^{\infty}\;\frac{(-1)^n e^{\pi(p+q)/4}}{\mu-\nu+b+n}
 \!\!\!\!\!\!\!\!\!\!\!\!&\;&{}_2F_1\left(\mu-\nu+b+n,\mu+\frac12+\frac{ip}{2},1+2\mu,2+i0\right)\\
 &\;&{}_2F_1\left(\mu-\nu+b+n,\mu+\frac12-\frac{iq}{2},1+2\mu,2-i0\right)\\
 &\;&\biggr.{}_2F_1\,\biggl(\mu-\nu+b+n,1+2\mu,\mu-\nu+b+n+1,-1\biggr)\biggl\}\,,
 \end{eqnarray*}
 It is worthwhile to note that both series in (\ref{ff02}) rapidly converge,
 since for large~$n$ their $n$-th terms tend to $0$ as at least  $n^{-2-2\mu}$.\vspace{0.1cm}\\
 \textbf{Remark}. We have obtained the representations
 (\ref{ff01})--(\ref{ff01b}) and (\ref{ff02}) assuming that
 $p,q\in\Rb$. However, they also give an analytic
 continuation of  $\dot{\Delta}_{-}^{(0,\nu)}(p,q)$ to the strips
 $|\mathrm{Im}\,p|,\;|\mathrm{Im}\,q|<1+2\mu$. The extension to the whole complex
 $p$- and $q$-plane may also be constructed: in particular, it can be shown that
 $\mathcal{F}_{\nu}(p,q)$ has only simple poles at $\ds p=\pm i(1+2\mu+2n_1)$,
 $\ds q=\pm i(1+2\mu+2n_2)$, where $n_{1,2}=0,1,2,\ldots$ This
 singularity structure will be used later in the study of the
 long-distance behaviour of the $\tau$-function.\vspace{0.2cm}

 The calculation
 of $\dot{\Delta}_{+}^{(0,\nu)}(p,q)$ is quite similar, the result being simply
 $\dot{\Delta}_{+}^{(0,\nu)}(p,q)=\dot{\Delta}_{-}^{(0,\nu)}(-p,-q)$.
 Finally, for another distinguished value of the SAE parameter,
 $\ds\Theta=\frac{\pi}{2}$, the representation (\ref{deltafanp})
 for the corresponding Green function
 implies that
 \be\label{ff03}
 R\,\dot{\Delta}_{\pm}^{(0,\nu)}(p,q)=
 \sqrt{\rho(p)\rho(q)}\;\mathcal{F}_{\nu-1}(\mp p,\mp q)\,.
 \eb
 Thus we can now compute the $\tau$-function of the Dirac hamiltonian for all four types of
 boundary conditions.\vspace{0.2cm}\\
 \textbf{Example}. Let us choose
 $\xi^{a_1}=0$, $\xi^{a_2}=l_s+i0$, as in the remark at the end of
 Subsection~\ref{bpr}. Recall that $l_s=\mathrm{arctanh}\sqrt{s}$ denotes rescaled geodesic distance
 between the points $a_1=0$ and $a_2=s$. We also assume that the upper component of the functions from
 the domain of the Dirac hamiltonian is regular as
 $\xi\rightarrow\xi^{a_{1}}, \xi^{a_{2}}$. Then the $\tau$-function
 (\ref{tauf2}) can be written as follows:
 \be\label{tauf3}
 \tau(s)=\mathrm{det}\Bigl(\mathbf{1}-L_{\nu_2,s}L'_{\nu_1,s}\Bigr),
 \eb
 where the kernels of the integral operators $L_{\nu,s}$, $L'_{\nu,s}$
 are
 \begin{eqnarray}
 \label{lnus1}
 L_{\nu,s}(p,q)&=&e^{i(p-q)l_s/2}\;\sqrt{\rho(p)\rho(q)}\;\mathcal{F}_{\nu}(p,q)\,,\\
 \label{lnus2}
  L'_{\nu,s}(p,q)&=&L_{\nu,s}(-p,-q)\,,
 \end{eqnarray}
 and $ p,q\in\Rb$. The functions $\rho(p)$, $\mathcal{F}_{\nu}(p,q)$ are given by
 (\ref{ff01a}), (\ref{ff01b}).

 \subsection{\label{pbtpvi}Relation to Painlev\'e~VI}
 Let us briefly describe the relation of the present paper to
 the PBT work \cite{beatty}. Recall that the hamiltonian
 $\hat{H}^{(0)}$ of a Dirac particle in the absence of the AB
 fluxes is given by the formula (\ref{dirac_ham}) with
 \begin{eqnarray*}
 K&=&R^{-1}\left[2(1-|z|^2)\partial_z+(1+2b)\bz\right],\\
 K^*&=&-R^{-1}\left[2(1-|z|^2)\partial_{\bz}+(1-2b)z\right].
 \end{eqnarray*}
 Consider the operator
 \ben
 \hat{A}=U(\hat{H}^{(0)}-E)U\sigma_z,\qquad
 U=\mathrm{diag}\left( \left(\frac{m+E}{m-E}\right)^{1/4},
  \left(\frac{m+E}{m-E}\right)^{-1/4}\right).
 \ebn
 It is straightforward to check that $\hat{A}$ coincides with the operator $m-D_k$
 studied by PBT (see, e.~g. the formulas (1.14)--(1.16) in \cite{beatty}) if we identify
 \ben
 m_{\text{PBT}}=\sqrt{m^2-E^2},\qquad k_{\text{PBT}}=-b.
 \ebn
 In the presence of branch points, one should only replace
 $\hat{H}^{(0)}$ in the definition of $\hat{A}$ by the operator $\hat{H}^{(a,\nu)}$, introduced
 in the beginning of this section (recall that $\hat{H}^{(a,\nu)}$
 is obtained from the Dirac hamiltonian with AB field by a singular gauge
 transformation).
 Thus there is a unique correspondence between the multivalued
 solutions of the Dirac equation considered in \cite{beatty} and the
 solutions of  $(\hat{H}^{(a,\nu)}-E)\psi=0$. Using this correspondence, we now reformulate the
 key steps of the PBT analysis in the context
 of the present work.

 \subsubsection{Symmetries and elementary solutions}
 The hamiltonian $\hat{H}^{(0)}$ transforms covariantly under the
 action of the isometry group of the Poincar\'e disk. In
 particular, if  $\psi(z)$ satisfies the equation
 $(\hat{H}^{(0)}-E)\psi=0$, then for any $g=\left(\begin{array}{cc}
 \alpha & \beta \\ \bar{\beta} & \bar{\alpha}
 \end{array}\right)\in SU(1,1)$ the function
 $\psi_g(z)$ defined by
 \be\label{transfsol}
 \psi_g(z)=\left(\begin{array}{cc}
 v(g,z)^{\frac{1-2b}{2}} & 0 \\ 0 & v(g,z)^{-\frac{1+2b}{2}}
 \end{array}\right)\psi(z_{g^{-1}}(z)),  \qquad v(g,z)=\frac{\bar{\alpha}-\beta
 \bar{z}}{\alpha-\bar{\beta}z},
 \eb
 is a solution of the same equation. A basis in the Lie algebra
 of the corresponding complexified infinitesimal symmetries can be chosen in
 the following way:
 \begin{eqnarray*}
  M_+&=&-z^2\partial_z+\partial_{\bz}+bz-\frac{z}{2}\;\sigma_z,\\
 M_-&=&\partial_z-\bz^2\partial_{\bz}-b\bz+\frac{\bz}{2}\;\sigma_z,\\
  M_3&=&z\partial_z-\bz\partial_{\bz}-b+\frac{1}{2}\,\sigma_z=\hat{L}-b.
 \end{eqnarray*}
 These generators satisfy ${sl}(2)$ commutation
 relations
 \ben
 [M_3,M_{\pm}]=\pm M_{\pm},\qquad [M_+,M_-]=2M_3,
 \ebn
 and, therefore, one may introduce two families of solutions $\{w_l\}$, $\{w^*_l\}$ of
 the Dirac equation on $D$ with one branch point at $z=0$, on
 which the symmetries act as follows:
 \be\label{sym_action}
 \begin{array}{lllllll}
 M_3 w_l&=&(l-b)w_l, & \qquad & M_3 w^*_l&=&-(l+b)w^*_l, \vspace{0.1cm}\\
  M_+ w_l&=&m_+(l,b)w_{l+1}, &\qquad & M_+ w^*_l&=&w^*_{l-1}, \vspace{0.1cm}\\
   M_- w_l&=&w_{l-1}, &\qquad & M_- w^*_l&=&m_-(l,b)w^*_{l+1}.
 \end{array}
 \eb
 These relations fix $\{w_l\}$, $\{w^*_l\}$ up to overall
 normalization. Since  $\{w_l\}$, $\{w^*_l\}$ diagonalize both the
 hamiltonian and angular momentum, they are related to the
 radial wave functions introduced in Section~2. It is convenient
 to choose
 \begin{eqnarray*}
 w_l(z)&=& \frac{e^{-i\pi
 l}}{2\pi}\,\frac{\Gamma(\mu-b+1)}{\Gamma(\mu-b+l+1/2)}\,\mathrm{w}^{(II,+)}_{l-1/2}(z),\\
 w^*_l(z)&=& \;\frac{e^{i\pi
 l}}{2\pi}\;\;\frac{\Gamma(\mu+b+1)}{\Gamma(\mu+b+l+1/2)}\,\mathrm{w}^{(II,-)}_{-l-1/2}(z),
 \end{eqnarray*}
 and it then follows that
 $ m_{\pm}(l,b)=\mu^2-(b\mp l\mp 1/2)^2$.
 Using (\ref{transfsol}), we also introduce the elementary
 solutions with one branch point at $z=a$:
 \ben
 w_l(z,a)=V(T[a],z)w_l(T[-a]z),\qquad  w^*_l(z,a)=V(T[a],z)w^*_l(T[-a]z),
 \ebn
 where
 \ben
  T[-a]z=\frac{z-a}{1-\bar{a}z},
  \qquad  V(T[a],z)=\mathrm{diag}\left(\left(\frac{1-a\bar{z}}{1-\bar{a}z}\right)^{\frac{1-2b}{2}},
 \left(\frac{1-a\bar{z}}{1-\bar{a}z}\right)^{-\frac{1+2b}{2}}\right).
 \ebn
 \subsubsection{Local expansions and deformation equations}
 Now consider multivalued solutions of the Dirac equation, which
 are branched at two points $a_1,a_2\in D$ with fixed monodromies
 $e^{2\pi i \nu_{1,2}}$. In a sufficiently small finite neighbourhood of
 each branch point any such solution can be represented by an
 expansion  of the form
 \ben
 \psi[a_j]=\sum\limits_{n\in\Zb+1/2}\alpha_n^j\,\omega_{n+\nu_j}(z,a_j)+
 \beta_n^j\,\omega^*_{n-\nu_j}(z,a_j),\qquad j=1,2.
 \ebn
 It is convenient to introduce instead of $\nu_j$ a new parameter
 $\tilde{\nu}_j$ ($j=1,2$), which can be equal to either $\nu_j$ or
 $\nu_j+1$. We assume that $-1<\nu_{1,2}<0$ and thus
 $0<|\tilde{\nu}_{1,2}|<1$. Consider the response functions $W_j(z,\tilde{\nu})$
 and $W^*_j(z,\tilde{\nu})$ ($j=1,2$), which
 satisfy the following conditions:
 \begin{itemize}
 \item  $W_j(z,\tilde{\nu})$
 and $W^*_j(z,\tilde{\nu})$ are multivalued solutions of the Dirac
 equation with the above monodromy which are square integrable (with the measure $d\mu$)
 as $|z|\rightarrow1$.
 \item $W_j(z,\tilde{\nu})$
 and $W^*_j(z,\tilde{\nu})$ have local expansions of the form
 \begin{eqnarray}
 \label{expp1}W_j(z,\tilde{\nu})[a_k]&=&\delta_{jk}\omega_{-1/2+\tilde{\nu}_k}(z,a_k)+\sum\limits_{n>0}
 \left[a_{n,j}^k\,\omega_{n+\tilde{\nu}_k}(z,a_k)+
  b_{n,j}^k\,\omega^*_{n-\tilde{\nu}_k}(z,a_k)\right],\\
 \label{expp2}W^*_j(z,\tilde{\nu})[a_k]&=&\delta_{jk}\omega^*_{-1/2-\tilde{\nu}_k}(z,a_k)+\sum\limits_{n>0}
 \left[c_{n,j}^k\,\omega_{n+\tilde{\nu}_k}(z,a_k)+
  d_{n,j}^k\,\omega^*_{n-\tilde{\nu}_k}(z,a_k)\right],
 \end{eqnarray}
 where $k=1,2$ and $n\in\Zb+1/2$.
 \end{itemize}
 It turns out that for real values of $E$ such that $|E|<m$ these requirements fix $W_j(z,\tilde{\nu})$
 and $W^*_j(z,\tilde{\nu})$ uniquely. Therefore, the expansions
 (\ref{expp1})--(\ref{expp2}) can be thought of as defining the coefficients $a_{n,j}^k$, $b_{n,j}^k$,
 $c_{n,j}^k$ and $d_{n,j}^k$ as functions of $a$ and $\tilde{\nu}$.

 The lowest order coefficients satisfy a set of deformation
 equations in $a$ (see Theorem~5.0 in \cite{beatty}). If we introduce
 the $2\times2$ matrices with the elements $A_{jk}=a_j\delta_{jk}$,
 $\bar{A}_{jk}=\bar{a}_j\delta_{jk}$,
 $\Lambda_{jk}=\tilde{\nu}_j\delta_{jk}$
 and also
 \begin{eqnarray}\label{relations}
 \begin{array}{lll}
 \left(\mathbf{a}_{1}\right)_{jk}&=&
 \ds a_{1/2,j}^k/{(1-|a_k|^2)},\qquad\qquad
 e=\Lambda-b\mathbf{1}+[\mathbf{a}_{1},A],\vspace{0.1cm}\\
  \left(\mathbf{b}_{1}\right)_{jk}&=&
 \ds b_{1/2,j}^k/{(1-|a_k|^2)},\qquad\qquad
 f=A\mathbf{b}_1\bar{A}-\mathbf{b}_1,\vspace{0.1cm}\\
  \left(\mathbf{c}_{1}\right)_{jk}&=&
 \ds c_{1/2,j}^k/{(1-|a_k|^2)},\qquad \qquad
 g=\mathbf{c}_1-\bar{A}\mathbf{c}_1A,\vspace{0.1cm}\\
  \left(\mathbf{d}_{1}\right)_{jk}&=&
 \ds d_{1/2,j}^k/{(1-|a_k|^2)},\qquad\qquad
 h=\Lambda-b\mathbf{1}-[\mathbf{d}_{1},\bar{A}],
 \end{array}
 \end{eqnarray}
 these equations are given by
 \begin{eqnarray}
 \label{iseqs1}
 de&=& fG+Fg+[E,e],\\
 \label{iseqs2}
 df&=& eF+Fh+Ef-fH,\\
 \label{iseqs3}
 dg&=& hG+Ge+Hg-gE,\\
 \label{iseqs4}
 dh&=& gF+Gf+[H,h],
 \end{eqnarray}
 where
 \begin{eqnarray*}
 E&=&\bigl(\Lambda-(b+1/2)\mathbf{1}\bigr)\,\frac{Ad\bar{A}+\bar{A}dA}{1-|A|^2}+[dA,\mathbf{a}_1],\\
 F&=&dA\mathbf{b}_1\bar{A}+A\mathbf{b}_1 d\bar{A},\\
 G&=&d\bar{A}\,\mathbf{c}_1A+\bar{A}\,\mathbf{c}_1dA,\\
 H&=&-\bigl(\Lambda-(b-1/2)\mathbf{1}\bigr)\,\frac{Ad\bar{A}+\bar{A}dA}{1-|A|^2}+[d\bar{A},\mathbf{d}_1].
 \end{eqnarray*}
 One also has symmetry relations
 \begin{eqnarray}
 \label{symrel1}  ef-fh\;=\;ge-hg&=&0,\\
  \label{symrel2}e^2-fg\;=\;h^2-gf&=&\mu^2\mathbf{1},\\
  \label{symrel3}\left[
 e\sin\pi\Lambda\;(1-|A|^2)\right]^{\dag}&=&h\sin\pi\Lambda\;(1-|A|^2),\\
  \label{symrel4}\left[f\sin\pi\Lambda\;(1-|A|^2)\right]^{\dag}&=&f\sin\pi\Lambda\;(1-|A|^2),\\
   \label{symrel5}\left[g\sin\pi\Lambda\;(1-|A|^2)\right]^{\dag}&=&g\sin\pi\Lambda\;(1-|A|^2).
 \end{eqnarray}
 In addition, the diagonal elements $ a_{1/2,j}^j$, $ d_{1/2,j}^j$
 may be expressed as follows:
 \begin{eqnarray}
 \label{diagajj}
  a_{1/2,j}^j&=&\bar{a}_j \;m_{+}(\tilde{\nu}_j-1/2,b)\;\;+\sum_{k\neq j}e_{jk}a^j_{1/2,k}+\sum_k f_{jk}\bar{a}_k c^j_{1/2,k},\\
 \label{diagdjj}
  d_{1/2,j}^j&=&{a}_j\, m_{-}(-\tilde{\nu}_j-1/2,b)-\sum_{k\neq j}h_{jk}d^j_{1/2,k}-\sum_k g_{jk}{a}_k
  b^j_{1/2,k}.
 \end{eqnarray}
 It is known that the system (\ref{iseqs1})--(\ref{iseqs4}) combined with the relations
 (\ref{symrel1})--(\ref{symrel5}) can be
 integrated in terms of a Painlev\'e~VI transcendent. In
 particular, if we choose $a_1=0$, $a_2=\sqrt{s}$ and set
 \be\label{pvidef}
 \frac{f_{12}f_{21}}{f_{11}f_{22}}=\frac{1-w}{1-s},
 \eb
 then for $\tilde{\nu}_{1,2}>0$ the function $w(s)$ satisfies Painlev\'e~VI equation (\ref{pvi})
 with parameters (\ref{pvipars}) (for the details of the proof, see \cite{beatty}).
 \subsubsection{\label{ssstau}Tau function}
 The link between the deformation equations and the tau function
 considered above is provided by a formula for the derivative of
 the Green function of $\hanu$:
 \begin{eqnarray}
 \label{facder1}(1-|a_j|^2)\,
 \partial_{a_j}\ddot{G}^{(a,\nu)}(z,z')&=&\frac{1}{2R\sin\pi\tilde{\nu}_j}\;
 W_j(z,\tilde{\nu})\otimes W_j^*(z',\tilde{\nu})^{\dag},\\
 \label{facder2} (1-|a_j|^2)\,
 \partial_{\bar{a}_j}\ddot{G}^{(a,\nu)}(z,z')&=&\frac{1}{2R\sin\pi\tilde{\nu}_j}\;
 W^*_j(z,\tilde{\nu})\otimes W_j(z',\tilde{\nu})^{\dag}.
 \end{eqnarray}
 Remarkable factorized form of these expressions follows from the
 fact that Green function $\ddot{G}^{(a,\nu)}(z,z')$ inverts
 the operator $\hanu-E$, and from the analysis of the local expansions of  $\ddot{G}^{(a,\nu)}(z,z')$
 near the branch points. Numerical factors in
 (\ref{facder1})--(\ref{facder2}) may be determined using a
 variant of
 Stokes theorem calculations from Subsection~\ref{taudef}.
 Different boundary conditions for $\hanu$ are encoded into the
 choice of $\{\tilde{\nu}_j\}$: the value $\tilde{\nu}_j=\nu_j+1$
 (or $\tilde{\nu}_j=\nu_j$)
 corresponds to the functions whose upper (resp. lower)
 component is regular at $a_j$.

 Using (\ref{facder1})--(\ref{facder2}) one can show (see
 Theorem~6.3 in \cite{beatty}) that the logarithmic derivative of
 the $\tau$-function (\ref{taudefin}) may be written in terms of
 the lowest order expansion coefficients of $W_j(z,\tilde{\nu})$
 and $W^*_j(z,\tilde{\nu})$:
 \be\label{taufinal}
 d\ln\tau(a,a^0)=\sum_j\frac{1}{1-|a_j|^2}\Bigl\{a^{j}_{1/2,j}da_j+d^j_{1/2,j}d\bar{a}_j\Bigr\}.
 \eb
 Specializing this formula to the case
 $a_1=0$, $a_2=\sqrt{s}$ and $\tilde{\nu}_{1,2}>0$ and using (\ref{diagajj})--(\ref{diagdjj}), PBT have obtained
 the relation~(\ref{taurelpvi}) between
 the corresponding $\tau$-function and the Painlev\'e~VI transcendent
 defined by (\ref{pvidef}). Analogous result holds for arbitrary branch point positions,
 as the $\tau$-function actually
 depends only on the geodesic distance between  $a_1$ and $a_2$.
 The proof of the last statement is given in the Appendix.
 \subsection{\label{ldass}Long distance asymptotics}
 Asymptotics of the PBT $\tau$-function as $l_s\rightarrow\infty$
 (i.~e. as $s\rightarrow1$) can be obtained by expanding the
 Fredholm determinant (\ref{tauf3}):
 \be\label{ldexp1}
 \ln\tau(s)=-\sum\limits_{n=1}^{\infty}\frac{\mathrm{Tr}\left(L_{\nu_2,s}L'_{\nu_1,s}\right)^n}{n}.
 \eb
 Leading order behaviour is determined by the first
 term in (\ref{ldexp1}),
 \be\label{integrand}
 \mathrm{Tr}\left(L_{\nu_2,s}L'_{\nu_1,s}\right)=\int\limits\limits_{-\infty}^{\infty}dp
 \int\limits_{-\infty}^{\infty}dq\;\;\rho(p)\rho(q)\,\mathcal{F}_{\nu_2}(p,q)\mathcal{F}_{\nu_1}(-p,-q)
 \,e^{i(p-q)l_s}.
 \eb
 The integrand in (\ref{integrand}) has particularly simple
 analytic properties in the complex $p$- and $q$-plane. It has
 already been noted above that $\mathcal{F}_{\nu_2}(p,q)$ and $\mathcal{F}_{\nu_1}(-p,-q)$
 have simple poles at $\ds p=\pm i(1+2\mu+2n_1)$, $\ds q=\pm i(1+2\mu+2n_2)$, where
 $n_{1,2}=0,1,2,\ldots$
 At the same points, the functions $\rho(p)$ and
 $\rho(q)$ have simple zeros and thus the whole integrand
 has simple poles. We can then close the contours of integration
 over $p$ and $q$ in the upper and lower half-plane, respectively
 (recall that $l_s>0$). Summing over the residues, one finds a
 long-distance expansion of the integral (\ref{integrand}). The leading term of
 this expansion corresponds to the poles at
 $p=i(1+2\mu)$ and $q=-i(1+2\mu)$ and yields
 \be\label{traceas}
 \mathrm{Tr}\left(L_{\nu_2,s}L'_{\nu_1,s}\right)\simeq \pi^2
 2^{4\mu}\Bigl[\mathrm{res}\,\mathcal{F}_{\nu_1}(p,q)\Bigr]_{\substack{p\,=\,i(1+2\mu) \\ q=-i(1+2\mu)}}
 \Bigl[\mathrm{res}\,
 \mathcal{F}_{\nu_2}(p,q)\Bigr]_{\substack{p\,=\,i(1+2\mu) \\ q=-i(1+2\mu)}}\;e^{-2(1+2\mu)l_s}+
 O\left(e^{-2(2+2\mu)l_s}\right).
 \eb
 The residues in the last formula can be extracted from, e.~g., the
 representation (\ref{ff02}). After somewhat cumbersome computation one finds
 \be\label{residue}
 \Bigl[\mathrm{res}\,\mathcal{F}_{\nu}(p,q)\Bigr]_{\substack{p\,=\,i(1+2\mu)\\
 q=-i(1+2\mu)}}=\frac{2\sin\pi\nu}{\pi^2}\frac{\Gamma(\mu+2+\nu-b)\Gamma(\mu-\nu+b)}{\Gamma(2+2\mu)}.
 \eb
 Combining (\ref{ldexp1})--(\ref{residue}) and using that $\ds 4e^{-2l_s}\simeq
 {1-s}$ as $s\rightarrow 1$, we finally obtain the asymptotics
 \ben
 \tau(s)\simeq 1-A_{\tau}(1-s)^{1+2\mu}+O\left((1-s)^{2+2\mu}\right)\;\text{ as
 }s\rightarrow1,
 \ebn
 with $A_{\tau}$ given by (\ref{tauas}). This finishes the proof
 of Theorem~\ref{mainthm}.
 \section{$PVI\rightarrow PV$: flat space limit}
 In the present section, the analogs of the above
 results in the limit of flat space are established. Since the self-adjoint extensions and
 the spectrum of the corresponding Dirac hamiltonian have already
 been discussed in \cite{falomir} (see also \cite{desousa} for the zero-field case),
 we will not dwell much on
 this point. One-vortex Green
 function was also computed in a closed form by Gavrilov \textit{et al}
 \cite{gavrilov}. However, the representation found  in
 \cite{gavrilov} is inconvenient for our purposes, so below we obtain another formula,
 in which the vortex-dependent
 contribution to the resolvent is manifestly separated from the ``free''
 part. This formula enables us to find Fredholm determinant representations for
 the two-point tau function of the
 Dirac hamiltonian on the plane, which turns out to be
 related to a class of Painlev\'e~V transcendents.

 \subsection{One-vortex Green function}
 Quantum motion of a Dirac particle on the plane in the presence
 of an external magnetic field is described by the hamiltonian
 \ben
 \hat{H}=\left(
 \begin{array}{cc}
 m & 2D_z \\ -2D_{\bz} & -m
 \end{array}\right),\qquad z\in\Cb.
 \ebn
 As above, we will consider a vector potential describing the
 superposition of a uniform magnetic field $B$ and of the field
 of an AB flux $\Phi=2\pi\nu$ ($-1<\nu<0$), situated at the origin.
 Analogously to (\ref{vect_b})--(\ref{vect_nu}), one can choose
 \begin{eqnarray}
 \label{ubfieldplane}
 \mathcal{A}^{(B)}&=&-\frac{i}{4}\,B\bigl(\bz\,
 dz-z\,d\bz\bigr)\;=\;\frac{Br^2}{2}\,d\varphi\,,
 \\
 \label{abfluxplane}
 \mathcal{A}^{(\nu)}&=&-\frac{i\nu}{2}\left(\frac{dz}{z}-\frac{d\bz}{\bz}\right)
 \;=\;\nu\,d\varphi\,,
 \end{eqnarray}
 and then the hamiltonian becomes
 \be\label{dirhamplane}
 \hat{H}=\left(
 \begin{array}{cc}
 m & \ds 2\partial_z+\frac{B\bz}{2}+\frac{\nu}{z} \\
 \ds -2\partial_{\bz}+\frac{Bz}{2}+\frac{\nu}{\bz} & -m
 \end{array}\right).
 \eb
 Radial hamiltonians $\hat{H}_{l_0+\nu}$ ($l_0\in\Zb$), corresponding to
 different angular momentum eigenvalues (equal to $\ds l_0+\frac12$), are defined by
 \ben
 \hat{H}_l=\left(\begin{array}{cc}
 m &
 \ds \partial_r+\frac{l+1}{r}+\frac{Br}{2} \\
 \ds
 -\partial_r+\frac{l}{r}+\frac{Br}{2}
 & -m
 \end{array}\right).
 \ebn
 The details of further calculation depend on the sign of $B$ and below we consider
 different cases separately.

 \subsubsection{$B>0$}
 First we introduce two families of solutions of the Dirac equation without
 AB field,
 $(\hat{H}^{(0)}-E)\psi=0$:
 \begin{eqnarray}
 \label{pwavep}
 \Psi_+(z,\theta)&=&
 e^{-\frac{B|z|^2}{4}\,-\sqrt{B/2}\;\bz\, e^{\theta}} \left(\begin{array}{c}
 \ds C_+^{-1}e^{-\theta/2}\left(1+\sqrt{B/2}\; z\, e^{-\theta}\right)^{-1-\frac{\lambda^2}{2B}} \\
 C_+e^{\,\theta/2}\left(1+\sqrt{B/2}\; z\, e^{-\theta}\right)^{-\frac{\lambda^2}{2B}}
 \end{array}\right),\\
 \label{pwavem}
 \Psi_-(z,\theta)&=&
 e^{\frac{B|z|^2}{4}\,+\sqrt{B/2}\;z\, e^{-\theta}} \left(\begin{array}{c}
 \ds C_-^{-1}e^{-\theta/2}\left(1+\sqrt{B/2}\; \bz\, e^{\theta}\right)^{\frac{\lambda^2}{2B}} \\
 -C_-e^{\,\theta/2}\left(1+\sqrt{B/2}\; \bz\, e^{\theta}\right)^{-1+\frac{\lambda^2}{2B}}
 \end{array}\right),
 \end{eqnarray}
 with
 \ben
 \lambda=\sqrt{m^2-E^2},
 \ebn
 \be\label{cpl01}
 C_{\pm}=\left(\frac{m-E}{m+E}\right)^{1/4}\left(\frac{2B}{\lambda^2}\right)^{\pm
 1/4}.
 \eb
 Define
 $\hat{\Psi}_{\pm}(z,\theta)=
 \Psi_{\pm}(z\longleftrightarrow\bz,\theta\longleftrightarrow-\theta)$,
 as in (\ref{psihat}). The functions $\Psi_+(z,\theta)$ and $\hat{\Psi}_-(z,\theta)$
 are delimited by the horizontal branch cuts $\Bigl(-\infty+i(\varphi+\pi+2\pi\Zb),
 \ln\left( \sqrt{B/2}\; r\right)+i(\varphi+\pi+2\pi\Zb)\Bigr]$ in the $\theta$-plane,
 whereas the branch cuts for
 $\Psi_-(z,\theta)$ and $\hat{\Psi}_+(z,\theta)$ are
 $\Bigl[-\ln \left( \sqrt{B/2}\; r\right) +i(\varphi+\pi+2\pi\Zb),\infty+i(\varphi+\pi+2\pi\Zb)\Bigr)$ (see Fig.~5).
 The arguments of both $1+\sqrt{B/2}\; z\, e^{-\theta}$ and $1+\sqrt{B/2}\; \bz\,
 e^{\theta}$ are fixed to be zero on the line
 $\mathrm{Im}\,\varphi=\pi$.
     \begin{figure}[h]
 \begin{center}
 \resizebox{7cm}{!}{
 \includegraphics{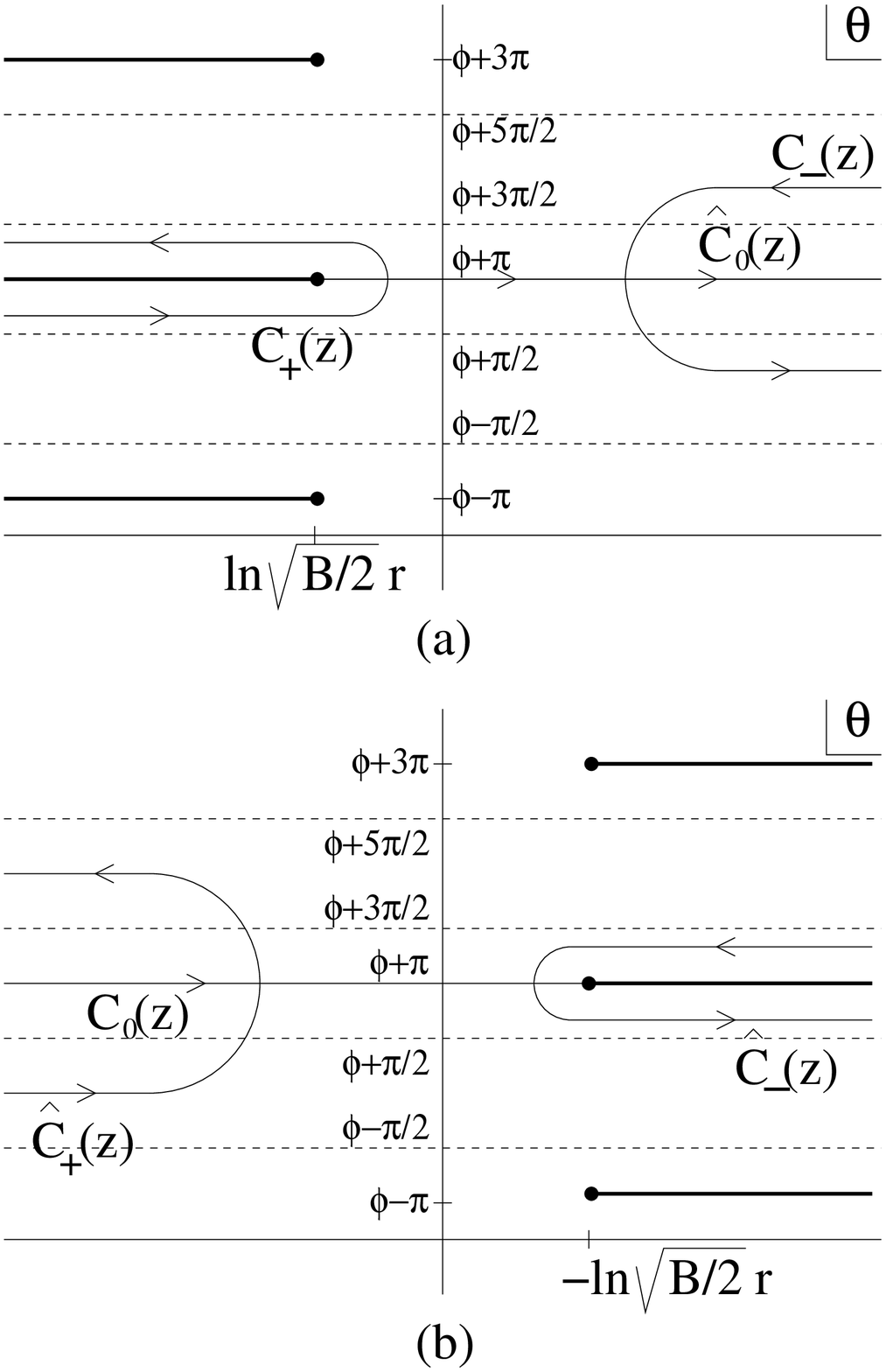}} \\ Fig. 5: Branch cuts and integration contours
 in the $\theta$-plane for \\  a) $\Psi_+(z,\theta)$ and $\hat{\Psi}_-(z,\theta)$
  b) $\Psi_-(z,\theta)$ and $\hat{\Psi}_+(z,\theta)$ \\
 \end{center}
 \end{figure}

 Next introduce the following functions:
  \be\label{planeradsol1}
 \mathrm{w}_l^{(I)}(z)=\int\nolimits_{C_0(z)}
 e^{\left(l+1/2\right)\theta}\,\Psi_-(z,\theta)\,d\theta=
 e^{i\pi l}
 \left(\begin{array}{cc}
 e^{il\varphi} & 0 \\ 0 & e^{i(l+1)\varphi}
 \end{array}\right) w_l^{(I)}(r),
 \eb
 \be\label{planeradsol2}
 \mathrm{w}_l^{(II,\pm)}(z)=\pm\int\nolimits_{C_\pm(z)}
 e^{\left(l+1/2\right)\theta}\,\Psi_+(z,\theta)\,d\theta=
 2\pi i\,e^{i\pi l}
 \left(\begin{array}{cc}
 e^{il\varphi} & 0 \\ 0 & e^{i(l+1)\varphi}
 \end{array}\right) w_l^{(II,\pm)}(r),
 \eb
 \be\label{planeradsol3}
 \hat{\mathrm{w}}_l^{(I)}(z)=\int\nolimits_{\hat{C}_0(z)}
 e^{-\left(l+1/2\right)\theta}\,\hat{\Psi}_-(z,\theta)\,d\theta=
 e^{-i\pi l}
 \left(\begin{array}{cc}
 e^{-il\varphi} & 0 \\ 0 & e^{-i(l+1)\varphi}
 \end{array}\right) w_l^{(I)}(r),
 \eb
  \be\label{planeradsol4}
 \hat{\mathrm{w}}_l^{(II,\pm)}(z)=\mp\int\nolimits_{\hat{C}_\mp(z)}
 e^{-\left(l+1/2\right)\theta}\,\hat{\Psi}_+(z,\theta)\,d\theta=
 2\pi i \,e^{-i\pi l}
 \left(\begin{array}{cc}
 e^{-il\varphi} & 0 \\ 0 & e^{-i(l+1)\varphi}
 \end{array}\right) w_l^{(II,\pm)}(r),
 \eb
 where the integration contours $C_0(z)$, $\hat{C}_0(z)$, $C_{\pm}(z)$ and $\hat{C}_{\pm}(z)$
 are shown in Fig.~5. Notice that $\mathrm{w}_l^{(II,+)}(z)$ and $\hat{\mathrm{w}}_l^{(II,+)}(z)$
 are well-defined for $l>-1$, while $\mathrm{w}_l^{(II,-)}(z)$, $\hat{\mathrm{w}}_l^{(II,-)}(z)$
 are well-defined for $l<0$. The functions $w_l^{(I)}(r)$,
 $w_l^{(II,\pm)}(r)$ satisfy radial Dirac equation
 $(\hat{H}_l-E)w_l=0$. They are explicitly given by
 \begin{eqnarray}
 \label{plsol1a}
 w_l^{(I)}(r)&=& \frac{\lambda}{\sqrt{2B}}\;\Gamma\left(\frac{\lambda^2}{2B}\right)
 e^{-Br^2/4}\left(\begin{array}{c}
 C_+^{-1}\bigl(\sqrt{B/2}\;r\bigr)^{-l}U\left(\frac{\lambda^2}{2B}+1,1-l,\frac{Br^2}{2}\right)\\
 C_+\bigl(\sqrt{B/2}\;r\bigr)^{-l-1}U\left(\frac{\lambda^2}{2B},-l,\frac{Br^2}{2}\right)
 \end{array}\right)=\\
 \label{plsol1b}
 &=&\frac{\lambda}{\sqrt{2B}}\;\Gamma\left(\frac{\lambda^2}{2B}\right)
 e^{-Br^2/4}\left(\begin{array}{c}
 C_+^{-1}\bigl(\sqrt{B/2}\;r\bigr)^{l}U\left(\frac{\lambda^2}{2B}+1+l,1+l,\frac{Br^2}{2}\right)\\
 C_+\bigl(\sqrt{B/2}\;r\bigr)^{l+1}U\left(\frac{\lambda^2}{2B}+1+l,2+l,\frac{Br^2}{2}\right)
 \end{array}\right),
 \end{eqnarray}
 \begin{eqnarray}
 \label{plsol2p}
 w_l^{(II,+)}(r)&=&
 \left(\begin{array}{c}
 \ds
 \frac{\Gamma\left(\frac{\lambda^2}{2B}+1+l\right)}{\Gamma\left(\frac{\lambda^2}{2B}+1\right)\Gamma(1+l)}
 \; C_+^{-1}e^{-Br^2/4}
 \bigl(\sqrt{B/2}\;r\bigr)^{l}\textstyle M\left(\frac{\lambda^2}{2B}+1+l,1+l,\frac{Br^2}{2}\right)\\
 \ds -
 \frac{\Gamma\left(\frac{\lambda^2}{2B}+1+l\right)}{\Gamma\left(\frac{\lambda^2}{2B}\right)\Gamma(2+l)}
 \; C_+e^{-Br^2/4}
 \bigl(\sqrt{B/2}\;r\bigr)^{l+1}\textstyle M\left(\frac{\lambda^2}{2B}+1+l,2+l,\frac{Br^2}{2}\right)
 \end{array}\right),\\
 \label{plsol2m}
  w_l^{(II,-)}(r)&=&
 \left(\begin{array}{c}
 \ds \frac{C_+^{-1}
  e^{-Br^2/4}}{\Gamma(1-l)}\;
 \bigl(\sqrt{B/2}\;r\bigr)^{-l}\textstyle M\left(\frac{\lambda^2}{2B}+1,1-l,\frac{Br^2}{2}\right)
 \\
 \ds - \frac{C_+e^{-Br^2/4}}{\Gamma(-l)}\;
 \bigl(\sqrt{B/2}\;r\bigr)^{-l-1}\textstyle M\left(\frac{\lambda^2}{2B},-l,\frac{Br^2}{2}\right)
 \end{array}\right),
 \end{eqnarray}
 where $M(\alpha,\beta,s)$ and $U(\alpha,\beta,s)$ denote Kummer's
 functions, see \cite{abst}. The solutions
 (\ref{plsol1a})--(\ref{plsol2m}) have the same integrability
 properties, as in the case of the disk. Namely,
 \begin{itemize}
 \item $w_l^{(I)}(r)$ is the only solution of the radial Dirac
 equation, square integrable as $r\rightarrow\infty$ with respect
 to the measure $d\mu_r=rdr$.
 \item For $l\geq 0$ ($l\leq-1$), $w_l^{(II,+)}(r)$ (resp. $w_l^{(II,-)}(r)$) is the only solution,
 square integrable as $r\rightarrow0$. For $l\in(-1,0)$ both
 $w_l^{(II,\pm)}(r)$ are square integrable as $r\rightarrow0$.
 \item  $w_l^{(I)}(r)$ and $w_l^{(II,+)}(r)$ are linearly
 independent for $l>-1$, $w_l^{(I)}(r)$ and $w_l^{(II,-)}(r)$ are
 linearly independent for $l<0$. The determinant of the
 fundamental matrix of solutions is given by
  \be
  \label{wrplane}
 \mathrm{det}\left(w_l^{(I)},w_l^{(II,\pm)}(r)\right)=
 -\frac{2}{\lambda r}\,.
 \eb
 \end{itemize}
 This implies
 \begin{prop}[\cite{falomir}] Let $\mathrm{dom}\,\hat{H}_l=C^{\infty}_0(\Rb^+)$.
 Then for
 $l\in(-\infty,-1]\cup[0,\infty)$ $\hat{H}_l$ is essentially self-adjoint.
 In the case $l\in(-1,0)$ it admits a one-parameter family of SAEs.
 \end{prop}
 We will consider two particular SAEs of  $\hat{H}_{\nu}$,
 whose domains are composed of functions with regular at $r=0$
 lower (or upper) component. Similarly to the above, these two cases will be
 referred to as corresponding to the SAE parameter
 $\displaystyle \Theta=\frac{\pi}{2}$ ($\displaystyle
 -\frac{\pi}{2}$).

 Using (\ref{wrplane}), one may show that for
 $l\in(-\infty,-1]\cup[0,\infty)$ the radial Green function
 $G_{E,l}(r,r')$ is given by
  \be\label{planegfans}
 G_{E,l}(r,r')=
  \begin{cases}
    \frac{\lambda}{2}\;w^{(II,\pm)}_l(r)\otimes
    \Bigl( w^{(I)}_l(r')\Bigr)^T & \text{for}\;\; 0<r<r'<1, \\
    \frac{\lambda}{2}\;w^{(I)}_l(r)\;\otimes
    \Bigl( w^{(II,\pm)}_l(r')\Bigr)^T & \text{for}\;\;
    0<r'<r<1.
  \end{cases}
 \eb
 As in Subsection~\ref{sect23}, the sign ``$+$'' (``$-$'') in (\ref{planegfans})
 corresponds to $l\geq0$ ($l\leq-1$). For $l\in(-1,0)$, the
 relation (\ref{planegfans}) gives the Green function of the SAE
 with $\ds\Theta=-\frac{\pi}{2}$ (when taken with the sign ``$-$'') and
 with $\ds\Theta=\frac{\pi}{2}$ (for the sign ``$+$'').

 We can now repeat word-for-word the calculation of the full one-vortex Green
 function
  $G(z,z')$ from Subsection~\ref{full1v}, using the formula
  (\ref{planegfans})
 and contour integral representations
 (\ref{planeradsol1})--(\ref{planeradsol4}). Setting for
 definiteness $\ds \Theta=-\frac{\pi}{2}$, one obtains the
 following representation for the Green function of the Dirac
 hamiltonian (\ref{dirhamplane}):
   \be\label{ansplane}
 G(z,z')=\left\{
 \begin{array}{rl}
 e^{- i \nu(\varphi-\varphi'+2\pi)}\,G^{(0)}(z,z')+\Delta(z,z') & \text{for }\varphi-\varphi'\in(-2\pi,-\pi),\\
 e^{- i \nu(\varphi-\varphi')}\; G^{(0)}(z,z')+\Delta(z,z') & \text{for }\varphi-\varphi'\in(-\pi,\pi),\\
 e^{- i \nu(\varphi-\varphi'-2\pi)}\,G^{(0)}(z,z')+\Delta(z,z') & \text{for
 }\varphi-\varphi'\in(\pi,2\pi)\,,
 \end{array}\right.
 \eb
 with
 \be\label{g0v1plane}
 G^{(0)}(z,z')=\frac{\lambda}{4\pi}\int\limits_{C_0(z)}\!\!d\theta\;
 \Psi_-(z,\theta)\otimes\hat{\Psi}^T_+(z',\theta),
 \eb
 \be\label{deltav1plane}
 \Delta(z,z')=\lambda\,e^{-i\nu(\varphi-\varphi')}
 \frac{1-e^{-2\pi i
 \nu}}{8i\pi^2}\;
  \int\limits_{C_0(z)}\!\!d\theta_1\!\!\!\!
 \int\limits_{\mathrm{Im}\,\theta_2=\varphi'}\!\!\!\!\!\!\!\! d\theta_2
 \;\;\;\Psi_-(z,\theta_1)\otimes\hat{\Psi}^T_+(z',\theta_2)\;
 \frac{e^{\left(1+\nu+\frac12\right)(\theta_1-\theta_2)}}{e^{\theta_1-\theta_2}-1}\,.
 \eb
 The integrals (\ref{g0v1plane}) and (\ref{deltav1plane})
 reduce to
 \begin{eqnarray}
 \label{g0planefan}
 G^{(0)}(z,z')&=&e^{\frac{B}{4}\,(\bz z'-z\bz')}
 \left(\begin{array}{cc}
 \displaystyle
 \zeta_{11}\Bigl(u(z,z')\Bigr) &
 \displaystyle
  \frac{\sqrt{u(z,z')}}{z'-z}\;\zeta_{12}\Bigl(u(z,z')\Bigr) \\
 \displaystyle
 - \frac{\sqrt{u(z,z')}}{\bz'-\bz}\;\zeta_{21}\Bigl(u(z,z')\Bigr) &
 \displaystyle
 -\zeta_{22}\Bigl(u(z,z')\Bigr)
 \end{array}\right),
 \\
 \label{deltafanmplane}
 \Delta(z,z')&=&\frac{\sin\pi\nu}{\pi}\int\limits_{-\infty}^{\infty}d\theta\;
 \frac{e^{(1+\nu)\theta+i(\varphi-\varphi')}}{e^{\theta+i(\varphi-\varphi')}+1}
 \;e^{-\frac{B}{2}\,rr'\sinh\theta}\;\times
 \\
 \nonumber &\times&\left(\begin{array}{cc}
 \displaystyle
 \zeta_{11}\Bigl(v(r,r',\theta)\Bigr) &
 \displaystyle
 \frac{e^{-i\varphi'}\sqrt{v(r,r',\theta)}}{re^{-\theta}+r'}\,
 \zeta_{12}\Bigl(v(r,r',\theta)\Bigr) \\
 \displaystyle
 \frac{ e^{\theta+i\varphi}\sqrt{v(r,r',\theta)}}{re^{\theta}+r'}\,
 \zeta_{21}\Bigl(v(r,r',\theta)\Bigr) &
 \displaystyle
 e^{\theta+i(\varphi-\varphi')}
 \zeta_{22}\Bigl(v(r,r',\theta)\Bigr)
 \end{array}\right),
 \end{eqnarray}
 where $\zeta(u)$, $u(z,z')$, and $v(r,r',\theta)$ are defined as
 follows:
 \be\label{zetadefplane}
 \zeta(u)=\frac{e^{-\frac{Bu}{4}}}{2\pi}\,
 \sqrt{\frac{B}{2}}\;\Gamma\left(\frac{\lambda^2}{2B}+1\right)
 \left(
 \begin{array}{cc}
 C_+^{-2}\,U\left(\frac{\lambda^2}{2B}+1,1,\frac{Bu}{2}\right) &
 \sqrt{\frac{Bu}{2}}\,U\left(\frac{\lambda^2}{2B}+1,2,\frac{Bu}{2}\right) \vspace{0.1cm} \\
 \sqrt{\frac{Bu}{2}}\,U\left(\frac{\lambda^2}{2B}+1,2,\frac{Bu}{2}\right) &
 C_+^{2}\,U\left(\frac{\lambda^2}{2B},1,\frac{Bu}{2}\right)
 \end{array}\right),
 \eb
 \be\label{vrrprplane}
 u(z,z')=|z-z'|^2,\qquad v(r,r',\theta)={r^2+r'^2+2rr'\cosh\theta}.
 \eb
 \textbf{Remark}. Green function for $\ds\Theta=\frac{\pi}{2}$ can be computed in a
 completely analogous manner. Final answer differs from
 (\ref{ansplane}), (\ref{g0planefan})--(\ref{vrrprplane}) in only
 one point: one has to replace $\ds
 e^{(1+\nu)\theta+i(\varphi-\varphi')}$ by $-e^{\nu\theta}$ in the
 first line of the integral representation (\ref{deltafanmplane})
 for $\Delta(z,z')$, just as in (\ref{deltafanp}) vs.
 (\ref{deltafanm}) in the disk case.

 \subsubsection{$B<0$}
 In the previous subsection, we have obtained $G(z,z')$ imitating
 the calculation made on the Poincar\'e disk. Alternatively, one can simply
 consider the limit
 \ben
 R\rightarrow\infty,\qquad r\rightarrow \frac{r}{R},\qquad
 r'\rightarrow \frac{r'}{R}
 \ebn
 of the relations (\ref{ans}), (\ref{g0fan})--(\ref{vrrpr}), using asymptotic properties of
 hypergeometric functions. For $B<0$ and $\ds
 \Theta=-\frac{\pi}{2}$ (regular upper component) this gives a
 representation of the one-vortex Green function, which has
 exactly the same form as (\ref{ansplane}),
 (\ref{g0planefan})--(\ref{deltafanmplane}), but with
  \be\label{zetadefplane2}
 \zeta(u)=\frac{e^{-\frac{|B|u}{4}}}{2\pi}\,
 \sqrt{\frac{|B|}{2}}\;\Gamma\left(\frac{\lambda^2}{2|B|}+1\right)
 \left(
 \begin{array}{cc}
 C_+^{-2}\,U\left(\frac{\lambda^2}{2|B|},1,\frac{|B|u}{2}\right) &
 \sqrt{\frac{|B|u}{2}}\,U\left(\frac{\lambda^2}{2|B|}+1,2,\frac{|B|u}{2}\right) \vspace{0.1cm} \\
 \sqrt{\frac{|B|u}{2}}\,U\left(\frac{\lambda^2}{2|B|}+1,2,\frac{|B|u}{2}\right) &
 C_+^{2}\,U\left(\frac{\lambda^2}{2|B|}+1,1,\frac{|B|u}{2}\right)
 \end{array}\right)
 \eb
 and
 \be
 \label{cplplane}
 C_+=\left(\frac{m-E}{m+E}\right)^{1/4}\left(\frac{\lambda^2}{2|B|}\right)^{
 1/4}.
 \eb

 \subsubsection{$B=0$}
 Taking further limit $B\rightarrow0$ in
 (\ref{g0planefan})--(\ref{zetadefplane}), we get (still for
 $\ds\Theta=-\frac{\pi}{2}$, the modification for $\ds\Theta=\frac{\pi}{2}$
 is as described above)
  \be
 \label{g0planefan00}
 G^{(0)}(z,z')=
 \left(\begin{array}{cc}
 \displaystyle
 \zeta_{11}\Bigl(u(z,z')\Bigr) &
 \displaystyle
  \frac{\sqrt{u(z,z')}}{z'-z}\;\zeta_{12}\Bigl(u(z,z')\Bigr) \\
 \displaystyle
 - \frac{\sqrt{u(z,z')}}{\bz'-\bz}\;\zeta_{21}\Bigl(u(z,z')\Bigr) &
 \displaystyle
 -\zeta_{22}\Bigl(u(z,z')\Bigr)
 \end{array}\right),
 \eb
 \begin{eqnarray}
 \label{deltafanmplane00}
 \Delta(z,z')&=&\frac{\sin\pi\nu}{\pi}\int\limits_{-\infty}^{\infty}d\theta\;
 \frac{e^{(1+\nu)\theta+i(\varphi-\varphi')}}{e^{\theta+i(\varphi-\varphi')}+1}
 \;\times\\
 \nonumber
 &\times&\left(\begin{array}{cc}
 \displaystyle
 \zeta_{11}\Bigl(v(r,r',\theta)\Bigr) &
 \displaystyle
 \frac{e^{-i\varphi'}\sqrt{v(r,r',\theta)}}{re^{-\theta}+r'}\,
 \zeta_{12}\Bigl(v(r,r',\theta)\Bigr) \\
 \displaystyle
 \frac{ e^{\theta+i\varphi}\sqrt{v(r,r',\theta)}}{re^{\theta}+r'}\,
 \zeta_{21}\Bigl(v(r,r',\theta)\Bigr) &
 \displaystyle
 e^{\theta+i(\varphi-\varphi')}
 \zeta_{22}\Bigl(v(r,r',\theta)\Bigr)
 \end{array}\right),
 \end{eqnarray}
 where the matrix $\zeta(u)$ is given by
 \be\label{zetab0}
 \zeta(u)=\frac{\lambda}{2\pi}
 \left(
 \begin{array}{cc}
 C_+^{-2}K_0\left(\lambda\sqrt{u}\right) &
 K_1\left(\lambda\sqrt{u}\right) \vspace{0.1cm} \\
 K_1\left(\lambda\sqrt{u}\right) &
 C_+^{2}K_0\left(\lambda\sqrt{u}\right)
 \end{array}\right),\qquad C_+=\left(\frac{m-E}{m+E}\right)^{1/4},
 \eb
 and $K_{0,1}(s)$ denote modified Bessel functions.

 Using integral representations for $K_{0,1}(s)$,
 one can write (\ref{deltafanmplane00}) in a different form. Namely, for any
 $\alpha\in\Rb$ such that $\ds |\varphi-\alpha|<\frac{\pi}{2}$, $\ds |\varphi'-\alpha|<\frac{\pi}{2}$
 we have
 \ben
 \Delta(z,z')=e^{-i\nu(\varphi-\varphi')}\frac{\lambda\sin\pi\nu}{4\pi^2}
 \int\limits_{-\infty}^{\infty}d\theta_1\int\limits_{-\infty}^{\infty}d\theta_2\;
 \;
 e^{-\lambda r\cosh(\theta_1+i(\alpha-\varphi))-\lambda
 r'\cosh(\theta_2+i(\alpha-\varphi'))}\times
 \ebn
 \ben
 \times\;\frac{e^{(3/2+\nu)(\theta_1-\theta_2)}}{e^{\theta_1-\theta_2}+1}
 \left(\begin{array}{c}
 C_+^{-1}e^{-\frac{\theta_1+i\alpha}{2}} \\
 C_+e^{\frac{\theta_1+i\alpha}{2}}
 \end{array}\right)\otimes
 \left(\begin{array}{c}
 C_+^{-1}e^{\frac{\theta_2+i\alpha}{2}} \\
 C_+e^{-\frac{\theta_2+i\alpha}{2}}
 \end{array}\right)^T.
 \ebn
 \textbf{Example}. For $0<\varphi<\pi$ and $0<\varphi'<\pi$, we may take
 $\ds \alpha=\frac{\pi}{2}$ so that
 \be\label{deltaplaneb0}
 \Delta(z,z')=e^{-i\nu(\varphi-\varphi')}\frac{\lambda\sin\pi\nu}{4\pi^2}
 \int\limits_{-\infty}^{\infty}d\theta_1\int\limits_{-\infty}^{\infty}d\theta_2\;
 \;
 e^{-\lambda\left(y\cosh\theta_1+ix\sinh\theta_1\right)-\lambda\left(y'\cosh\theta_2+ix'\sinh\theta_2\right)}\times
 \eb
 \ben
 \times\;\frac{e^{(3/2+\nu)(\theta_1-\theta_2)}}{e^{\theta_1-\theta_2}+1}
 \left(\begin{array}{c}
 C_+^{-1}e^{-{\theta_1}/{2}} \\
 i\,C_+e^{\,{\theta_1}/{2}}
 \end{array}\right)\otimes
 \left(\begin{array}{c}
 C_+^{-1}e^{\,{\theta_2}/{2}} \\
 -i\,C_+e^{-{\theta_2}/{2}}
 \end{array}\right)^T.
 \ebn
 As we will see a bit later, this formula makes the computation of zero-field form
 factors particularly simple.
 Note that similar expressions for the one-vortex Green function on the plane
 have already appeared in different papers (see, e.~g., \cite{marino,pacific}).

 \subsection{Two-point tau function and Painlev\'e V}
 In order to write the tau function as a Fredholm determinant,
 we will consider  free Dirac equation  in another
 gauge. Set the potential of the uniform magnetic field to be
 \be\label{ubfieldplane2}
 \mathcal{A}^{(B)}=-By\,dx\,,
 \eb
 so that the corresponding Dirac hamiltonian
 \ben
 \hat{H}^{(0)}_{tr}=\left(\begin{array}{cc}
 m & \!\!\!\!\!\!\!\!\partial_x-i\partial_y-iBy \\
 -\partial_x-i\partial_y+iBy  & \!\!\!\!\!\!\!\!-m
 \end{array}\right)
 \ebn
 commutes with the $x$-momentum operator $\hat{P}_x=-i\partial_x$.
 The eigenspace of $\hat{P}_x$ with momentum $p$ is spanned by the
 functions $g(p,y)e^{ipx}$. Let us look at the solutions of the
 partial Dirac equation
 \be\label{dirplanetr}
 \qquad(\hat{H}_p-E)g(p,y)=0,\qquad
 \hat{H}_p=\left(\begin{array}{cc}
 m &  \!\!\!\!\!\!\!\!-i(\partial_y-p+By) \\
 -i(\partial_y+p-By)  &  \!\!\!\!\!\!\!\!-m
 \end{array}\right).
 \eb
 As above, we assume that $E$ is real and $|E|<m$. It is
 convenient to choose two linearly independent solutions of
 (\ref{dirplanetr}) as follows:
 \begin{eqnarray}
 \label{phipmplane1}
 & B>0:\qquad &
 {\Phi}^{(\pm)}(p,y)=
 {\textstyle\left[\frac{1}{\sqrt{2\pi}}\,\Gamma\left(\frac{\lambda^2}{2B}+1\right)\right]^{1/2}}
 \left(\begin{array}{c}
 C_+^{-1}D_{-\frac{\lambda^2}{2B}-1}\left(\pm\sqrt{2B}\left(y-\frac{p}{B}\right)\right)
 \vspace{0.1cm}
 \\
 \pm iC_+\,D_{-\frac{\lambda^2}{2B}}\left(\pm\sqrt{2B}\left(y-\frac{p}{B}\right)\right)
 \end{array}
 \right),\vspace{0.1cm}\\
 \label{phipmplane2}
 &B<0: \qquad & {\Phi}^{(\pm)}(p,y)=
 {\textstyle\left[\frac{1}{\sqrt{2\pi}}\,\Gamma\left(\frac{\lambda^2}{2|B|}+1\right)\right]^{1/2}}
 \left(\begin{array}{c}
 C_+^{-1}D_{-\frac{\lambda^2}{2|B|}}\left(\pm\sqrt{2|B|}\left(y+\frac{p}{|B|}\right)\right)
 \vspace{0.1cm}
 \\
 \pm iC_+\,D_{-\frac{\lambda^2}{2|B|}-1}\left(\pm\sqrt{2|B|}\left(y+\frac{p}{|B|}\right)\right)
 \end{array}
 \right),\qquad\vspace{0.1cm}\\
 \label{phipmplane3}
 &B=0:\qquad &
 {\Phi}^{(\pm)}(p,y)=\frac{e^{\mp\sqrt{\lambda^2+p^2}\,y}}{\sqrt{2}}
 \left(\begin{array}{c}
 C_+^{-1} \left(1\pm \frac{p}{\sqrt{\lambda^2+p^2}}\right)^{1/2} \\
 \pm iC_+ \left(1\mp \frac{p}{\sqrt{\lambda^2+p^2}}\right)^{1/2}
 \end{array}\right).
 \end{eqnarray}
 Here, $D_{\alpha}(s)$  denotes the parabolic cylinder function
 and the constant $C_+$ in (\ref{phipmplane1}), (\ref{phipmplane2}) and (\ref{phipmplane3})
 is determined by (\ref{cpl01}), (\ref{cplplane}) and
 (\ref{zetab0}), correspondingly. Note that
 ${\Phi}^{(+)}(p,y)$ (${\Phi}^{(-)}(p,y)$) is square integrable as
 $y\rightarrow\infty$ (resp. $y\rightarrow-\infty$). These two
 solutions satisfy symmetry relations, analogous to
 (\ref{relsgstrip}):
  \be\label{relsgplane}
  \Phi^{(+)}(p,y)=\sigma_z  \,{\Phi}^{(-)}(-p,-y),\qquad\qquad
   \Phi^{(\pm)}(p,y)=\sigma_z  \,\overline{\Phi^{(\pm)}(p,y)}\,.
 \eb
 The normalization in (\ref{phipmplane1})--(\ref{phipmplane3}) was chosen
 so that in all three cases
  \be\label{detgplane}
 \mathrm{det}\left(\Phi^{(+)}(p,y),\Phi^{(-)}(p,y)\right)= -i\,.
 \eb

 We now adapt the reasoning of Subsection~\ref{bpr} to flat
 space. Consider a line $\mathcal{L}_{y^{(0)}}=\left\{(x,y)\in\Rb^2\,|\,y=y^{(0)}\right\}$
 and an arbitrary $\Cb^2$-valued function $g_{y^{(0)}}(x)\in
 H^{1/2}(\mathcal{L}_{y^{(0)}})$,
 written as Fourier integral
  \be\label{fourier_dec_plane}
 g_{y^{(0)}}(x)=\int\nolimits_{-\infty}^{\infty}dp\;\;
 g(p,y^{(0)})\,e^{ipx}.
 \eb
 Decompose Fourier transform $g(p,y^{(0)})$ as follows:
 \ben
 g(p,y^{(0)})=\tilde{g}_+(p,y^{(0)})\Phi^{(+)}(p,y^{(0)})+
 \tilde{g}_-(p,y^{(0)})\Phi^{(-)}(p,y^{(0)}),
 \ebn
 where $\Phi^{(\pm)}(p,y^{(0)})$ denote the functions
 defined by (\ref{phipmplane1}), (\ref{phipmplane2}) or
 (\ref{phipmplane3}), depending on the value of $B$. The formula
 ({\ref{detgplane}) and symmetry relations (\ref{relsgplane}) imply
 that
  \be\label{hpmcoordsplane}
 \tilde{g}_{\pm}(p,y^{(0)})=
 \mp\,{i}
 \left(\Phi^{(\mp)}(p,y^{(0)})\right)^{\dag}\sigma_x\;
 g(p,y^{(0)})\,.
 \eb
 Recall that $\tilde{g}_{+}(p,y^{(0)})$ and $\tilde{g}_{-}(p,y^{(0)})$  can be thought of as
 coordinates in the spaces of boundary values of solutions of the
 free Dirac equation $(\hat{H}^{(0)}_{tr}-E)\psi=0$ in the half planes $y>y^{(0)}$ and
 $y<y^{(0)}$.

 It is now straightforward to write down the analogs of the
 Propositions~\ref{ppp1} and \ref{ppp2}:
  \begin{prop}\label{ppp3}
 Let us consider a strip $\ds \mathcal{S}=\left\{(x,y)\in\Rb^2\,|\,
 y^{(L)}<y<y^{(R)}\right\}$. Suppose that 
 $\psi\in H^{1/2}(\partial\mathcal{S})$ can  be continued to
 $\mathcal{S}$ as a solution of the free Dirac equation
 $\left(\hat{H}^{(0)}_{\text{tr}}-E\right)\psi=0$. Then
 \be\label{stripprojwithoutplane}
    \left(
  \begin{array}{c}
 \tilde{\psi}_{L,-}(p)  \\ \tilde{\psi}_{R,+}(p)
  \end{array}\right)=
  \left(
  \begin{array}{cc}
  0 & 1 \\
  1 & 0
  \end{array}
  \right)
  \left(
  \begin{array}{c}
 \tilde{\psi}_{L,+}(p)  \\ \tilde{\psi}_{R,-}(p)
  \end{array}\right).
 \eb
 \end{prop}
  \begin{prop}\label{ppp4}
 Assume that the strip $\mathcal{S}$ contains one branching
 point $a_0$ (i.~e. $y^{(L)}<a_{0y}<y^{(R)}$) and introduce a horizontal branch cut
 $\ell=\bigl(-\infty+ia_{0y},a_{0x}+ia_{0y}\bigr]$. Suppose
 that $\psi\in  H^{1/2}(\partial\mathcal{S})$ is the boundary value
 of a multivalued solution of the free Dirac equation on
 $\mathcal{S}\backslash\ell$, which is characterized by the monodromy $e^{2\pi i
 \nu}$ at the point ${a_0}$. Then
 \ben
    \left(
  \begin{array}{c}
 \tilde{\psi}_{L,-}(p)  \\ \tilde{\psi}_{R,+}(p)
  \end{array}\right)=
  \left(
  \begin{array}{cc}
  \hat{\alpha}_{\mathcal{S}}({a_0}) & \hat{\beta}_{\mathcal{S}}({a_0}) \\
  \hat{\gamma}_{\mathcal{S}}({a_0}) & \hat{\delta}_{\mathcal{S}}({a_0})
  \end{array}
  \right)
  \left(
  \begin{array}{c}
 \tilde{\psi}_{L,+}(p)  \\ \tilde{\psi}_{R,-}(p)
  \end{array}\right),
 \ebn
 where
 \begin{eqnarray}
 \label{alphaplane}
 \left(\hat{\alpha}_{\mathcal{S}}({a_0})\tilde{\psi}_{L,+}\right)(p)&=&
 \int\nolimits_{-\infty}^{\infty}
 \dot{\Delta}_-^{({a_0},\nu)}(p,q)\;
 \tilde{\psi}_{L,+}(q)\;dq,\\
 \label{betaplane}
 \left(\hat{\beta}_{\mathcal{S}}({a_0})\tilde{\psi}_{R,-}\right)(p)&=&
 \int\nolimits_{-\infty}^{\infty}
 \dot{G}^{({a_0},\nu)}_{+}(p,q)\;
 \tilde{\psi}_{R,-}(q)\;dq,\\
 \label{gammaplane}
 \left(\hat{\gamma}_{\mathcal{S}}({a_0})\tilde{\psi}_{L,+}\right)(p)&=&
 \int\nolimits_{-\infty}^{\infty}
 \dot{G}^{({a_0},\nu)}_{-}(p,q)\;
 \tilde{\psi}_{L,+}(q)\;dq,\\
 \label{deltaplane}
 \left(\hat{\delta}_{\mathcal{S}}({a_0})\tilde{\psi}_{R,-}\right)(p)&=&
 \int\nolimits_{-\infty}^{\infty}
 \dot{\Delta}_+^{({a_0},\nu)}(p,q)\;
 \tilde{\psi}_{R,-}(q)\;dq.
 \end{eqnarray}
 and
 \begin{eqnarray}
 \label{deltapmplane1}
 \dot{\Delta}_{\pm}^{({a_0},\nu)}(p,q)&=&
 \frac{1}{2\pi}
 \int\limits_{-\infty}^{\infty}\int\limits_{-\infty}^{\infty}dx\,dx'\;
 e^{-ipx+iqx'}\left(\Phi^{(\mp)}(p,y)\right)^{\dag}\sigma_x\;
 \dot{\Delta}^{(a_0,\nu)}_{{tr}}(z,z')\Bigl|_{y,y'\gtrless a_{0y}}\sigma_x
 \Phi^{(\mp)}(q,y'),\\
  \label{gpmplane2}
 \dot{G}_{\pm}^{({a_0},\nu)}(p,q)&=&
 -\frac{1}{2\pi}
 \int\limits_{-\infty}^{\infty}\int\limits_{-\infty}^{\infty}dx\,dx'\;
 e^{-ipx+iqx'}\left(\Phi^{(\pm)}(p,y)\right)^{\dag}\sigma_x\;
 \dot{G}^{(a_0,\nu)}_{{tr}}(z,z')\Bigl|_{
 \substack{ \scriptstyle y\lessgtr a_{0y},\\ \scriptstyle y'\gtrless a_{0y} }}\sigma_x
 \Phi^{(\mp)}(q,y').\qquad
 \end{eqnarray}
 Here, $ \dot{G}^{(a_0,\nu)}_{{tr}}(z,z')$ denotes the
 Green function of the Dirac hamiltonian $\hat{H}^{(0)}_{tr}$ on the plane with one branching point
 $a_0$, and $\dot{\Delta}^{(a_0,\nu)}_{{tr}}(z,z')=
  \dot{G}^{(a_0,\nu)}_{{tr}}(z,z')-
  {G}^{(0)}_{{tr}}(z,z')$.
 \end{prop}

 The definition of the tau function of the Dirac hamiltonian on
 the plane with two branch points $a_1$ and $a_2$ is also completely
 analogous to the disk case. Repeating the arguments of
 Subsection~\ref{taudef}, one ends up with the following Fredholm
 determinant representation:
 \be\label{tauplane}
 \tau(a)=\mathrm{det}\left(\mathbf{1}-\hat{\alpha}({a_2})\hat{\delta}({a_1})\right),
 \eb
 where $\hat{\alpha}(a_2)$ and $\hat{\delta}({a_1})$ are given by
 (\ref{alphaplane}), (\ref{deltaplane}). As above, the fact that
 the tau function depends only on the distance between the points
 $a_1$ and $a_2$ allows us to choose $a_1=0$, $a_2=t+i0$ ($t\in\Rb^+$), and the
 invariance of $\hat{H}^{(0)}_{tr}$ with respect to $x$-translations reduces the problem
 of calculation of $\tau(a)$ to finding the form factors
 $\dot{\Delta}_{\pm}^{({0},\nu)}(p,q)$. Finally, the symmetry of the free Dirac
 hamiltonian $\hat{H}^{(0)}_{tr}$ combined with the relations (\ref{relsgplane})
 implies that
 \be\label{ffsymmetryplane}
 \dot{\Delta}_{\pm}^{(0,\nu)}(p,q)=
 \overline{\dot{\Delta}_{\pm}^{(0,\nu)}(p,q)}=
 \dot{\Delta}_{\pm}^{(0,\nu)}(q,p)=
 \dot{\Delta}_{\mp}^{(0,\nu)}(-p,-q).
 \eb
 The form factors $\dot{\Delta}_{\pm}^{(0,\nu)}(p,q)$ are determined by
 the relation (\ref{deltapmplane1}) or by the equivalent formula
  \be\label{deltapmplane2}
 \frac{1}{2\pi}\int\limits_{-\infty}^{\infty}
  \int\limits_{-\infty}^{\infty}dx\, dx'\;e^{-ipx+iqx'}
  \dot{\Delta}^{(0,\nu)}_{{tr}}(z,z')\biggl|_{y,y'\gtrless 0}=
  \dot{\Delta}_{\pm}^{(0,\nu)}(p,q)\;
  \Phi^{(\pm)}(p,y)\otimes
  \left(\Phi^{(\pm)}(q,y')\right)^{\dag}.
 \eb
 We remark that $\dot{\Delta}^{(0,\nu)}_{{tr}}(z,z')$ and the
 function $\Delta(z,z')$ defined by
 (\ref{deltafanmplane})--(\ref{cplplane}) (for $B\neq0$) or by
 (\ref{deltafanmplane00})--(\ref{zetab0}) (for $B=0$) are related
 by
 \be\label{gtrplane}
 \dot{\Delta}^{(0,\nu)}_{{tr}}(z,z')=e^{i\nu(\varphi-\varphi')}\times
 e^{\frac{iB}{2}\left(xy-x'y'\right)}\times
 \Delta(z,z'),\qquad
 \varphi\in (-\pi,\pi).
 \eb
 Here, the first factor corresponds to a singular gauge
 transformation removing the AB field, and the second one
 comes from the change of the vector potential of the uniform
 magnetic field from (\ref{ubfieldplane}) to
 (\ref{ubfieldplane2}).

 Also note that $y$ and $y'$ in (\ref{deltapmplane1}) and
 (\ref{deltapmplane2}) can be chosen arbitrarily. Analogous
 observation  in the disk case
 allowed us to obtain a more
 explicit representation for $\dot{\Delta}_{\pm}^{(0,\nu)}(p,q)$ by analyzing the asymptotics
 of a relation similar to (\ref{deltapmplane2}) near the disk
 boundary. For $B\neq0$ the asymptotic analysis of the LHS of (\ref{deltapmplane2})
 as $y,y'\rightarrow\pm\infty$ becomes rather complicated and we have not managed to repeat
 the above trick  in this case.
 However, when both $p$, $q$ are positive or negative,
 one can choose $y$ and $y'$ in such a way
 that the arguments of parabolic cylinder functions in the RHS  of
 one of the relations
 (\ref{deltapmplane2}) are equal to zero. This leads to a simpler
 (than (\ref{deltapmplane2}) for general $y,y'$) representation of
 $\dot{\Delta}_{\pm}^{(0,\nu)}(p,q)$.\vspace{0.1cm}\\
 \textbf{Example}. Assume that $B>0$, $p>0$, $q>0$ and $\ds\Theta=-\frac{\pi}{2}$. Then, setting
  in (\ref{deltapmplane2}) $\ds y=\frac{p}{B}$,  $\ds
  y'=\frac{q}{B}$ and taking into account (\ref{gtrplane}), one finds
  a triple integral representation for
  $\dot{\Delta}_{+}^{(0,\nu)}(p,q)$:
  \ben
  \dot{\Delta}_{+}^{(0,\nu)}(p,q)=\sqrt{2B}\cdot\frac{pq}{B^2}\;\frac{{2}^{1+{\lambda^2}/{2B}}
  \left[\Gamma\left(\frac{\lambda^2}{4B}+1\right)\right]^2}{(2\pi)^{5/2}}
  \frac{\sin\pi\nu}{\pi} \int\limits_{-\infty}^{\infty}d\theta
  \int\limits_{0}^{\pi}d\varphi\int\limits_0^{\pi}d\varphi'\;
  \frac{e^{(1+\nu)(\theta+i(\varphi-\varphi'))}}{e^{\theta+i(\varphi-\varphi')}+1}\;\times
  \ebn
  \ben
  \times\;\frac{1}{\sin^2\varphi\sin^2\varphi'}\;\exp\left\{-\frac{1}{4B}\left(\frac{p^2}{\sin^2\!\varphi}+
  \frac{q^2}{\sin^2\!\varphi'}+\frac{2{p}{q}\,e^{\theta}}{\sin\varphi\sin\varphi'}
  +2ip^2\mathrm{ctg}\,\varphi-2iq^2\mathrm{ctg}\,\varphi'\right)\right\}\times
  \ebn
  \ben
  \times\;U\left(\frac{\lambda^2}{2B}+1,1,\frac{1}{2B}\left(\frac{p^2}{\sin^2\!\varphi}+
  \frac{q^2}{\sin^2\!\varphi'}+\frac{2{p}{q}\cosh{\theta}}{\sin\varphi\sin\varphi'}\right)\right).
  \ebn\vspace{0.2cm}

  Much simpler results can be obtained for $B=0$. In this case,
  it is convenient to introduce instead of the momentum $p$ a rapidity variable $\theta_p$
  defined by
  \ben
  p=\lambda\sinh\theta_p,\qquad
  \sqrt{\lambda^2+p^2}=\lambda\cosh\theta_p\,.
  \ebn
  Partial waves (\ref{phipmplane3}) can then be written as
  \ben
  {\Phi}^{(\pm)}(p,y)=\frac{e^{\mp\lambda y\cosh\theta_p}}{\sqrt{2\cosh\theta_p}}
 \left(\begin{array}{c}
 C_+^{-1} e^{\pm\theta_p/2} \\
 \pm iC_+ e^{\mp\theta_p/2}
 \end{array}\right).
  \ebn
  Set $\ds \Theta=-\frac{\pi}{2}$ and substitute the
  representation (\ref{deltaplaneb0}) for $\Delta(z,z')$ (recall that it is valid
  for $y,y'>0$) into (\ref{deltapmplane2})--(\ref{gtrplane}).
  After elementary integration over $x$ and $x'$ in the LHS of
  (\ref{deltapmplane2}) we find
  \be\label{ffplanemp2}
  \dot{\Delta}_{\pm}^{(0,\nu)}(p,q)=\frac{1}{\lambda\sqrt{\cosh\theta_p\cosh\theta_q}}\;
  \frac{\sin\pi\nu}{\pi}\;\frac{e^{\mp(1+\nu)(\theta_p+\theta_q)}}{2\cosh\frac{\theta_p+\theta_q}{2}}\,.
  \eb
  Similarly, for $\ds \Theta=\frac{\pi}{2}$ one obtains
  \be\label{ffplanepp2}
  \dot{\Delta}_{\pm}^{(0,\nu)}(p,q)=-\frac{1}{\lambda\sqrt{\cosh\theta_p\cosh\theta_q}}\;
  \frac{\sin\pi\nu}{\pi}\;\frac{e^{\mp\nu(\theta_p+\theta_q)}}{2\cosh\frac{\theta_p+\theta_q}{2}}\,.
  \eb
  It should be mentioned that the formulas equivalent to (\ref{ffplanemp2})--(\ref{ffplanepp2})
  were first obtained by Schroer and Truong \cite{schroer}. In a context similar to ours,
  they were rediscovered  by Palmer in \cite{pacific}.

  We finally comment on the limiting form of the equation
  (\ref{pvi}) in flat space. Introduce
  $s=R^{-2}{t}$ and let $R\rightarrow \infty$.
  Then, setting $w(s)=1-y(t)$ and
  using asymptotic behaviour  of the parameters $\beta$ and $\delta$ in (\ref{pvipars})
  \ben
  \beta-\delta\simeq R^2
  \gamma',\qquad\delta\simeq R^4\delta',
  \ebn
  \be\label{gammadelta}
  \gamma'=\frac{m^2-E^2+B(1+\nu_1+\nu_2)}{2},\qquad
  \delta'=-\frac{B^2}{8}\,,
  \eb
   it is straightforward
  to check that $y(t)$ satisfies Painlev\'e~V equation
  \be\label{pv}
  \frac{d^2y}{dt^2}=\left(\frac{1}{2y}+\frac{1}{y-1}\right)\left(\frac{dy}{dt}\right)^2-
  \frac{1}{t}\frac{dy}{dt}+\frac{(y-1)^2}{t^2}\left(\alpha'y+\frac{\beta'}{y}\right)+
  \frac{\gamma'y}{t}+\frac{\delta' y(y+1)}{y-1}
  \eb
  with parameters $\ds\alpha'=\alpha=\frac{\lambda^2}{2}$, $\beta'=0$ and $\gamma'$,
  $\delta'$ defined by (\ref{gammadelta}). The equation
  (\ref{taurelpvi}) in the planar limit transforms into
  \be\label{zetasmj}
  t\frac{d}{dt}\ln\tau(t)=\frac{t^2}{4y(y-1)^2}\left(\frac{dy}{dt}\right)^2-\frac{\lambda^2y}{4}
 +\frac{\eta\theta}{2}\frac{ty}{y-1}-\frac{\eta^2}{4}\frac{t^2y}{(y-1)^2},
  \eb
  where $\ds\eta=-\frac{B}{2}$ and $\eta(\theta+1)=\gamma'$.
  The RHS of (\ref{zetasmj}) coincides,
  up to addition of a constant,
  with the Okamoto hamiltonian for Painlev\'e~V equation (\ref{pv}) and
  thus the $\tau$-function (\ref{tauplane}) associated to the
  Dirac operator in flat space is very simply related to  Painlev\'e~V
  $\tau$-function. Although one could expect a similar relation in
  the case of Painlev\'e~VI and Dirac operator on the
  hyperbolic disk, this appears not to be the case \cite{beatty}.

 \section{Concluding remarks}
 An important problem left outside the scope of this paper
 concerns the short-distance ($s\rightarrow0$) behaviour of the
 PBT $\tau$-function. Extending the conjecture of \cite{doyon}
 to the case $b\neq0$,
 one could assume that $\tau(s)$  coincides with the
 two-point correlator of twist fields in the Dirac theory in
 a more general classical background (Poincar\'e metric $+$
 uniform magnetic field). External fields
 drastically change the infrared asymptotics of the correlation
 function, but they should not affect the exponent $\sigma$ in its conformal behaviour
  $\tau(s\rightarrow0)\simeq C s^{\sigma}$. To
 prove this rigorously, one would require a generalization of the
 connection formulas for Painlev\'e~VI \cite{guzzetti2,jimbo} to non-generic values of parameters (recall that
 in our case $\gamma=0$). These formulas, however, are not known except for
 the special case of PVI with $\beta=\gamma=0$, $\delta=\frac{1}{2}$
 \cite{dubrovin,guzzetti}.

 The  theory of
 Painlev\'e equations does not provide any answer for the
 coefficient $C$. For the $\tau$-function arising
 in the scaling limit of the two-dimensional Ising model, this constant was extracted from a
 careful asymptotic analysis of the corresponding Fredholm determinant \cite{tracy}.
 Although it seems hopeless to repeat such an analysis with the determinant (\ref{frere}), one
 could try to obtain  $C$ using QFT arguments: it may
 be expressed in terms of the vacuum expectation values of twist
 fields, which can be computed by the method of angular quantization \cite{doyon,lukyanov}.

 It is curious to note
 that another one-parameter class of solutions of the
 Painlev\'e~VI equation with one singular parameter arises in the
 representation theory of the infinite-dimensional unitary group
 \cite{borodin}. However, its relation to PVI transcendents studied in the present
 paper remains rather obscure.

 \renewcommand{\theequation}{A.\arabic{equation}}
 \setcounter{equation}{0}
 \addcontentsline{toc}{section}{Appendix}
 \section*{Appendix}
 We wish to show that the logarithmic derivative of the
 $\tau$-function is invariant under the joint
 $SU(1,1)$-transformation of the branch point positions:
 \ben
 a_j\mapsto \frac{\alpha
 a_j+\beta}{\bar{\beta}a_j+\bar{\alpha}},\qquad
 \left(\begin{array}{cc}
 \alpha & \beta \\
 \bar{\beta} & \bar{\alpha}
 \end{array}\right)\in SU(1,1).
 \ebn
 It is sufficient to prove that the 1-form (\ref{taufinal}) is annihilated
 by two vector fields
 \begin{eqnarray*}
 X&=&\sum_j
 \left(a_j\partial_{a_j}-\bar{a}_j\partial_{\bar{a}_j}\right),\\
 Y&=&\sum_j\left[(1-a_j^2)\partial_{a_j}+(1-\bar{a}_j^2)\partial_{\bar{a}_j}\right],
 \end{eqnarray*}
 whose integral curves are orbits of a compact and a noncompact subgroup of
 $SU(1,1)$.

 Let us first apply $X$ to $d\ln\tau$. Substituting
 (\ref{diagajj})--(\ref{diagdjj}) into the resulting expression, one finds after some
 simplification
 \be\label{twosums}
 X(d\ln\tau)=\sum_j\sum_{k\neq j}\frac{1}{1-|a_j|^2}\left(a_j e_{jk} a^j_{1/2,k}+
 \bar{a}_j h_{jk} d^j_{1/2,k}\right)+
 \sum_{j,k}\frac{1}{1-|a_j|^2}\left(a_j\bar{a}_k f_{jk} c^j_{1/2,k}+
 \bar{a}_j a_k g_{jk} b^j_{1/2,k}\right).
 \eb
 Two sums  in (\ref{twosums}) are separately equal to zero. For example, the
 first one can be rewritten as
 \begin{eqnarray*}
 &\;&\sum_j\sum_{k\neq j}\Bigl\{e_{jk}(\mathbf{a}_1
 A)_{kj}+h_{jk}(\mathbf{d}_1\bar{A})_{kj}\Bigr\}=\sum_{j,k}\Bigl\{
 [\mathbf{a}_1,A]_{jk}(\mathbf{a}_1
 A)_{kj}-[\mathbf{d}_1,\bar{A}]_{jk}(\mathbf{d}_1
 \bar{A})_{kj}\Bigr\}=\\
 &=&\frac{1}{2}\,\mathrm{Tr}\Bigl\{[\mathbf{a}_1,A]^2-[\mathbf{d}_1,\bar{A}]^2\Bigr\}=
 \frac{1}{2}\,\mathrm{Tr}\Bigl\{(e-\Lambda+b\mathbf{1})^2-(h-\Lambda+b\mathbf{1})^2\Bigr\}=\\
 &=&
 \frac{1}{2}\,\mathrm{Tr}\Bigl\{fg-gf-2(\Lambda-b\mathbf{1})(e-h)\Bigr\}=0.
 \end{eqnarray*}
 Besides the relations (\ref{relations}) and (\ref{symrel2}), in the above we have used the fact that
 the diagonal parts of the commutators $[\mathbf{a}_1,A]$ and $[\mathbf{d}_1,\bar{A}]$
 and of the difference $e-h$
 are equal to zero. Similarly, the second sum in (\ref{twosums})
 gives
 \ben
 \mathrm{Tr}\Bigl(f\bar{A}\mathbf{c}_1
 A+gA\mathbf{b}_1\bar{A}\Bigr)=
 \mathrm{Tr}\Bigl((A\mathbf{b}_1\bar{A}-\mathbf{b}_1)\bar{A}\mathbf{c}_1A+
 (\mathbf{c}_1-\bar{A}\mathbf{c}_1A)A\mathbf{b}_1\bar{A}\Bigr)=0.
 \ebn

 Next consider the action of $Y$. We get
 \begin{eqnarray}
 \nonumber Y(d\ln\tau)&=&\sum_j\sum_{k\neq j}\Bigl\{
 e_{jk}\left(\mathbf{a}_1(1-A^2)\right)_{kj}-
 h_{jk}\left(\mathbf{d}_1(1-\bar{A}^2)\right)_{kj}\Bigr\}+\sum_j(a_j+\bar{a}_j)m_+(\tilde{\nu}_j-1/2,b)+\\
 \label{threesums}&+&\sum_{j,k}\Bigl\{f_{jk}(\bar{A}\mathbf{c}_1-\bar{A}\mathbf{c}_1A^2)_{kj}-
 g_{jk}(A\mathbf{b}_1-A\mathbf{b}_1\bar{A}^2)_{kj}\Bigr\}.
 \end{eqnarray}
 The first sum in (\ref{threesums}) can be transformed into
 \begin{eqnarray*}
 &\;&\sum_{j}\sum_{k\neq
 j}\Bigl\{-e_{jk}(\mathbf{a}_1A^2)_{kj}+h_{jk}(\mathbf{d}_1\bar{A}^2)_{kj}\Bigr\}=
 -\mathrm{Tr}\Bigl([\mathbf{a}_1,A]^2 A+[\mathbf{d}_1,\bar{A}]^2
 \bar{A}\Bigr)=
 \\
 &=&-\mathrm{Tr}\Bigl(e^2A+h^2\bar{A}-2(\Lambda-b\mathbf{1})(eA+h\bar{A})+(\Lambda-b\mathbf{1})^2(A+\bar{A})\Bigr)=\\
 &=&-\mathrm{Tr}\Bigl(e^2A+h^2\bar{A}-(\Lambda-b\mathbf{1})^2(A+\bar{A})\Bigr),
 \end{eqnarray*}
 while the third one gives
 \ben
 \mathrm{Tr}\Bigl(f(\bar{A}\mathbf{c}_1-\mathbf{c}_1A+gA)-g(A\mathbf{b}_1-\mathbf{b}_1\bar{A}-f\bar{A})\Bigr)=
 \mathrm{Tr}\bigl(fgA+gf\bar{A}\bigr).
 \ebn
 Summing up the three contributions in (\ref{threesums}) and using (\ref{symrel2}) once
 again, one finds that $d\ln\tau$ is invariant under the flow  of $Y$.
 
\end{document}